\documentclass[fleqn,10pt]{wlscirep}
\usepackage[utf8]{inputenc}
\usepackage[T1]{fontenc}
\usepackage{graphicx}% Include figure files
\usepackage{dcolumn}% Align table columns on decimal point
\usepackage{bm}% bold math
\usepackage{siunitx}
\usepackage{subcaption}
\usepackage{multirow}
\usepackage{xcolor}
\usepackage{makecell}
\usepackage{float}
\usepackage{comment}
\usepackage[normalem]{ulem}
\usepackage{braket}
\usepackage[version=4]{mhchem}

\title{Control and Readout of a 13-level Trapped Ion Qudit}

\author[1]{Pei Jiang Low}
\author[1]{Brendan White}
\author[1,*]{Crystal Senko}
\affil[1]{University of Waterloo, Physics and Astronomy, Waterloo, N2L3G1, Canada}
\affil[*]{csenko@uwaterloo.ca}

\begin{abstract}
To implement useful quantum algorithms which demonstrate quantum advantage, we must scale currently demonstrated quantum computers up significantly.
Leading platforms such as trapped ions face physical challenges in including more information carriers.
A less explored avenue for scaling up the computational space involves utilizing the rich energy level structure of a trapped ion to encode multi-level qudits rather than two-level qubits.
Here we show control and single-shot readout of qudits with up to 13 computational states, using protocols which can be extended directly to manipulate qudits of up to 25 levels in our chosen information host, \ce{^{137}Ba^{+}}.
This represents more than twice as many computational states per qudit compared with prior work in trapped ions\cite{Ringbauer2022}.
In addition to the preparation and readout protocols we demonstrate, universal quantum computation requires other quantum logic primitives such as entangling gates. These primitives have been demonstrated for lower qudit dimensions and can be directly generalized to the higher dimensions we employ. Hence, our advance opens an avenue towards using high-dimensional qudits for large-scale quantum computation.
We anticipate efficiently utilizing available energy states in a trapped ion to play a significant and complementary role in tackling the challenge in scaling up the computational space of a trapped ion quantum computer.
A qudit architecture also offers other practical benefits, which include affording relaxed fault tolerance thresholds for quantum error correction \cite{Campbell-Anwar-Browne-2012, Campbell-2014, Andrist-Wootton-Katzgraber-2015, Hutter2015, Watson2015}, providing an avenue for efficient quantum simulation of higher spin systems \cite{Senko2015, Barbara2022}, and more efficient qubit gates \cite{Lanyon-et-al-2008, Ralph-Resch-Gilchrist-2007}.

\end{abstract}
\begin{document}

\flushbottom
\maketitle
\thispagestyle{empty}

\section*{Main}

Realizing quantum computation involves encoding quantum information in a physical system, such as atomic energy states, quantized states of superconducting circuits, and polarization states of photons \cite{Ladd2010}.
Trapped ions are a promising platform because of the high fidelities of the quantum operations \cite{Bruzewicz2019}, where gate errors below the quantum error correcting code threshold have been empirically achieved \cite{Gaebler2016}.
Another attractive feature of trapped ions is the certainty that all ions are identical to each other by nature.
This mitigates the increasing complexities of hardware calibrations required when scaling up a system with non-homogeneous quantum information carriers \cite{Philips2022}.

Conventional quantum computing approaches encode two computational states in a quantum information unit, called a qubit.
However, most platforms naturally exhibit more than two quantum states per information carrier, and it is not obvious why only two states should be utilized.
By encoding a general number of computational states, $d$, one defines a qudit.
Scaling up to a higher value of $d$ provides a direct increase of the computational Hilbert space given a constraint on the number of information carriers.
This constraint is relevant to trapped ions, as there are technical limitations to the number of ions which can be coherently controlled in a single ion chain \cite{Bruzewicz2019}.
High-dimensional qudit encodings can therefore complement other trapped ion scaling efforts, such as the use of ion trap arrays \cite{Kielpinski2002} or photonic interconnects \cite{Monroe2014}, to further increase the computational Hilbert space.
There are other advantages of a qudit quantum system, which include a more relaxed quantum error correction threshold \cite{Campbell-Anwar-Browne-2012, Campbell-2014, Andrist-Wootton-Katzgraber-2015, Hutter2015, Watson2015}, more direct higher spin quantum simulations \cite{Senko2015, Barbara2022}, and more efficient qubit gates \cite{Lanyon-et-al-2008, Ralph-Resch-Gilchrist-2007}.

Higher-dimensional atomic qudits have received increasing attention in recent years \cite{Smith2013, Senko2015, Leupold2018, Malinowski2018, Hrmo2023}, including a demonstration of universal quantum computing with 5-level trapped ion qudits \cite{Ringbauer2022}.
In this work, we demonstrate state preparation and measurement (SPAM) of a 13-level qudit, which more than doubles the number of qudit levels implemented in a single trapped ion \cite{Ringbauer2022}.
This advance is made possible by exploiting the abundant stable and metastable energy levels in \ce{^{137}Ba^{+}}, which enables qudit encodings of up to 25 levels with the protocol presented in this work.
The richer energy level structure of \ce{^{137}Ba^{+}} comes with non-trivial complexities in understanding the transition frequencies and strengths, which we resolve by constructing predictive theoretical models that match empirical data.
This understanding is crucial for any researchers aspiring to use \ce{^{137}Ba^{+}} or other ion species with similar energy level structures for high-dimensional qudit work.
We demonstrate an average 13-level SPAM fidelity of $91.7 \pm 0.3 \%$, and the major sources of errors are technical with known solutions.
The calibration times for the SPAM parameters do not increase with $d$ for $d \ge 6$ with the protocols that we present in this work.
Our results demonstrate the feasibility of efficiently utilizing high numbers of trapped ion energy levels, and are a stepping stone towards high-dimensional qudit quantum computing.

\subsection*{\label{sec:Ba-137_ion}Energy Levels and Qudit Encoding}

\begin{figure*}
    \centering
    \tabskip=0pt
\valign{#\cr
  \hbox{%
    \begin{subfigure}{.53\textwidth}
    \centering
    \includegraphics[width=\textwidth]{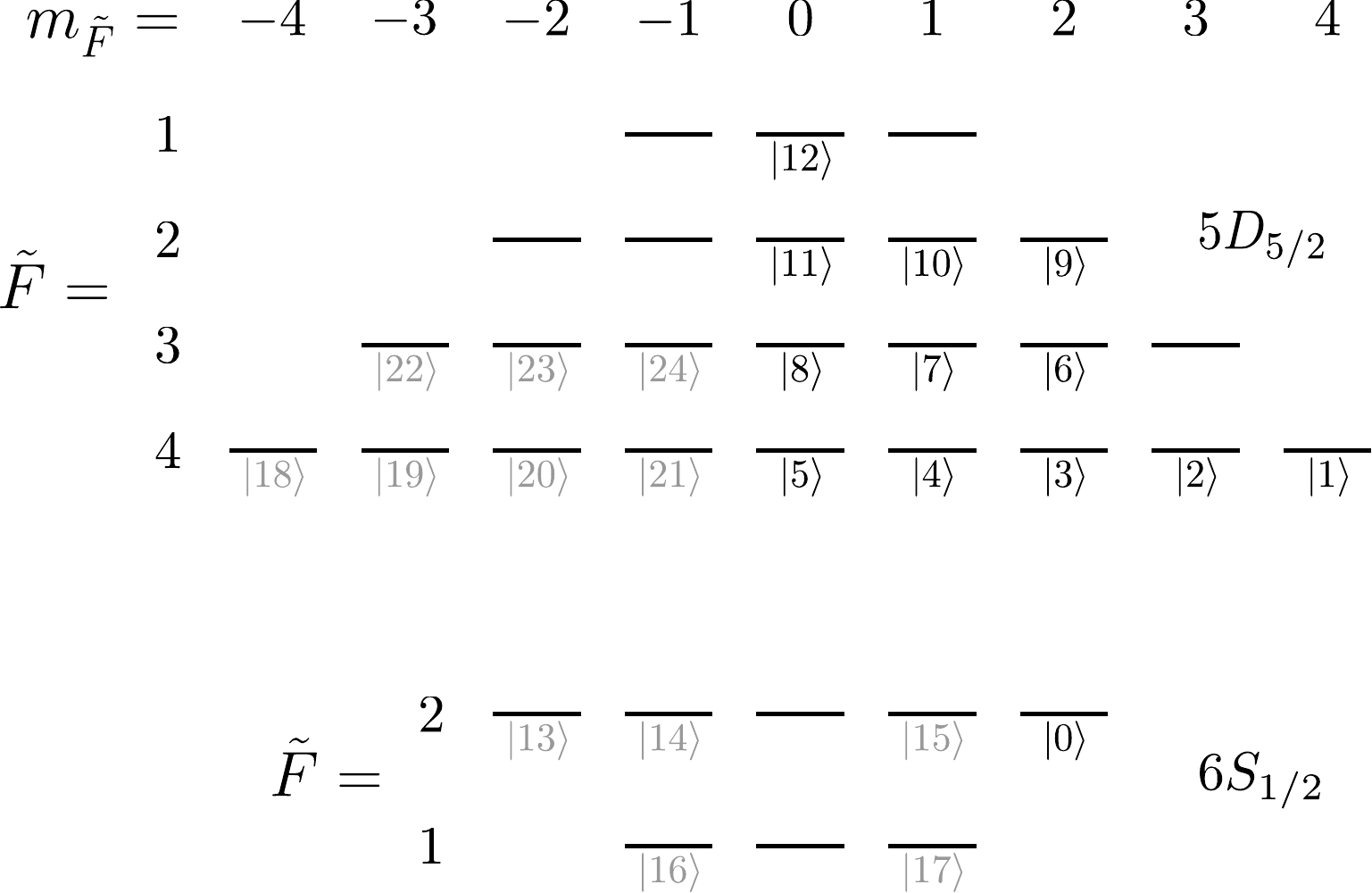}
    \caption{}
    \label{fig:QuditEncoding}
    \end{subfigure}%
  }\vfill
  \hbox{%
    \begin{subfigure}{.55\textwidth}
    \centering
    \includegraphics[width=0.75\textwidth]{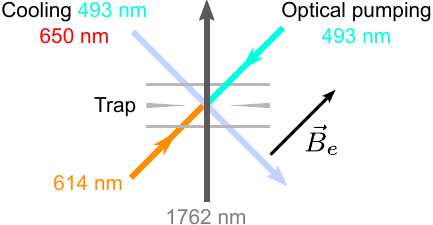}
    \caption{}
    \label{fig:BeamOrientationSimple}  
    \end{subfigure}%\
  }\cr
  \hbox{%
    \begin{subfigure}[b]{.45\textwidth}
    \centering
    \includegraphics[width=\textwidth]{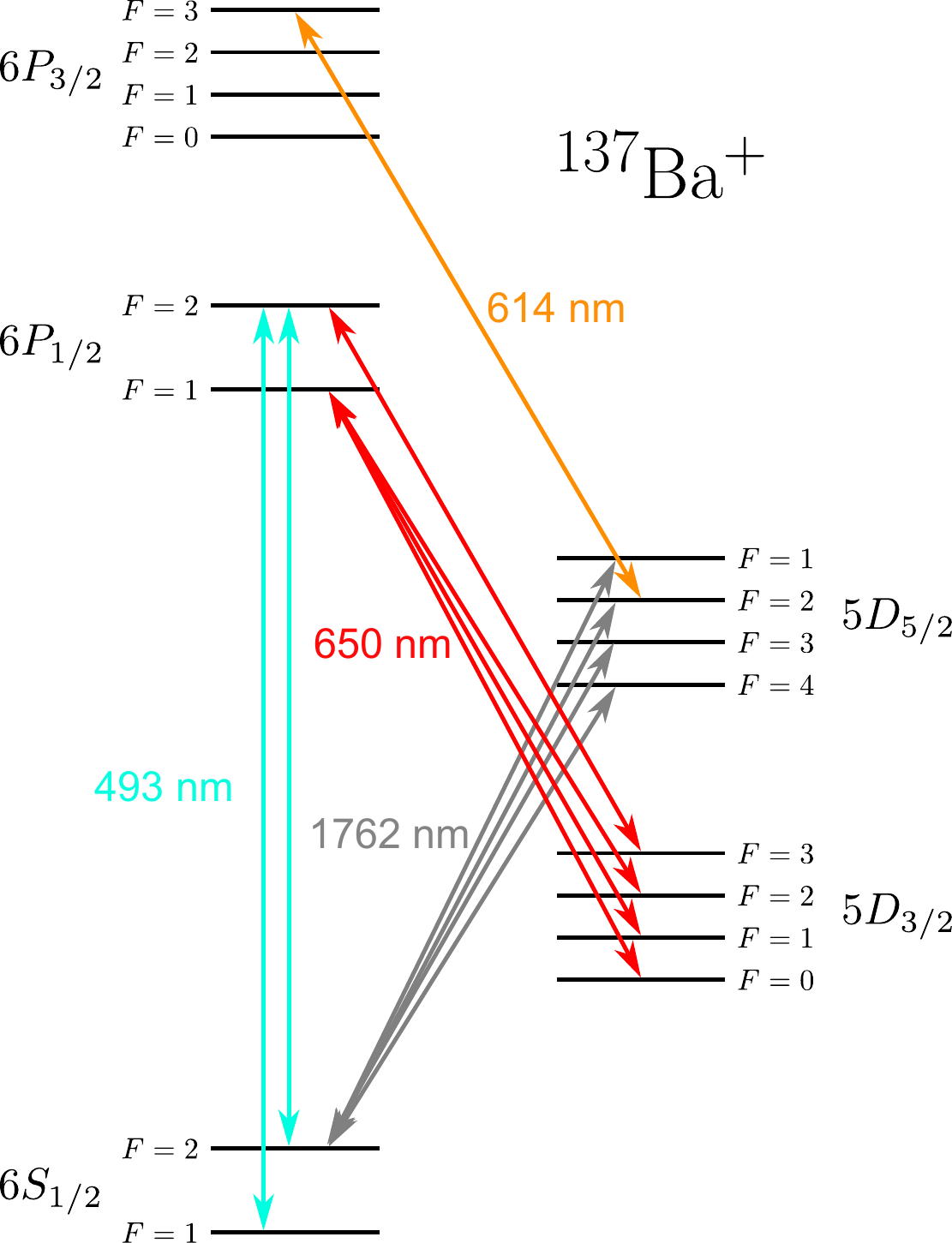}
    \caption{}
    \label{fig:Ba_General_Energy_Levels}
    \end{subfigure}%
  }\cr
  \noalign{\hfill}
}
    
    \caption{(a) Qudit encoding schemes for \ce{^{137}Ba^{+}}. 
    Black texts indicate the encoding scheme employed in this work for \ce{^{137}Ba^{+}}. 
    All states with sufficiently strong (resulting in $\pi$-pulse transition fidelity of $\ge 75\%$, discussed later in main text) allowed transitions to the $5D_{5/2}$ level from $\lvert 0 \rangle$ are encoded.
    Gray texts indicate a possible extension of the encoding scheme to up to 25 levels, where the additional states are chosen arbitrarily.
    (b) Illustration of the optical setup in this work.
    $\vec{B}_e$ denotes the magnetic field.
    See Extended Fig. \ref{fig:BeamOrientation} for detailed illustration of the setup.
    (c) The \ce{^{137}Ba^{+}} energy level structure of the energy levels relevant to this work.
    $\qty{493}{\nano\meter}$ and $\qty{650}{\nano\meter}$ lasers are used for laser cooling and fluorescent readout of the ion.
    Metastable states in the $5D_{5/2}$ level are accessed from the $6S_{1/2}$ ground level with a \qty{1762}{\nano \meter} laser.
    A \qty{614}{\nano \meter} laser is used to reset the experiments by pumping states out of the $5D_{5/2}$ level to the $6P_{3/2}$ level, which undergoes fast decay mostly to the ground level.
    }
\end{figure*}

In a static non-zero magnetic field, \ce{^{137}Ba^+} has 8 distinct stable energy states in the $6S_{1/2}$ level and 24 distinct metastable energy states (with a lifetime of $\qty{35}{\second}$ \cite{Madej1990}) in the $5D_{5/2}$ level (see Fig. \ref{fig:QuditEncoding}).
The lifetime of the $5D_{5/2}$ metastable states is orders of magnitude larger than a typical quantum operation time scale \cite{Gaebler2016, Ringbauer2022}, so this abundance of stable or metastable states in \ce{^{137}Ba^{+}} makes it an excellent candidate for high-dimensional qudit encoding.

The measurement procedures used in this article allow 25 of the 32 states in the $6S_{1/2}$ and $5D_{5/2}$ levels to be distinguishable in a single shot (discussed in next sections), thus allowing qudit encodings of up to 25 levels in principle.
In this work, we encode computational states in the $\lvert 6S_{1/2}, F = 2, m_{F} = 2 \rangle$ state and the subset of $5D_{5/2}$ states accessible from the $\lvert 6S_{1/2}, F = 2, m_{F} = 2 \rangle$ state using quadrupole-allowed \qty{1762}{\nano \meter} transitions. 
We exclude $5D_{5/2}$ states with $\pi$-pulse transition fidelities of $\le 75\%$ (see Supplementary Information), resulting in a 13-level encoding, illustrated in Fig. \ref{fig:QuditEncoding}.

Fig. \ref{fig:BeamOrientationSimple} illustrates the setup used in this work for optical control of the \ce{^{137}Ba^+} energy levels (see Methods and Extended Fig. \ref{fig:BeamOrientation} for more details).
To attain sufficient experimental controls to perform SPAM, other energy levels in \ce{^{137}Ba^+} are utilized. 
The relevant energy levels and their corresponding laser frequencies for control are summarized in Fig. \ref{fig:Ba_General_Energy_Levels}.

\begin{figure*}
\centering
    \begin{subfigure}{0.45\linewidth}
        \includegraphics[width=\linewidth]{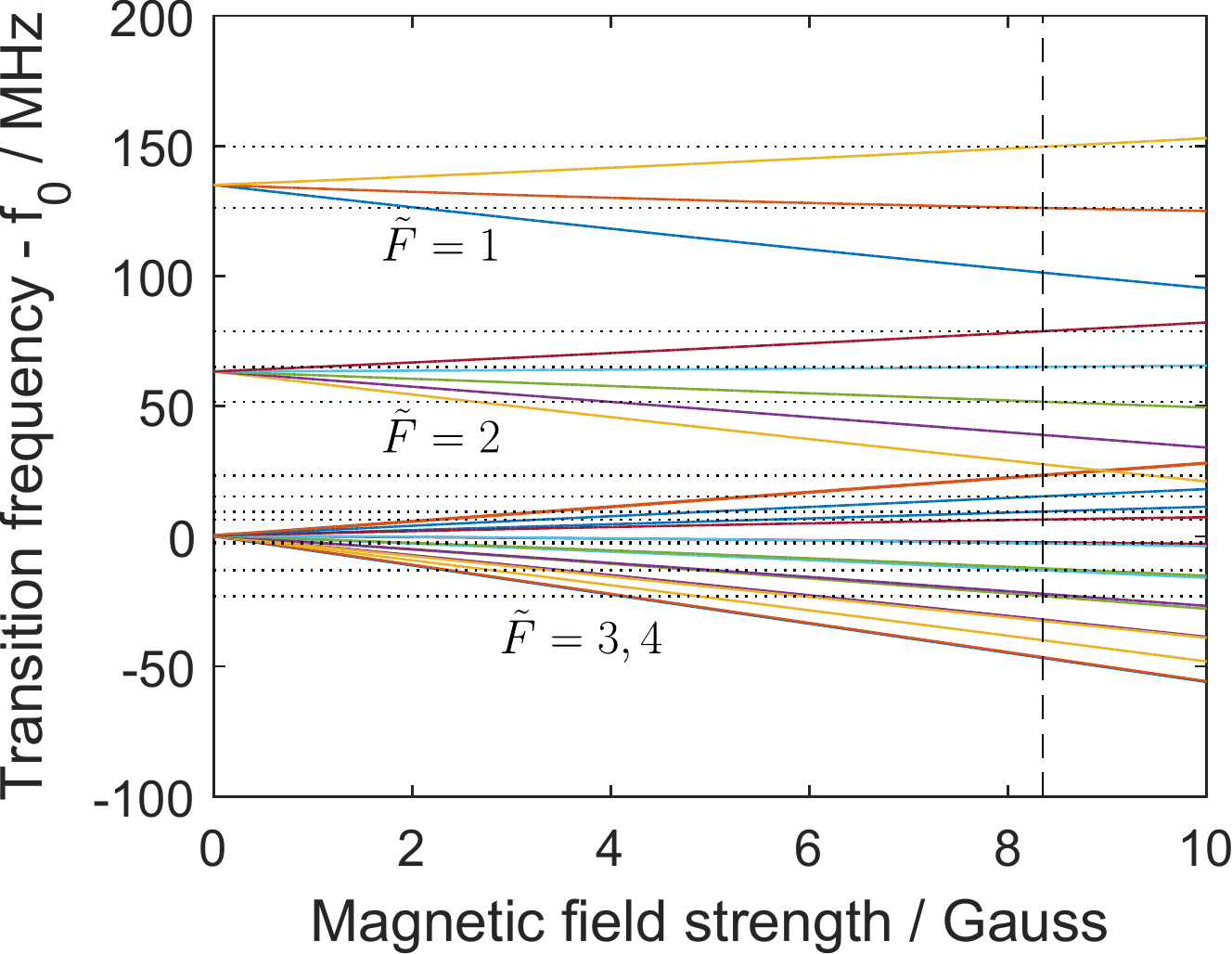}
        \caption{}
        \label{fig:TransitionFreq_vs_B_Sim_Ba137}
    \end{subfigure}
    \begin{subfigure}{0.45\linewidth}
        \includegraphics[width=\linewidth]{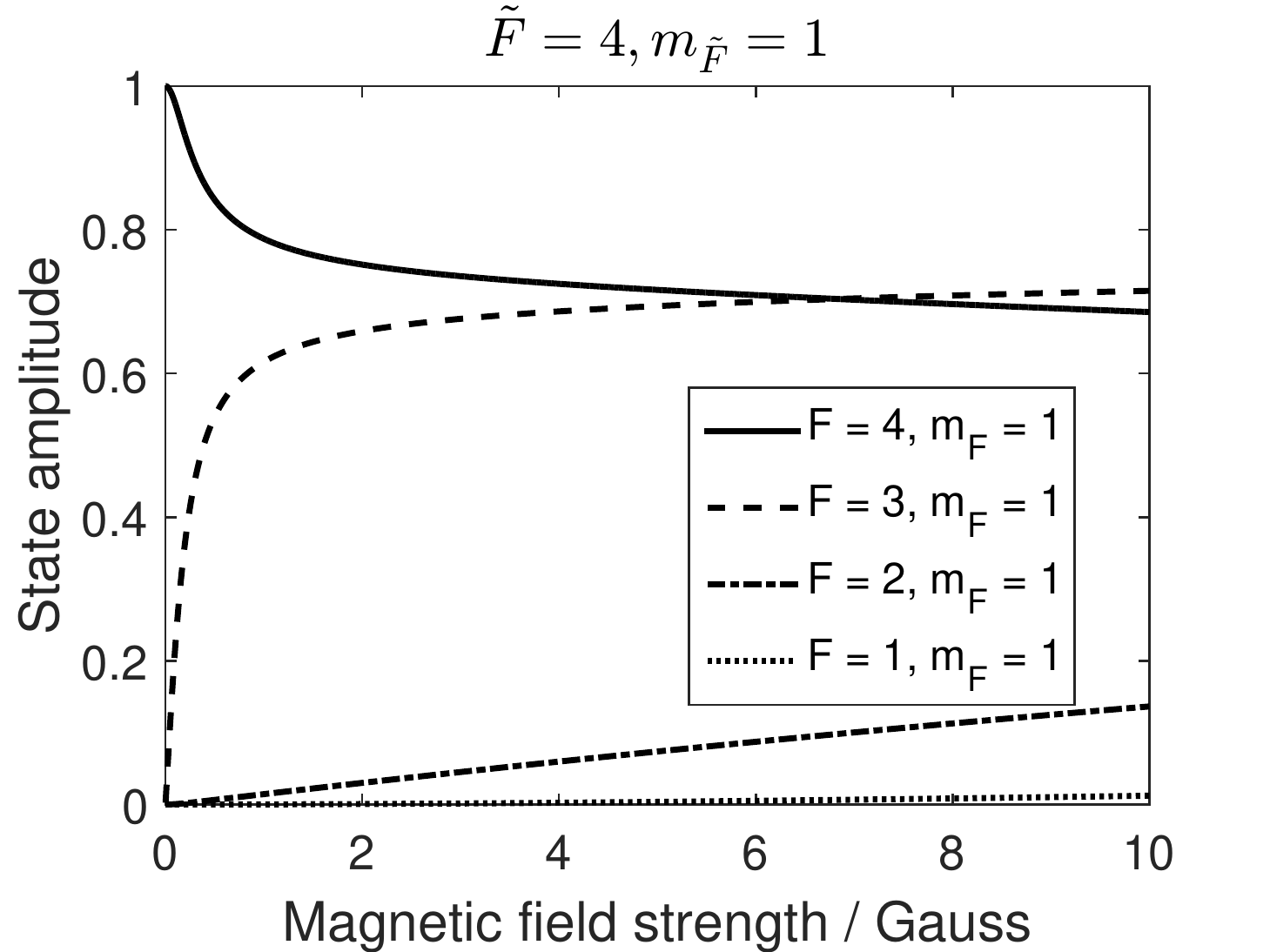}
        \caption{}
        \label{fig:StateEvol_F4m1}
    \end{subfigure}
    \begin{subfigure}{0.9\linewidth}
        \includegraphics[width=\linewidth]{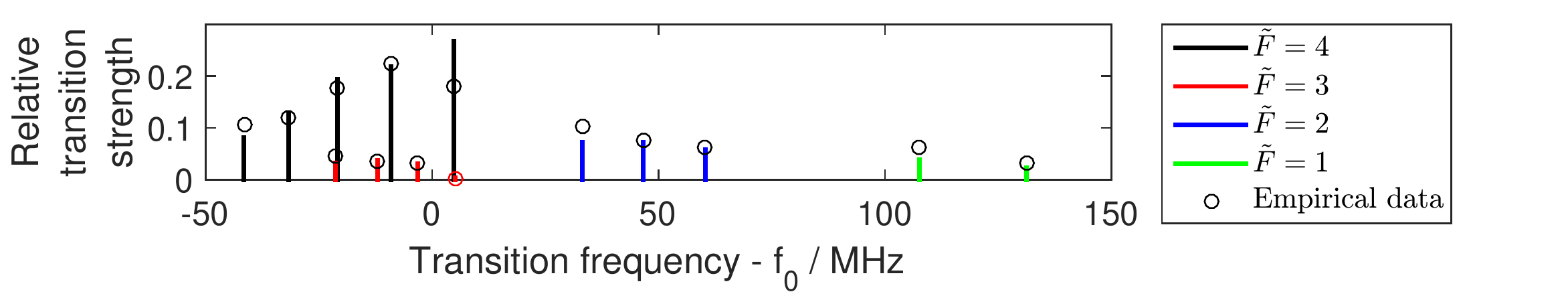}
        \caption{}
        \label{fig:TransitionStrengthPlot}
    \end{subfigure}
\caption{
(a) Simulated transition energy levels to the $5D_{5/2}$ states from the $\lvert \Tilde{F}=2, m_{\Tilde{F}}=2 \rangle$ state in the $6S_{1/2}$ level.
$f_0$ is the transition frequency to the $F=4$ level at zero magnetic field strength.
The horizontal dotted lines indicate the empirically measured transition frequencies, which match well with the simulated transition frequency splittings at \qty{8.35}{G}, marked by the vertical dashed line.
(b) A representative plot of the simulated energy eigenstates, illustrating the mixing of the zero-field eigenstates. The $\lvert \Tilde{F}=4, m_{\Tilde{F}}=1 \rangle$ energy eigenstate is expressed in the $\lvert F, m_F \rangle$ basis for magnetic field strengths from \qty{0} to \qty{10}{G}. 
Components with zero amplitudes are not plotted.
(c) (Color online) Theoretically estimated non-zero transition strengths to the $5D_{5/2}$ states from the $\lvert 6S_{1/2}, \Tilde{F}=2, m_{\Tilde{F}}=2 \rangle$ state, relative to the reduced transition matrix element (see Methods). 
The empirical data are scaled values of the measured Rabi frequencies.
The error bars of the empirical data are smaller than the plot markers and are not plotted.
The $\lvert \Tilde{F}=3, m_{\Tilde{F}}=3 \rangle$ relative transition strength is too weak to be clearly visible on the plot or accurately estimated experimentally due to coherence time limitations.
The red marker is an upper bound of the empirically measured Rabi frequency for this transition.
}
\end{figure*}

Note that we have labelled the states using $\Tilde{F}$ and $m_{\Tilde{F}}$ notations in Fig. \ref{fig:QuditEncoding}, which we define as the energy eigenstates of the ion in a general magnetic field strength, $B_e$. Each $\lvert \Tilde{F}, m_{\Tilde{F}} \rangle$ state approaches the corresponding $\lvert F, m_F \rangle$ state at low magnetic field strengths, i.e. $\lvert \Tilde{F}, m_{\Tilde{F}} \rangle \approx \lvert F, m_{F} \rangle$ as $B_e \rightarrow 0$.
This distinction is necessary as the hyperfine energy level splitting between the $F=3$ and $F=4$ states in the $5D_{5/2}$ level is small, at $\qty{486}{\kilo \hertz}$ \cite{Silverans1986}, and the linear Zeeman approximation does not hold for typical values of $B_e$, as indicated by the strong overlap between $F=3$ and $F=4$ energies in Fig. \ref{fig:TransitionFreq_vs_B_Sim_Ba137}.
Fig. \ref{fig:StateEvol_F4m1} illustrates this effect further; it can be seen that $\lvert 5D_{5/2}, \Tilde{F}=4, m_{\Tilde{F}=1} \rangle$ state differs significantly from $\ket{F=4,m_F=1}$ for external fields as low as \qty{0.2}{G}.
This trend applies to other states with $\Tilde{F}=3$ and $\Tilde{F}=4$, except for states with $m_{\Tilde{F}}=\pm 4$ (see Extended Figs. \ref{fig:SuppMat_EigenstatesEvol_F1F2}, \ref{fig:SuppMat_EigenstatesEvol_F3}, \ref{fig:SuppMat_EigenstatesEvol_F4}).
To suppress coherent dark states \cite{Berkeland2002}, which is a prerequisite for ion cooling and fluorescence readout, it is necessary to apply an external field of more than \qty{0.2}{G}, and therefore necessary to work in the regime where the energy eigenstates of the $5D_{5/2}$ level differ significantly from the $\lvert F, m_F \rangle$ states. In this work, the magnetic field strength is estimated to be $B_e = \qty{8.35}{G}$ (see Fig. \ref{fig:TransitionFreq_vs_B_Sim_Ba137}), well within the regime where the linear Zeeman approximation breaks down.

The discrepancies between the $5D_{5/2}$ energy eigenstates and the $\lvert F, m_F \rangle$ states result in $6S_{1/2} \leftrightarrow 5D_{5/2}$ transition strengths that differ significantly from those calculated in the $\lvert F, m_F \rangle$ basis.
From Fig. \ref{fig:TransitionStrengthPlot}, the numerically simulated (see Methods) and empirically measured relative transition strengths at \qty{8.35}{G} are much weaker for $\Tilde{F}=3$ states as compared to $\Tilde{F}=4$ states, starting from the $\ket{6S_{1/2}, \Tilde{F}=2, \Tilde{m}=2}$ state.
This deviates significantly from what is predicted assuming $\lvert F, m_F \rangle$ states (not shown).
Despite these changes in transition strengths, we nevertheless calculate that all states in the $5D_{5/2}$ level are practically accessible (with transition strengths within an order of magnitude of the strongest transitions, see Extended Data Table \ref{tab:TransitionStrengths}).

\subsection*{\label{sec:State_Preparation} State Preparation}

In this work, we initialize the \ce{^{137}Ba^{+}} in the $\lvert 6S_{1/2},F=2,m_F=2 \rangle$ state, which we encode as the $\lvert 0 \rangle$ state.
This is done via optical pumping by sending $\sigma^{+}$-polarized \qty{493}{\nano \meter} light to the ion, together with \qty{650}{\nano \meter} light (see Fig. \ref{fig:BeamOrientationSimple}), with laser frequencies as shown in Fig. \ref{fig:Ba_General_Energy_Levels}.
In this work, to prepare the ion in any of the $\lvert n \ne 0 \rangle$ states, a $\pi$-pulse of $\qty{1762}{\nano\meter}$ light with the corresponding frequency to drive the $\lvert 0 \rangle \leftrightarrow \lvert n \rangle$ transition is applied to the ion.
In general, to encode up to 25 levels, any desired state can be prepared with at most three sequential \qty{1762}{\nano \meter} laser transitions.

\subsection*{\label{sec:State_measurement}State Measurement}

\ce{^{137}Ba^{+}} emits fluorescence when it is driven by $\qty{493}{\nano\meter}$ and $\qty{650}{\nano\meter}$ lasers if the ion is in the $6S_{1/2}$ state.
It does not fluoresce if the ion is in any of the encoded $5D_{5/2}$ states.
We make use of this property to construct a single-shot qudit measurement process, which is described in the sequence below and sketched in Fig. \ref{fig:SimplePulseSequence}.
The process consists of multiple steps, but we characterize it as single-shot in the sense that the  projected quantum state can be determined definitively in a single run of the measurement protocol.
\begin{enumerate}
    \item Without loss of generality, the $\lvert 0 \rangle$ state is assigned as one of the $6S_{1/2}$ states. 
    Any population of the encoded states in the $6S_{1/2}$ level other than state $\lvert 0 \rangle$ is brought up to a corresponding unencoded $5D_{5/2}$ state by sending $\pi$-pulses of \qty{1762}{\nano \meter} laser with frequencies resonant to the desired transitions in sequence. We call this a shelving process.
    \item The \qty{493}{\nano \meter} and \qty{650}{\nano \meter} lasers are turned on to check for fluorescence. The ion is measured to be in the qudit state $\lvert 0 \rangle$ if fluorescence is observed at this step.
    \item The population corresponding to the next computational state $\lvert n \rangle$ in $5D_{5/2}$ is brought down to one of the $6S_{1/2}$ states by sending a $\pi$-pulse of $\qty{1762}{\nano\meter}$ laser with the corresponding transition frequency. We call this a de-shelving process.
    \item Step 2 is repeated to check for fluorescence. The ion is measured to be in the qudit state $\lvert n \rangle$ if fluorescence is observed for the first time at this step, as exemplified in Fig. \ref{fig:Table_bright_dark}.
    \item Steps 3 and 4 are repeated until all states are de-shelved and checked for fluorescence.
\end{enumerate}

For the qudit encoding demonstrated in this article, as shown in Fig. \ref{fig:QuditEncoding}, the shelving process in Step 1 is unnecessary. 
The $6S_{1/2}$ ground state chosen for the de-shelving process is also fixed to be $\lvert 6S_{1/2},F=2,m_{F}=2 \rangle$ for convenience.
Fig. \ref{fig:SimplePulseSequence} summarizes the simplified pulse sequence for performing SPAM as described here.
From this measurement protocol, it can be seen from Step 1 that 7 out of the 32 stable/metastable states need to be left unencoded to achieve full distinguishability, resulting in a maximum encoding of 25 states.

\subsection*{SPAM Experimental Results and Discussion}

\begin{figure*}
\centering
    \tabskip=0pt
\valign{#\cr
  \hbox{%
    \begin{subfigure}{.55\textwidth}
    \centering
    \includegraphics[width=\textwidth]{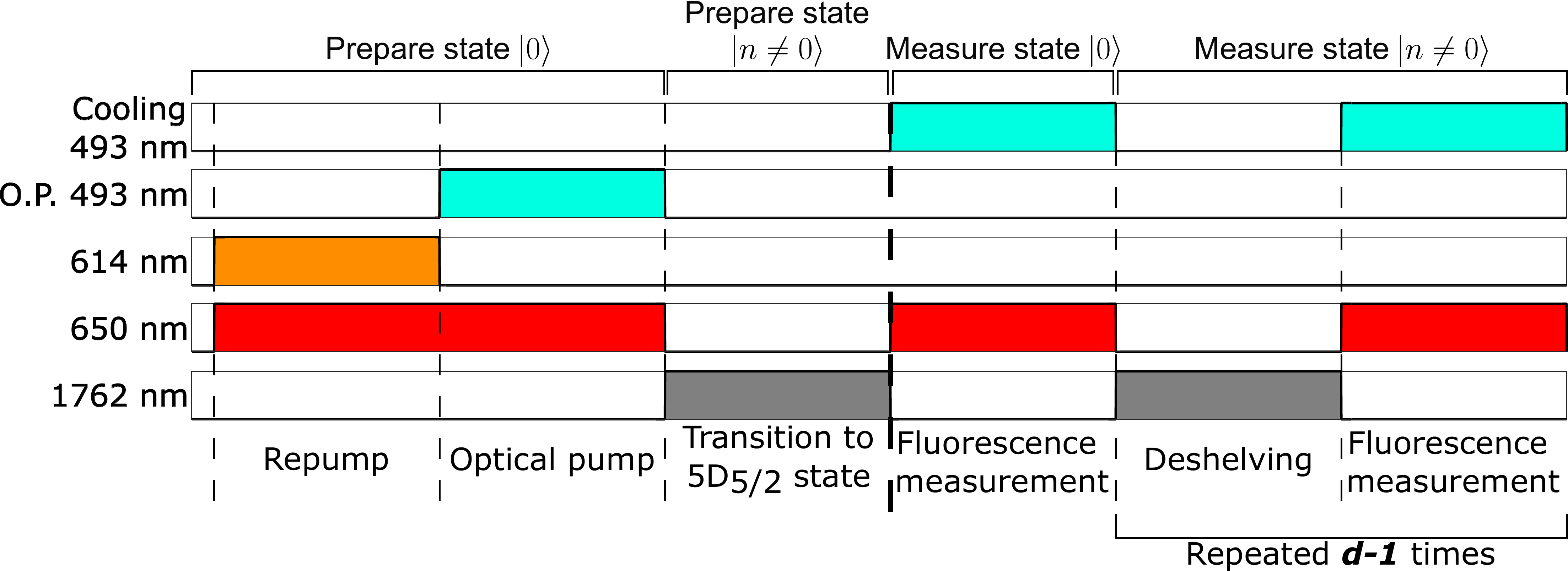}
    \caption{}
    \label{fig:SimplePulseSequence}
    \end{subfigure}%
  }\vfill
  \hbox{%
    \begin{subfigure}{.55\textwidth}
    \centering
    \includegraphics[width=\textwidth]{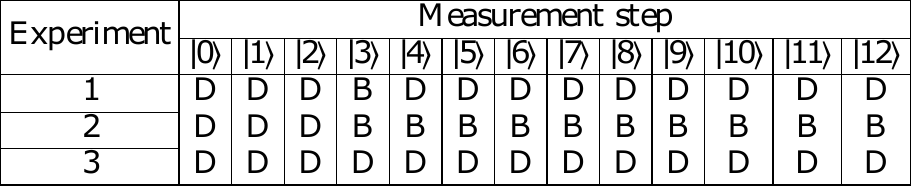}
    \caption{}
    \label{fig:Table_bright_dark}
    \end{subfigure}%
  }\cr
  \hbox{%
    \begin{subfigure}[b]{.45\textwidth}
    \centering
    \includegraphics[width=\textwidth]{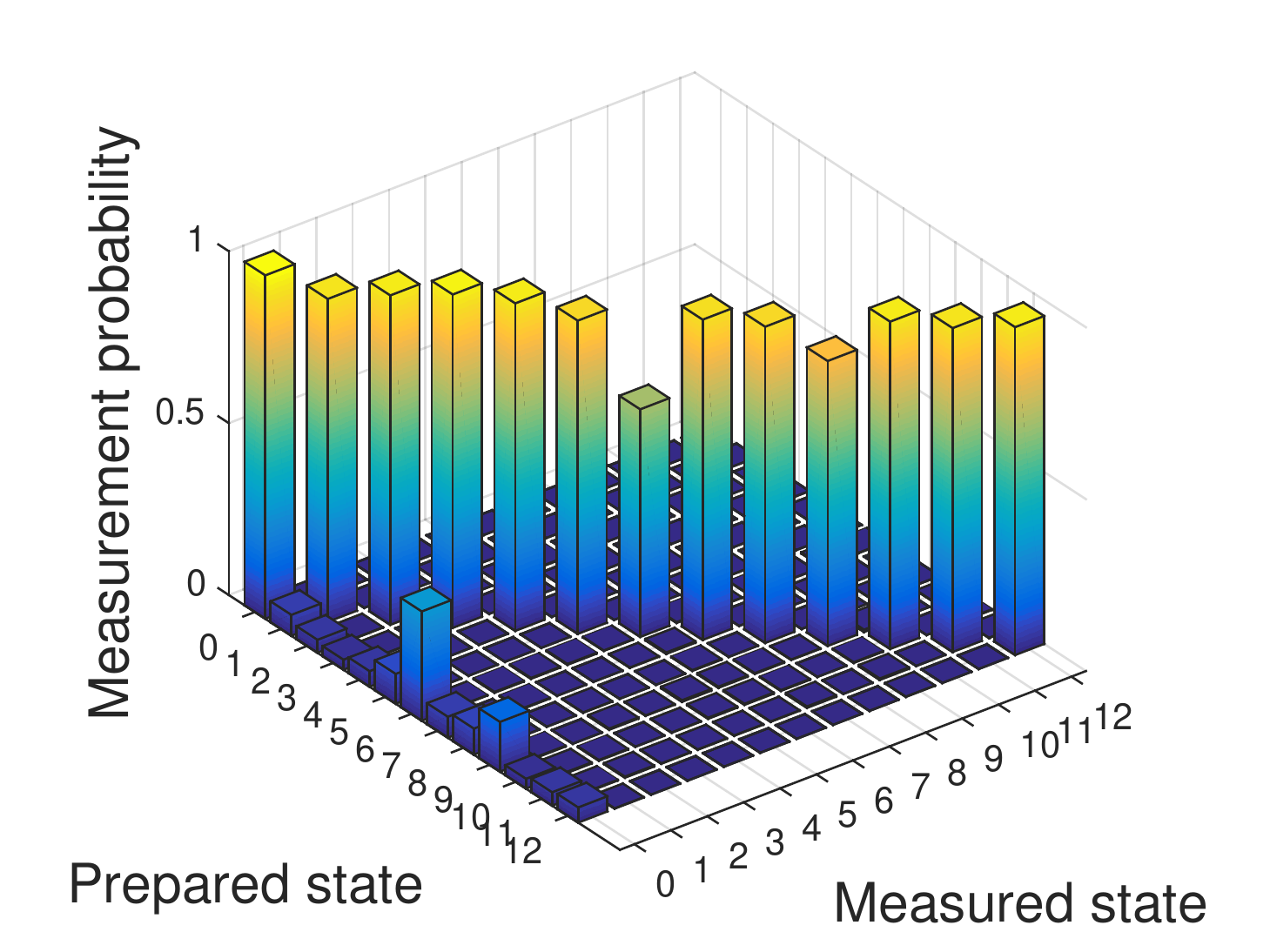}
    \caption{}
    \label{fig:Fidel3Dbar}
    \end{subfigure}%
  }\cr
  \noalign{\hfill}
}
\caption{
(a) Simplified pulse sequence for one SPAM experiment for a prepared state $\lvert n \rangle$ of a $d$-dimensional qudit, demonstrated in this work for $d=13$. 
O.P. denotes optical pumping.
The time axis is not drawn to scale.
The pulse sequences to the left and right of the bold dashed line are the state preparation and state measurement processes respectively.
The \qty{1762}{\nano \meter} laser pulse for the state preparation step is only needed if the prepared state is not $\lvert 0 \rangle$.
For a detailed pulse sequence relevant to the experimental apparatus, see extended Fig. \ref{fig:PulseSequence}.
(b) Examples of measurement outcomes. 
B and D denote that the ion is detected as bright and dark respectively.  
The de-shelving process in the measurement sequence as described in the main text is performed in ascending order of the encoded states.
Experiments 1 and 2 are interpreted as measuring the system to be in state $\lvert 3 \rangle$.
Experiment 3 is regarded as a measurement failure and is removed from the data set for the post-selected SPAM results.
In this experiment, the deshelving process is simultaneously a reshelving process (see Extended Fig. \ref{fig:PulseSequence}).
An alternative interpretation that treats Experiment 2 as a measurement failure rather than a measurement of the state $\ket{3}$ is discussed in the Supplementary Information.
(c) Post-selected measurement probability of the 13-level qudit SPAM experiment.
}
\end{figure*}

Fig. \ref{fig:Table_bright_dark} shows representative examples of possible measurement outcomes.
Experiment 3 in Fig. \ref{fig:Table_bright_dark} is a directly detectable failure of the measurement procedure.
This is arguably a less critical error than misdiagnosing the quantum state, as the user directly knows an error has occurred and can rerun the computation.
Fig. \ref{fig:Fidel3Dbar} summarizes the post-selected SPAM experimental results, where the cases when no bright state is detected throughout the measurement sequence are removed from the data set.
For the raw SPAM results, where the cases with no bright states detected are counted as errors, see Extended Data Table \ref{tab:SPAMDataRaw}.
The average raw and post-selected SPAM errors are computed to be $13.1 \pm 0.3 \%$ and $8.3 \pm 0.3 \%$ respectively for a 13-level qudit.
The raw data sets and analysis scripts for this work can be found in the repository linked at Ref. \cite{GIT}.

\begin{figure*}
\centering
    \begin{subfigure}{0.48\linewidth}
        \includegraphics[width=\linewidth]{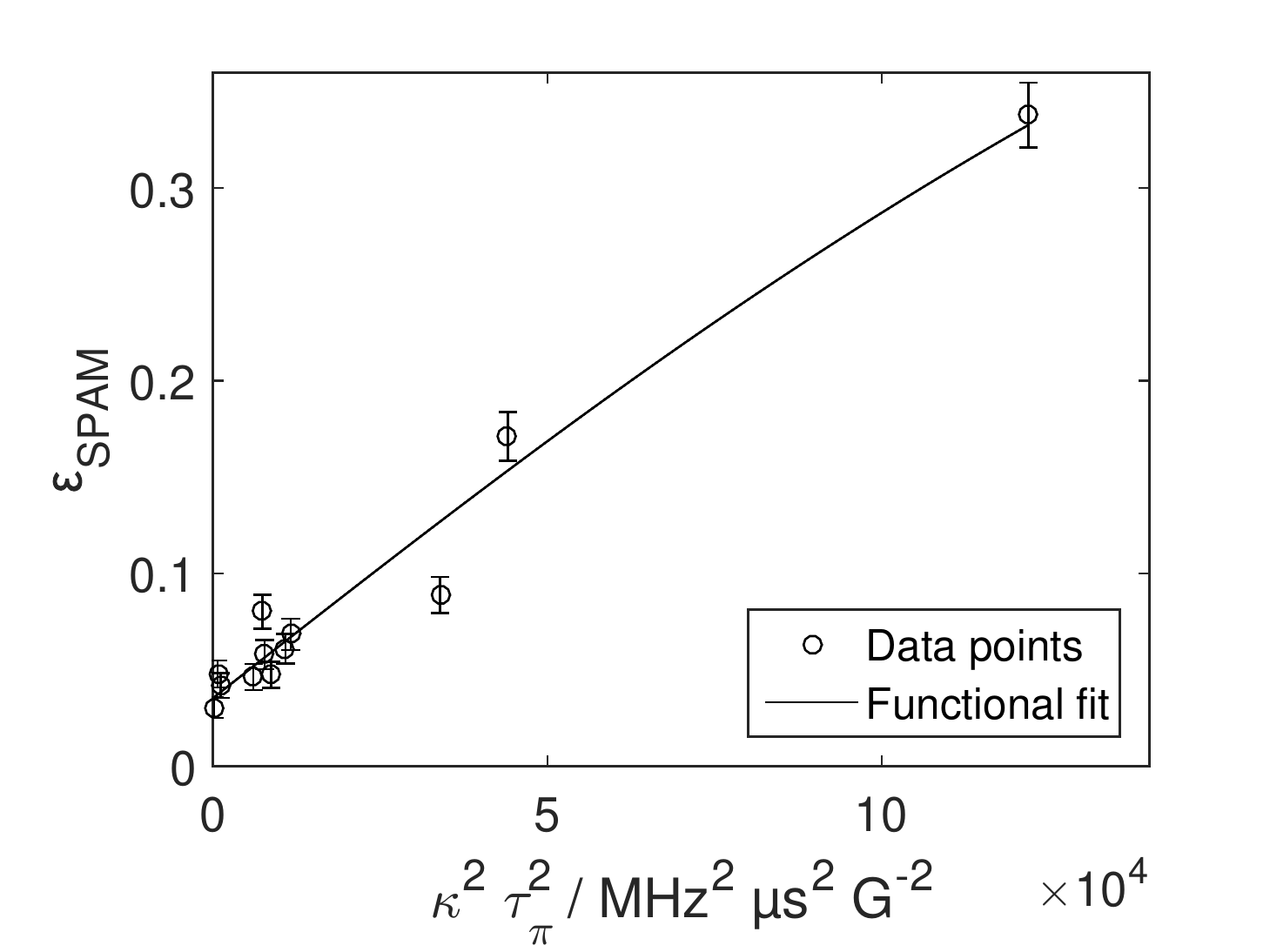}
        \caption{}
        \label{fig:Errorvsktausquared}
    \end{subfigure}
    \begin{subfigure}{0.48\linewidth}
        \includegraphics[width=\linewidth]{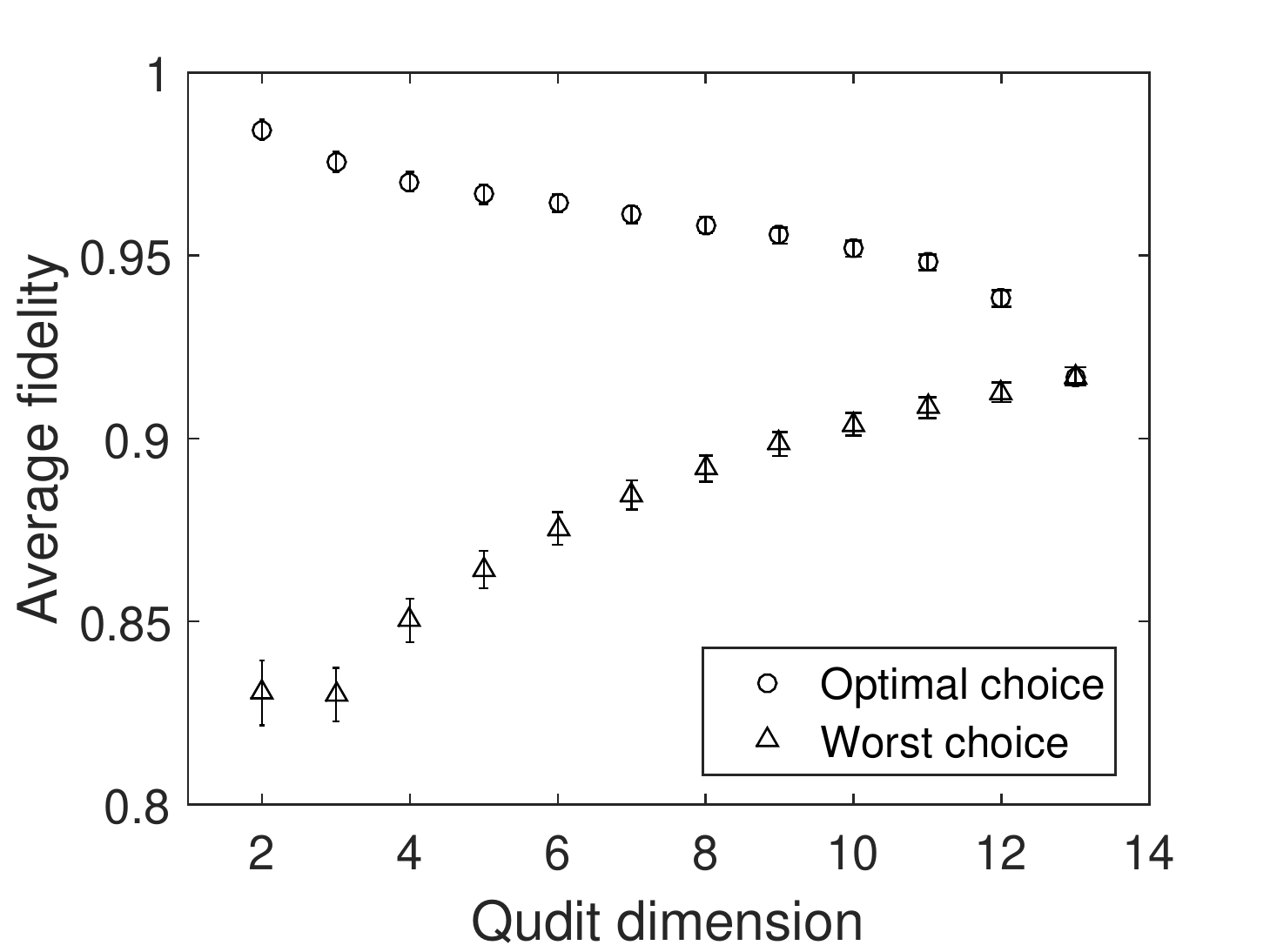}
        \caption{}
        \label{fig:QuditScaling}
    \end{subfigure}
\caption{
(a) Plot of the post-selected SPAM error for each prepared state from Fig. \ref{fig:Fidel3Dbar} against $\kappa^2 \tau_{\pi}^2$, which we identify as the relevant figure of merit for errors induced by magnetic field fluctuations.
$\kappa$ is the magnetic field sensitivity and $\tau_{\pi}$ is the $\pi$-pulse time.
An obvious positive correlation of $\epsilon_{SPAM}$ with $\kappa^2 \tau_{\pi}^2$ is observed, indicating agreement with Eq. \ref{eq:Chi_Propto}.
Eqs. \ref{eq:SPAM_Error} and \ref{eq:Error_General} with an additional freely varying parameter for the vertical intercept are used to fit the data.
(b) Scaling of average SPAM fidelity with qudit dimension, and its dependence on the choice of qudit states, computed using the data set in Fig. \ref{fig:Fidel3Dbar}. 
For the 'optimal choice' of qudit states, the $\lvert 6S_{1/2}, F=2, m_F=2 \rangle$ state and the $d-1$ states with the best $\pi$-pulse transition fidelities are chosen as the computational basis.
For the 'worst choice', the $d-1$ states with the worst $\pi$-pulse transition fidelities are chosen instead. 
}
\end{figure*}

We conclude that the magnetic field noise is a major source of error in this work, using the following error analysis.
The post-selected SPAM error for a given prepared state $\ket{n \ne 0}$ due to decoherence from magnetic field noise is 
\begin{equation}
    \epsilon_{SPAM} = \frac{\epsilon_{\pi}}{\epsilon_{\pi} + \left(1 - \epsilon_{\pi} \right)^2},
    \label{eq:SPAM_Error}
\end{equation}
where $\epsilon_{\pi}$ is the error for a single $\pi$-pulse transition.
Using filter function theory \cite{Ball2015, Day2022}, $\epsilon_{\pi}$ can be expressed as
\begin{equation}
    \epsilon_{\pi} = \frac{1}{2} \left( 1 - e^{-\chi} \right),
    \label{eq:Error_General}
\end{equation}
where $\chi$ is a spectral overlap of the transition frequency noise power spectral density (PSD), $S\left( \omega \right)$, with the filter function of the target operation, $F \left( \omega \right)$
\begin{equation}
    \chi = \frac{1}{\pi} \int_0^{\infty} \frac{1}{\omega^2} S\left( \omega \right) F \left( \omega \right) d\omega.
    \label{eq:Chi_General}
\end{equation}
Assuming the magnetic field noise to be a $1/f$ noise with a spectral peak at the mains electricity frequency and a baseline white noise, $\chi$ can be derived to scale as
\begin{equation}
    \chi \propto \kappa^2\tau_{\pi}^2,
    \label{eq:Chi_Propto}
\end{equation}
where $\kappa$ is the magnetic field sensitivity of the transition frequency and $\tau_{\pi}$ is the $\pi$-pulse time.
See Supplementary Information for the derivation of Eq. \ref{eq:Chi_Propto}.
The scaling of $\epsilon_{SPAM}$ with $\kappa^2 \tau_{\pi}^2$ as shown in Fig. \ref{fig:Errorvsktausquared} shows agreement with this error model, which supports the notion that magnetic field noise is a major source of error in this work. 
This simplified model may also be useful for predicting which of the 126 allowed quadrupole transitions between $6S_{1/2}$ and $5D_{5/2}$ states will be most useful for quantum operations as the 1762 nm laser orientation and polarization are varied.

The vertical intercept of $\epsilon_{SPAM} = 0.04 \pm 0.01$ in Fig. \ref{fig:Errorvsktausquared} may indicate that around 4\% of the error comes from other sources.
Around $1.5 \pm 2 \%$ of the remaining error is estimated to be from drifts of the experimental parameters from the time of calibration to the time SPAM experiments are done (see Supplementary Information).
We speculate that the majority of the rest of the remaining error to be from unclean laser polarization from the optical pumping step, but the available data in this work does not allow accurate quantitative estimations of the remaining errors.
A more complex state preparation protocol \cite{An2022} has been shown to reduce optical pumping errors to below $10^{-4}$, and can be applied in the future extension of this work with sufficient hardware upgrades.
We estimated errors from other known error sources (from spontaneous decay from the $5D_{5/2}$ level, off-resonant transition error and bright/dark state discrimination error) and found that they contribute less than $0.5 \%$ of error (see Supplementary Information).

Because the magnetic field noise is a dominant source of error, the fidelity is highly dependent on the choice of qudit states. 
We study the best- and worst-case SPAM fidelities for qudits of differing dimension in Fig. \ref{fig:QuditScaling}, always including the state $\ket{0}$. 
The fact that it is possible to choose qudit encodings where the errors improve with higher dimension indicate that our results are not limited by any effects that intrinsically depend on the qudit dimension.

It should be noted that we have not made any efforts to mitigate magnetic field noise in this work.
To reduce error from magnetic field noise, passive \cite{Ruster2016} and active \cite{Hu2022} methods have been empirically demonstrated, with the former estimated to contribute to an error level lower than light scattering error from hyperfine Raman transitions \cite{Low2020}.
Thus, it is expected that magnetic field noise error is not a fundamental limiting factor in the qudit protocol of this work, despite the lack of magnetically insensitive encodings for high qudit dimensions.

Another practical merit to this qudit scheme is that the calibration times of the experimental parameters (laser frequencies and $\pi$-pulse times) for SPAM do not scale with the qudit dimension with certain protocols, which are presented in Methods and Supplementary Information.
Such a protocol is employed for the laser frequencies calibrations in this work.
The efficient $\pi$-pulse times calibrations protocol is not used in this work due to a limitation of our experimental apparatus (see Supplementary Information).

The total measurement time of around \qty{100}{\milli \second} in this work is relatively long for typical trapped ion quantum operations \cite{Gaebler2016}.
However, this is an artificial limitation from our waveform generation methods and lack of frequency modulation for the \qty{614}{\nano \meter} laser (see Supplementary Information for details).
A fluorescence collection time of \qty{350}{\micro \second} has been demonstrated with \ce{^{137}Ba^{+}} \cite{An2022}.
In principle, the measurement time scales linearly with qudit levels in the absence of these limitations.
Thus, qudit measurement times on the order of \qty{1}{\milli \second} or better should be possible.

In this work, we have demonstrated high-level qudit encoding and SPAM of up to 13 levels using a \ce{^{137}Ba^{+}} ion with an average post-selected SPAM error of $8.3 \pm 0.3\%$.
The major source of SPAM error in this work is magnetic field noise.
However, this should not be a major roadblock for this qudit protocol, as the methods to rectify magnetic field noise are known.
With improved technical control, it is possible to extend this work and encode up to 25 levels in a single \ce{^{137}Ba^{+}} ion with good SPAM fidelities.
To build a functioning quantum computer, the ability to perform single qudit gates and entangling gates are required. 
Such procedures have been demonstrated in Ref. \cite{Ringbauer2022} for a qudit dimension of 5 with \ce{^{40}Ca^+}, with a similar state encoding scheme and manipulation, and could be straightforwardly generalized for \ce{^{137}Ba^{+}}.
Thus, our work opens prospects for a trapped-ion-based universal quantum computer with more than double the number of qudit states per ion.

\section*{Methods} \label{sec:Methods}

\subsection*{\ce{^{137}Ba^{+}} Energy Levels Simulations} \label{sec:Ba137EnergyLevelsSimulations}

To calculate the energy eigenvalues and eigenstates for an arbitrary magnetic field strength, the Hamiltonian for the corresponding energy orbital is constructed in the $\lvert I, m_I; J, m_J \rangle$ basis as follows
\begin{equation}
    \begin{aligned}
        \hat{H} &= h A_D \vec{I}\cdot\vec{J} \\
        &+ h B_Q \frac{3\left( \vec{I}\cdot\vec{J} \right)^2 + \frac{3}{2}\vec{I}\cdot\vec{J}-I \left( I + 1 \right) J \left( J + 1 \right)}{2 I \left(2I - 1\right) J \left(2J - 1 \right)} \\
        &+ B_e\mu_B \left( g_J m_J + g_I m_I \right)
    \end{aligned}
    \label{eq:EnergyLevelHamiltonian}
\end{equation}
where $h$ is the Planck constant, $A_D$ is the magnetic dipole hyperfine structure constant, $B_Q$ is the electric quadrupole hyperfine structure constant, $B_e$ is the magnetic field strength, $\vec{I}$ and $\vec{J}$ are the nuclear and electron angular momentum vectors respectively, $I$ and $J$ are the nuclear and electron angular momentum numbers respectively, $m_I$ and $m_J$ are the projection of the nuclear and electron angular momenta along the magnetic field axis respectively, $g_I$ and $g_J$ denote the nuclear and electron g-factor respectively, and $\mu_B$ is the Bohr magneton.
The Hamiltonian matrix is then solved numerically to obtain its eigenvalues and eigenvectors.
In this work, we have set $g_I = 0$ as it is negligible compared to $g_J$.
For the $6S_{1/2}$ level, $A_D = \qty{4018.871}{\mega\hertz}$ and $B_Q = 0$ \cite{Blatt1982}.
For the $5D_{5/2}$ level, $A_D = \qty{-12.028}{\mega\hertz}$ and $B_Q = \qty{59.533}{\mega\hertz}$ \cite{Silverans1986}.
With these values of the hyperfine constants for the $5D_{5/2}$ level, at a magnetic field strength of $B_e = \qty{8.35}{G}$, the Zeeman splitting term in Eq. \ref{eq:EnergyLevelHamiltonian} is comparable to the first 2 hyperfine energy splitting terms, and the $\lvert F, m_F \rangle$ states cease to be good approximations of the energy eigenstates of the ion.
Figs. \ref{fig:TransitionFreq_vs_B_Sim_Ba137} and \ref{fig:StateEvol_F4m1} show the simulation results of the energy eigenvalues and eigenvectors.

\subsection*{Quadrupole Transition Strength Geometric Factor}
\label{sec:QuadrupoleTransitionGeometricFactor}
In a static magnetic field, when a quadrupole transition of $\Delta m = q$ is driven with a laser, the component of the laser field that is driving the transition can be defined by a factor $g^{(q)}\left( \gamma, \phi \right)$.
The $g^{(q)}\left( \gamma, \phi \right)$ factor is dependent on the angle between the laser wavevector $\vec{k}$ and the magnetic field vector $\vec{B_e}$, which we define as $\phi$, and the polarization angle of the laser electric field with respect to the plane formed by $\vec{k}$ and $\vec{B_e}$, which we define as $\gamma$.
The expressions of the $g^{(q)}\left( \gamma, \phi \right)$ factors can be derived to be \cite{Roos2000, Bramman2019}
\begin{equation}
    \begin{aligned}
    g^{(0)}\left( \gamma, \phi \right) &= \frac{1}{2}|\cos\gamma\sin 2\phi | \\
    g^{(\pm 1)}\left( \gamma, \phi \right) &= \frac{1}{\sqrt{6}}|\mp\cos\gamma\cos 2\phi + i\sin\gamma\cos\phi| \\
    g^{(\pm 2)}\left( \gamma, \phi \right) &= \frac{1}{\sqrt{6}}\bigg|\frac{1}{2} \cos\gamma\sin 2\phi \mp i\sin\gamma\sin\phi\bigg|.
    \end{aligned}
\label{eq:Geometrical}
\end{equation}

\subsection*{\ce{^{137}Ba^{+}} Transition Strengths in the Intermediate Magnetic Field Regime} \label{sec:Ba137TransitionStrengthsSimulations}

The reduction of $6S_{1/2} \leftrightarrow 5D_{5/2}$ transition strengths for a general magnetic field strength in a linearly polarized laser perturbation can be expressed as
\begin{equation}
    \begin{aligned}
        &g^{(q)}\left(\gamma, \phi\right)\langle \Tilde{F}_{D}, m_{\Tilde{F},D} \rvert \hat{Q}_{q=m_{\Tilde{F},D}-m_{\Tilde{F},S}} \lvert \Tilde{F}_{S}, m_{\Tilde{F},S} \rangle \\
        &= g^{(q)}\left(\gamma, \phi\right) \langle J_D = 5/2 \rvert \lvert \hat{Q} \rvert \lvert J_S = 1/2 \rangle \langle \Tilde{F}_{S}, m_{\Tilde{F},S}; k = 2, q = m_{\Tilde{F},D} - m_{\Tilde{F},S} \lvert \Tilde{F}_{D}, m_{\Tilde{F},D} \rangle,
    \end{aligned}
    \label{eq:TransitionStrengthFormula}
\end{equation}
where the subscripts $S$ and $D$ denote the $6S_{1/2}$ and $5D_{5/2}$ levels respectively, $\hat{Q}$ is the electric quadrupole energy operator, $k=2$ is the tensor rank of the electric quadrupole energy operator, $\langle J_D = 5/2 \rvert \lvert \hat{Q} \rvert \lvert J_S = 1/2 \rangle$ is the reduced transition matrix element for the $6S_{1/2} \leftrightarrow 5D_{5/2}$ transition. 
The dimensionless prefactors $g^{(q)}\left(\gamma, \phi\right) \langle \Tilde{F}_{S}, m_{\Tilde{F},S}; k = 2, q = m_{\Tilde{F},D} - m_{\Tilde{F},S} \lvert \Tilde{F}_{D}, m_{\Tilde{F},D} \rangle$ are effectively the relative $6S_{1/2} \leftrightarrow 5D_{5/2}$ transition strengths, which are the values plotted in Fig. \ref{fig:TransitionStrengthPlot}.
The details of the calculations for the $g^{(q)}\left(\gamma, \phi\right) \langle \Tilde{F}_{S}, m_{\Tilde{F},S}; k = 2, q = m_{\Tilde{F},D} - m_{\Tilde{F},S} \lvert \Tilde{F}_{D}, m_{\Tilde{F},D} \rangle$ prefactors and the numerical results can be found in the Supplementary Information.

\subsection*{\label{sec:Experimental_setup}Experimental Setup}

In a vacuum chamber with an air pressure of $\qty{1e{-10}}{\milli \bar}$, a 4-rod linear Paul trap is used to trap \ce{^{137}Ba^{+}}. 
Radiofrequency (RF) voltages with an estimated amplitude of $\qty{240}{\volt}$ and a frequency of $\qty{20.772}{\mega\hertz}$ are sent to the 4-rod electrodes, with the diagonal pairs of the electrodes being out of phase with each other.
Static voltages of $\qty{3}{\volt}$ are sent to one pair of the diagonal rods to break the degeneracy of the ion radial secular motional frequencies.
$\qty{10}{\volt}$ of static voltage is applied to the needle electrodes for axial confinement.
This setup results in a trapped \ce{^{137}Ba^{+}} with radial secular motional frequencies of $\qty{1.2}{\mega \hertz}$ and $\qty{1.4}{\mega \hertz}$, and an axial motional frequency of $\qty{10}{\kilo \hertz}$.

The magnetic field orientation is as shown in Fig. \ref{fig:BeamOrientationSimple}. 
It is generated by a pair of solenoids attached to the viewports of the vacuum chamber, with an estimated magnetic field strength of $\qty{8.35}{G}$.

The $\qty{493}{\nano\meter}$ laser is sourced from a commercial Toptica DL Pro external cavity diode laser (ECDL).
An electro-optic modulator (EOM) is used to generate $\pm \qty{4012}{\mega\hertz}$ sidebands for the $\qty{493}{\nano\meter}$ laser, where the red and blue sidebands are used to drive the $\lvert 6S_{1/2},F=2 \rangle \leftrightarrow \lvert 6P_{1/2},F=2 \rangle$ and $\lvert 6S_{1/2},F=1 \rangle \leftrightarrow \lvert 6P_{1/2},F=2 \rangle$ transitions respectively.
The $\qty{493}{\nano\meter}$ laser is split to 2 paths and sent to the trap as shown in Fig. \ref{fig:BeamOrientationSimple}.
Each path goes through an acousto-optic modulator (AOM), which acts as a switch for the laser beam.
The $\qty{493}{\nano\meter}$ beam that is parallel to the magnetic field is circularly-polarized, $\sigma^{+}$-polarized, for optical pumping the \ce{^{137}Ba^{+}} ion.
The $\qty{493}{\nano\meter}$ beam that is perpendicular to the magnetic field is used for cooling and fluorescent readout.
This beam is linearly polarized and the polarization is tuned to maximize ion fluorescence.
This beam orientation cools the ion along all three principal trap axes.
The fluorescence and optical pumping $\qty{493}{\nano\meter}$ beams have powers of $\qty{35}{\micro\watt}$ and $\qty{6.5}{\micro\watt}$ at the ion trap respectively.
Both beams are focused to a beam diameter of approximately $\qty{70}{\micro\meter}$ at the ion.

The $\qty{650}{\nano\meter}$ laser is also sourced from a Toptica DL Pro ECDL.
An EOM is used to generate the necessary sidebands to drive the transitions as described in Fig. \ref{fig:Ba_General_Energy_Levels} and an AOM is used as a switch for the beam going to the ion trap.
The power of this beam at the trap is $\qty{230}{\micro\watt}$ and the beam is focused down to a diameter of $\qty{70}{\micro\meter}$ at the ion.

The $\qty{614}{\nano\meter}$ laser is frequency doubled from a $\qty{1228}{\nano\meter}$ laser.
The $\qty{1228}{\nano\meter}$ laser is sourced from Moglabs CEL and the frequency doubler is from NTT Electronics model WH-0614-000-F-B-C.
An AOM is used as a switch for this laser.
The $\qty{614}{\nano\meter}$ laser power at the ion trap is $\qty{15}{\micro\watt}$ and is focused down to a beam diameter of $\qty{53}{\micro \meter}$ at the ion. 
The $\qty{614}{\nano\meter}$ laser frequency is set to be resonant to the transition from the $\lvert 5D_{5/2}, \Tilde{F}=4, m_{\Tilde{F}}=2 \rangle$ state to the $\lvert 6P_{3/2}, F=3 \rangle$ state.

The $\qty{1762}{\nano\meter}$ laser is from a Toptica DL Pro ECDL.
The frequency stabilization system is built by Stable Laser Systems, where the laser frequency is referenced and locked to a temperature controlled Fabry–Pérot cavity in vacuum using the Pound-Drever-Hall (PDH) method.
The $\qty{1762}{\nano\meter}$ laser carrier frequency is chosen to be locked to a frequency around $\qty{600}{\mega\hertz}$ higher than the $\lvert 6S_{1/2},F=2 \rangle \leftrightarrow \lvert 5D_{5/2} \rangle$ transitions.
The laser is then passed through an EOM, where sideband frequencies are generated, and the red-sideband frequency is used to drive a specific chosen $\lvert 6S_{1/2},F=2 \rangle \leftrightarrow \lvert 5D_{5/2} \rangle$ transition.
The RF source for the $\qty{1762}{\nano\meter}$ EOM comes from an arbitrary waveform generator (AWG), which allows us to quickly change the sideband frequency to drive the desired transitions for the SPAM process.
The \qty{1762}{\nano \meter} laser is sent to the ion from an angle that minimizes coupling to the weakly confined axial motional mode, which is $\phi = \qty{45}{\degree}$ from the magnetic field axis.
The laser power is \qty{2.0}{\milli \watt} and is focused down to a beam diameter of \qty{23}{\micro \meter} at the centre of the ion trap.
The \qty{1762}{\nano \meter} laser is linearly polarized and is rotated to an angle that is $\gamma = +\qty{58}{\degree}$ from the angle that is optimal for $\Delta m = 0$ transitions, in order to be able to drive $\Delta m = \pm 1$ transitions as well.

A custom-built imaging objective of numerical aperture, $NA = 0.26$, is used to image the trapped ion(s). An imaging system sends the light collected from the ion to a photo-multiplier tube (PMT) which gives us the photon counts for the experiments.

\subsubsection*{\label{sec:Freq_calibration} Transition Frequency Calibration}

The frequency calibration process employed in this work requires empirically determining $6S_{1/2} \leftrightarrow 5D_{5/2}$ transition frequencies for only 3 of the encoded states, regardless of the dimension of the qudit.
The first transition frequency is the transition with the lowest magnetic field strength sensitivity, $f_{offset}$, which is the $\lvert 5D_{5/2}, \Tilde{F}=2, m_{\Tilde{F}}=1 \rangle$ state in this work.
The other 2 transition frequencies are transitions with the largest magnetic fields strength sensitivity with respect to each other, $f_{low}$ and $f_{up}$, which are the $\lvert 5D_{5/2}, \Tilde{F}=4, m_{\Tilde{F}}=-4 \rangle$ and $\lvert 5D_{5/2}, \Tilde{F}=4, m_{\Tilde{F}}=4 \rangle$ states in this work.
The transition frequencies for some other state $\lvert n \rangle$ encoded in the $5D_{5/2}$ level, $f_n$, are determined via Eq. \ref{eq:TransitionFreqCal},
\begin{equation}
    f_n = h\left( \boldsymbol{a_n}, \Delta f, f_{offset} \right)
    \label{eq:TransitionFreqCal}
\end{equation}
where $\boldsymbol{a_n} = \left(a_{n,1}, a_{n,2}, \ldots a_{n,N} \right)$ is a list of parameters for the function $h\left( \boldsymbol{a_n}, \Delta f, f_{offset} \right)$ and $\Delta f = f_{up} - f_{low}$.
We find that setting $h\left( \boldsymbol{a_n}, \Delta f, f_{offset} \right)$ as a linear function is sufficient for calibrating the transition frequencies for the drifts that we experience.
So, we use
\begin{equation}
    h\left( \boldsymbol{a_n}, \Delta f, f_{offset} \right) = a_{n,1} \Delta f + f_{offset} + a_{n,2}.
    \label{eq:TransitionFreqLinearCal}
\end{equation}
The parameters $a_{n,1}$ and $a_{n,2}$ are determined from prior experiments (see Supplementary Information).
In principle, it is possible to empirically determine only 2 transition frequencies, $f_{low}$ and $f_{up}$, using this technique, and set $f_{offset}=f_{low}$.
This leads to a different set of parameters $\boldsymbol{a_n}$ that are still sufficient information to determine $f_n$.
However, since $f_{low}$ is magnetically sensitive, it can drift during the calibration process and lead to offset errors for other calibrated frequencies $f_n$.
Empirically, we observe that determining a transition frequency that is insensitive to magnetic field as the offset frequency, $f_{offset}$, is required for optimal results.

\subsubsection*{\label{sec:SPAM_Procedure} Experimental SPAM Procedure}

The AWG waveform frequencies are calibrated to the resonant frequencies of the $\lvert 0 \rangle \leftrightarrow \lvert n \neq 0 \rangle$ transitions, to the precision of $\qty{1}{\kilo\hertz}$. 
The calibrated frequencies are then set as individual waveforms in a waveform sequence table in the AWG. 
A waveform in the AWG waveform sequence table repeats itself indefinitely, until an external trigger signal is sent to the AWG, which causes the AWG to switch to the subsequent waveform in the sequence table.
This allows fast switching of the $\qty{1762}{\nano\meter}$ laser frequency that is sent to the ion.
Then, the pulse time required to drive a $\pi$-pulse transition for each $\lvert 0 \rangle \leftrightarrow \lvert n \rangle$ transition is calibrated.
The detailed methods for calibrating the AWG frequencies and $\pi$-pulse times can be found in Supplementary Information.
With the AWG frequencies and $\pi$-pulse times calibrated, the pulse sequence as shown in Extended Fig. \ref{fig:PulseSequence} is performed for collecting SPAM data.
The experiment is repeated 1000 times for each prepared state to obtain a sample size of 1000.

\subsection*{Extended Data Figures and Tables}

\setcounter{figure}{0}
\setcounter{table}{0}
\renewcommand{\thefigure}{E\arabic{figure}}
\renewcommand{\thetable}{E\arabic{table}}

\begin{figure}[H]
\centering
    \includegraphics[width=\linewidth]{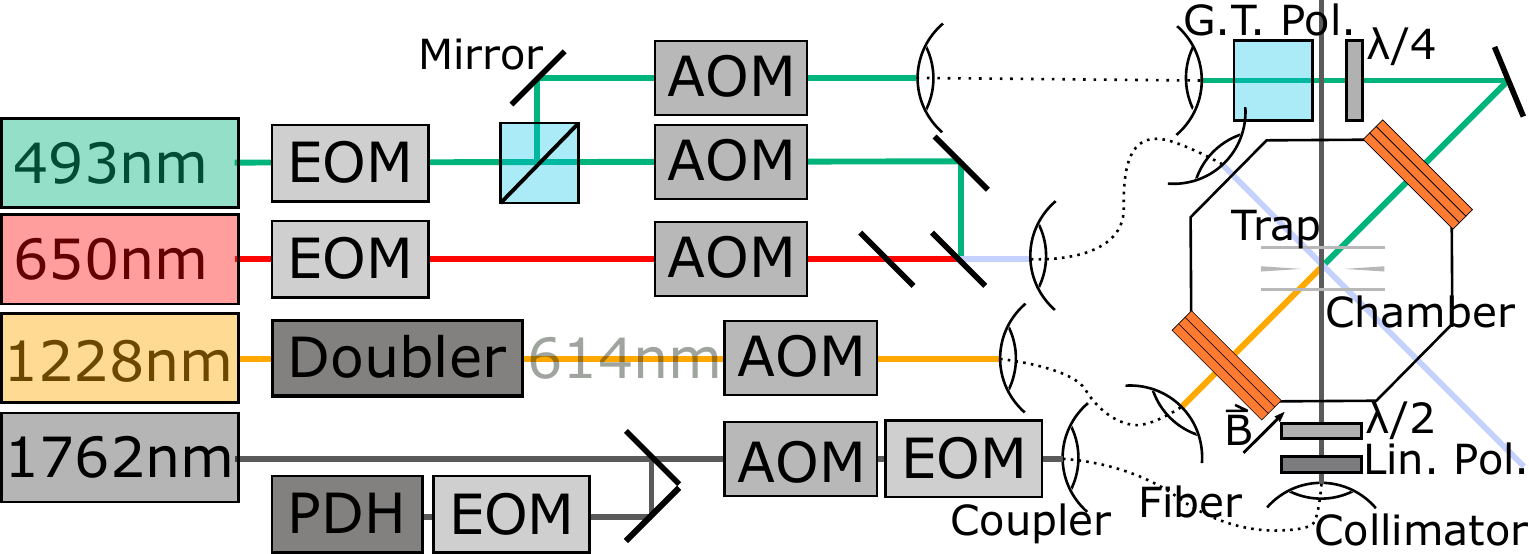}
    \caption{Detailed illustration of the optical setup used in this work.
    G.T. Pol. and Lin. Pol. denotes Glan-Thompson polarizer and linear polarizer respectively.
    $\lambda/2$ and $\lambda/4$ denote half waveplate and quarter waveplate respectively.}
    \label{fig:BeamOrientation}
\end{figure}

\begin{figure}[H]
\centering
\begin{subfigure}{0.3\linewidth}
    \includegraphics[width=\linewidth]{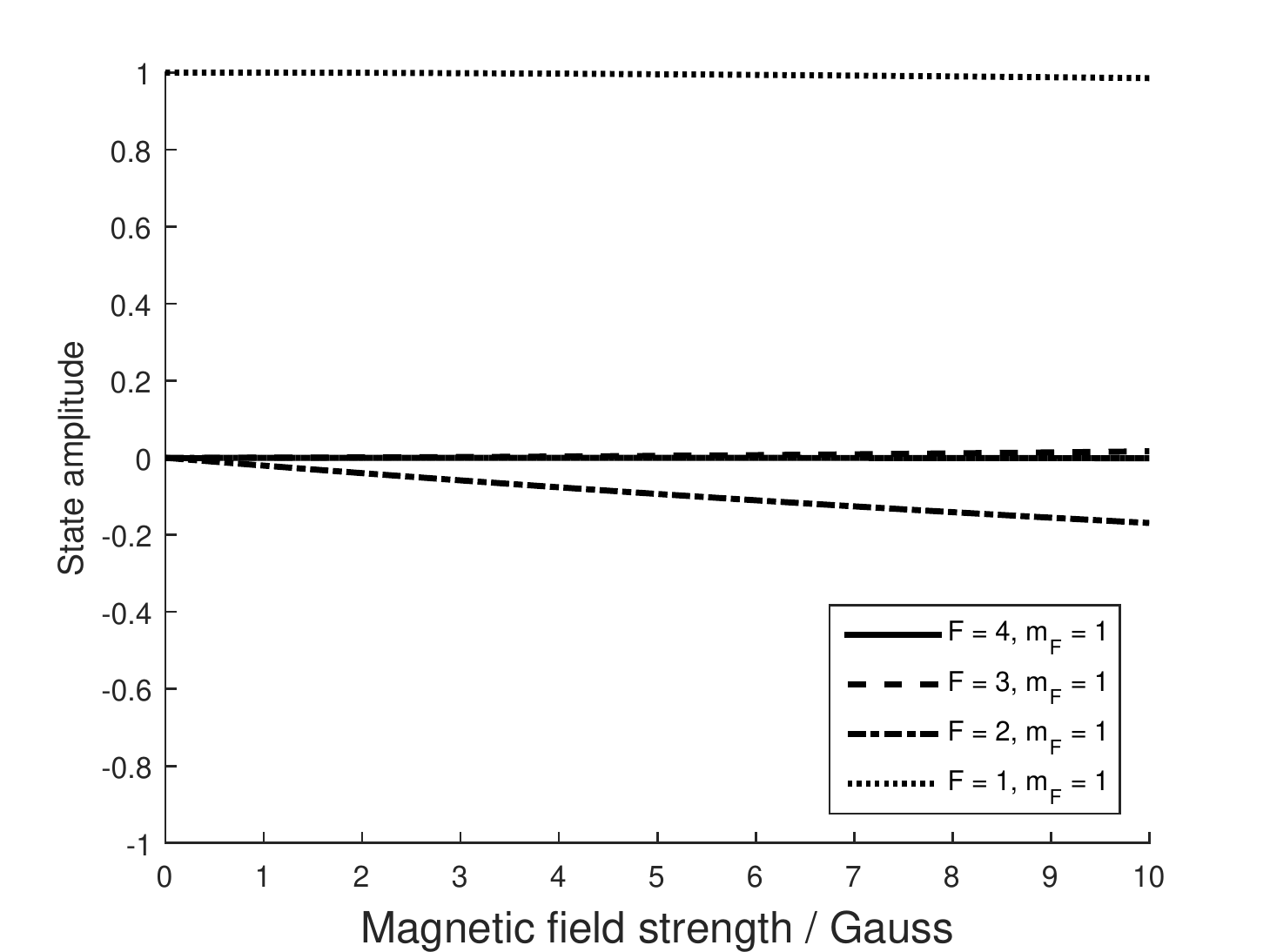}
    \caption{$\Tilde{F}=1, m_{\Tilde{F}}=1$}
\end{subfigure}
\begin{subfigure}{0.3\linewidth}
    \includegraphics[width=\linewidth]{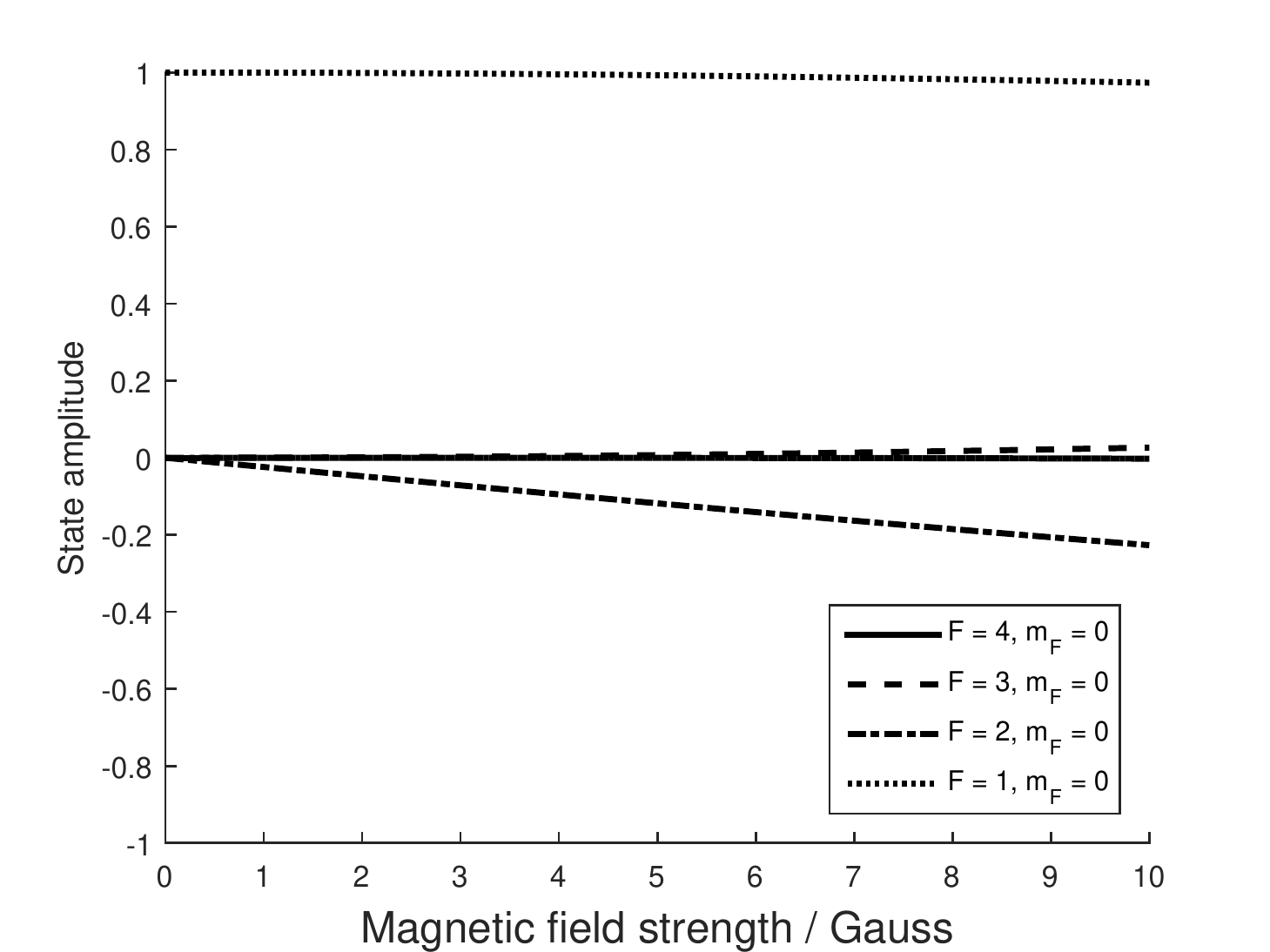}
    \caption{$\Tilde{F}=1, m_{\Tilde{F}}=0$}
\end{subfigure}
\begin{subfigure}{0.3\linewidth}
    \includegraphics[width=\linewidth]{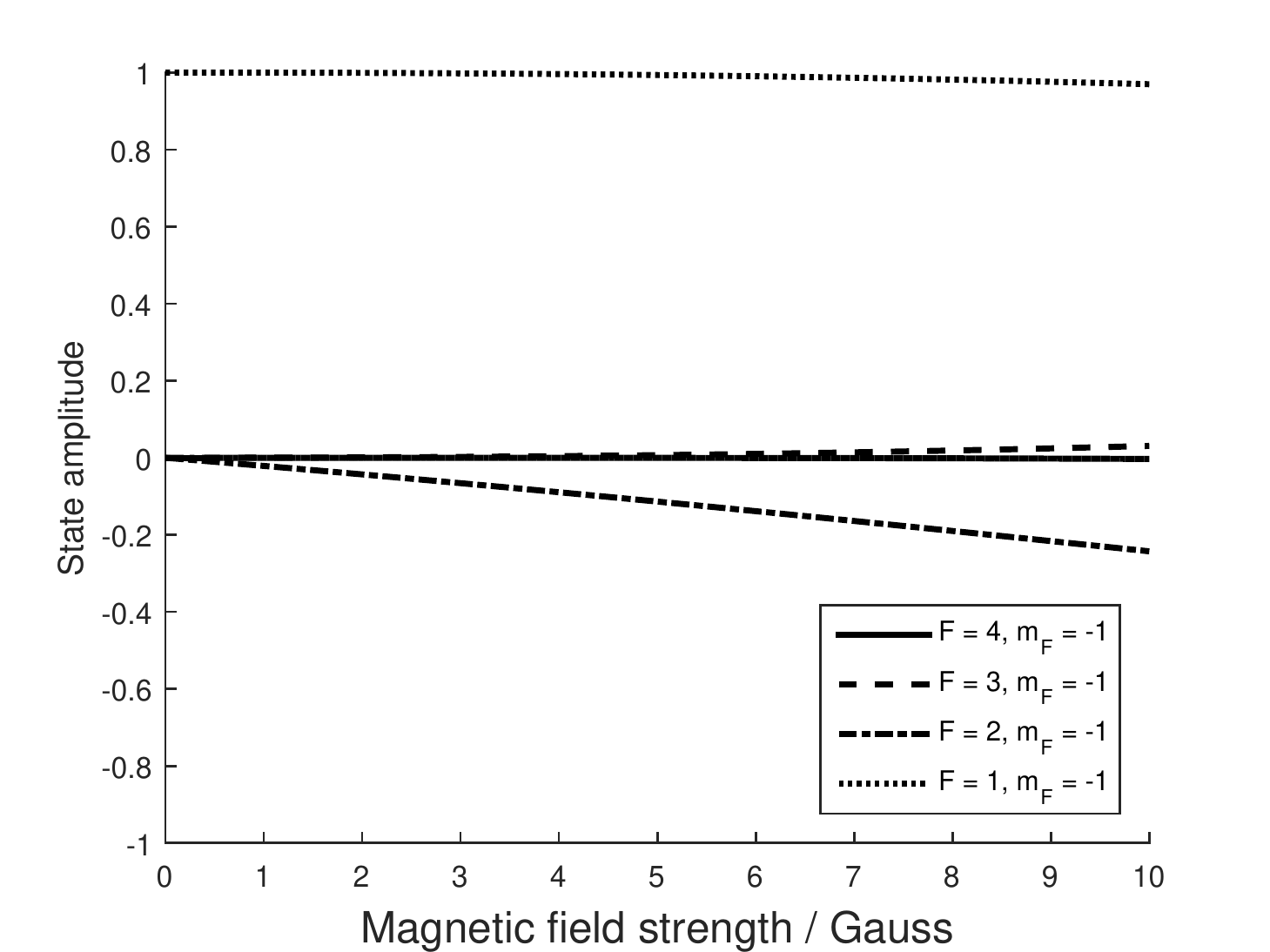}
    \caption{$\Tilde{F}=1, m_{\Tilde{F}}=-1$}
\end{subfigure}
\begin{subfigure}{0.3\linewidth}
    \includegraphics[width=\linewidth]{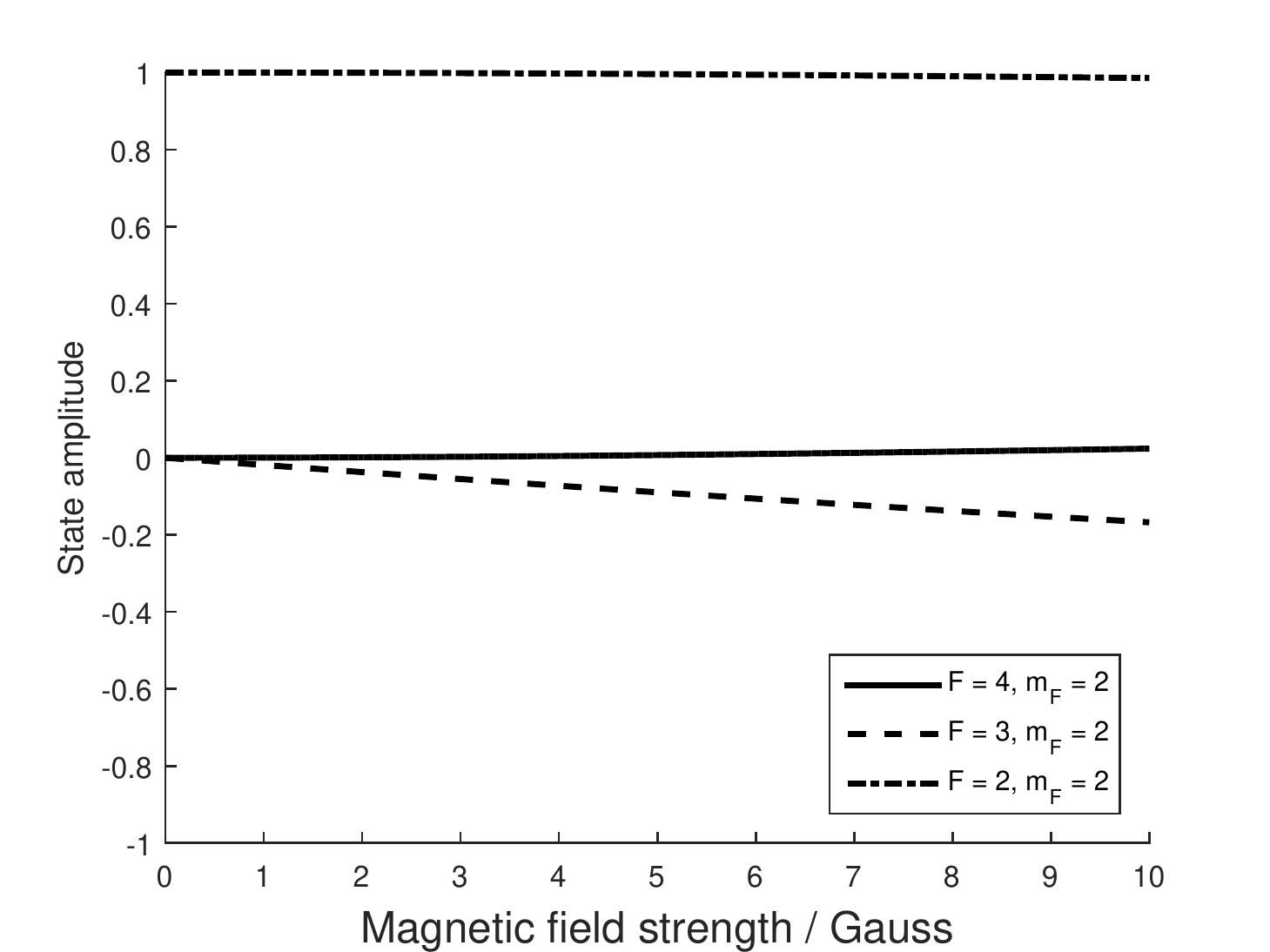}
    \caption{$\Tilde{F}=2, m_{\Tilde{F}}=2$}
\end{subfigure}
\begin{subfigure}{0.3\linewidth}
    \includegraphics[width=\linewidth]{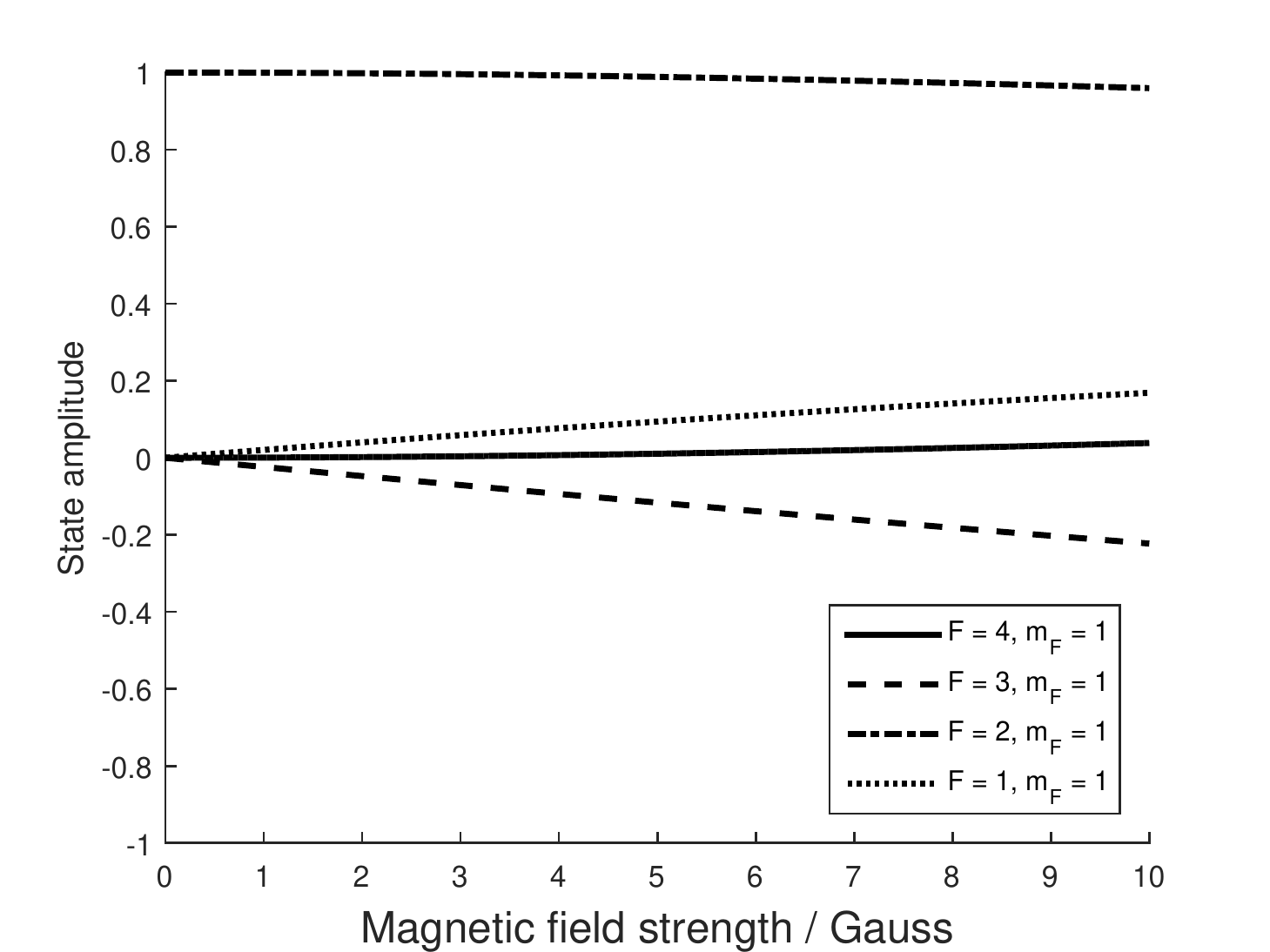}
    \caption{$\Tilde{F}=2, m_{\Tilde{F}}=1$}
\end{subfigure}
\begin{subfigure}{0.3\linewidth}
    \includegraphics[width=\linewidth]{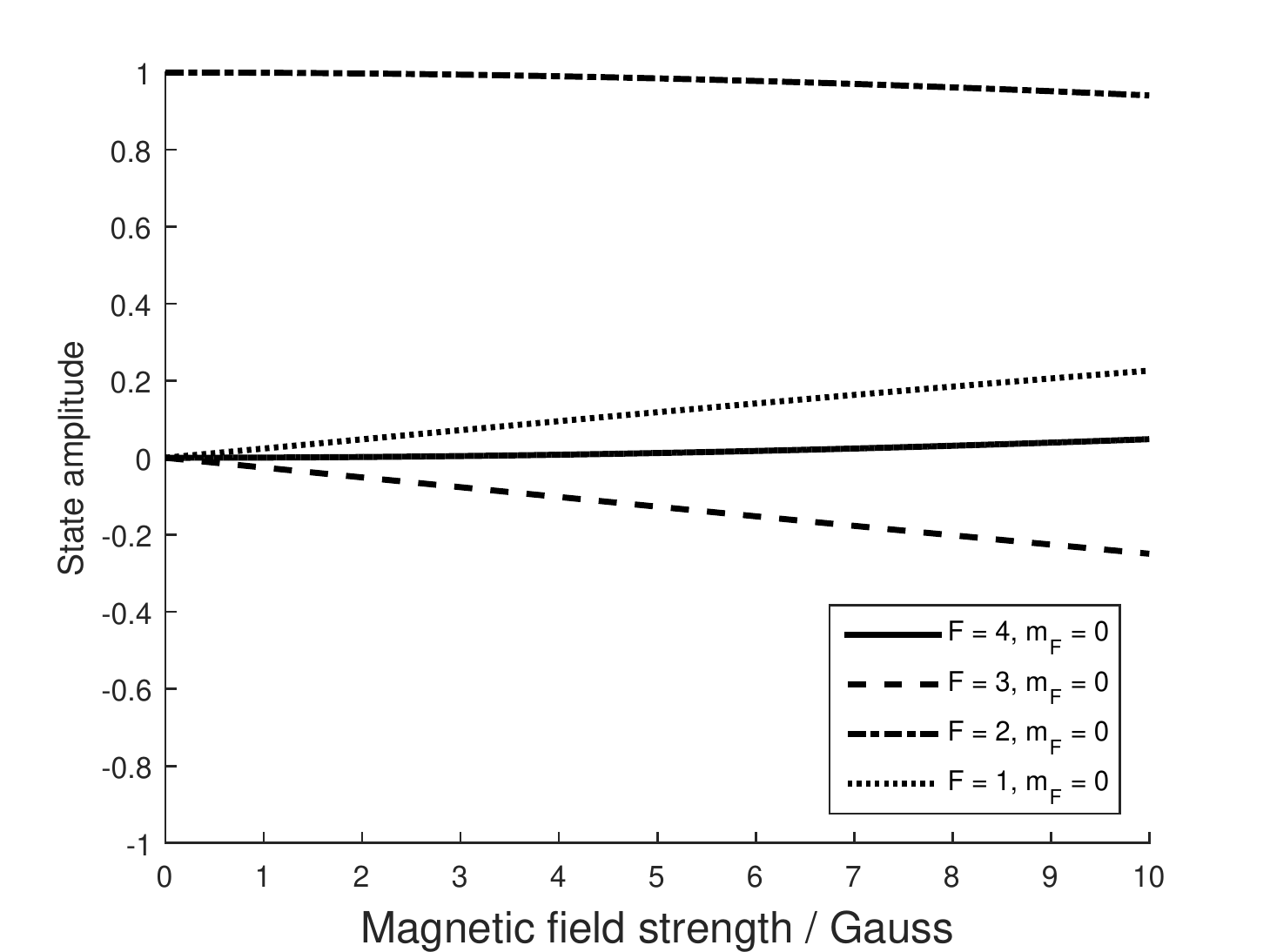}
    \caption{$\Tilde{F}=2, m_{\Tilde{F}}=0$}
\end{subfigure}
\begin{subfigure}{0.3\linewidth}
    \includegraphics[width=\linewidth]{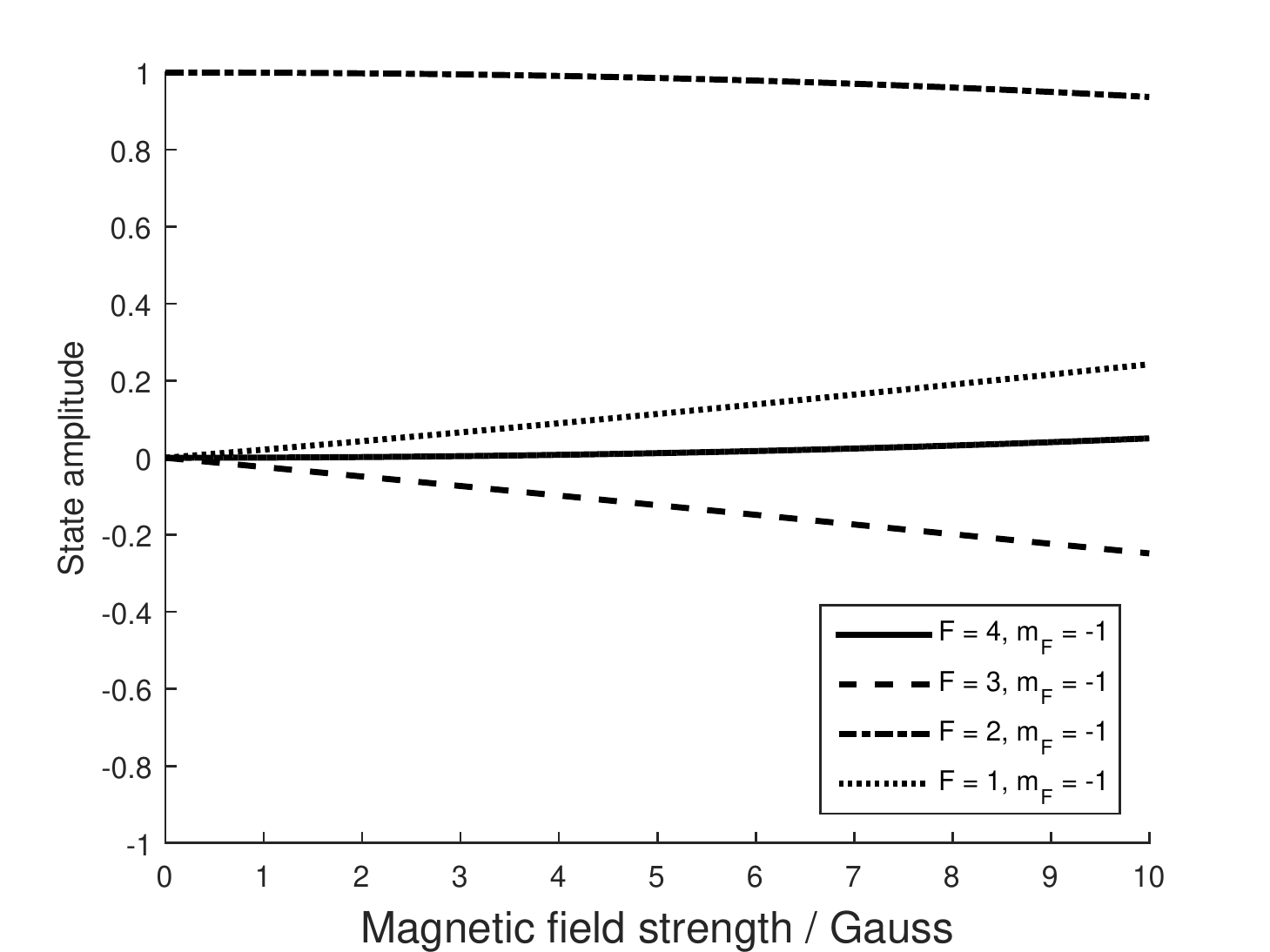}
    \caption{$\Tilde{F}=2, m_{\Tilde{F}}=-1$}
\end{subfigure}
\begin{subfigure}{0.3\linewidth}
    \includegraphics[width=\linewidth]{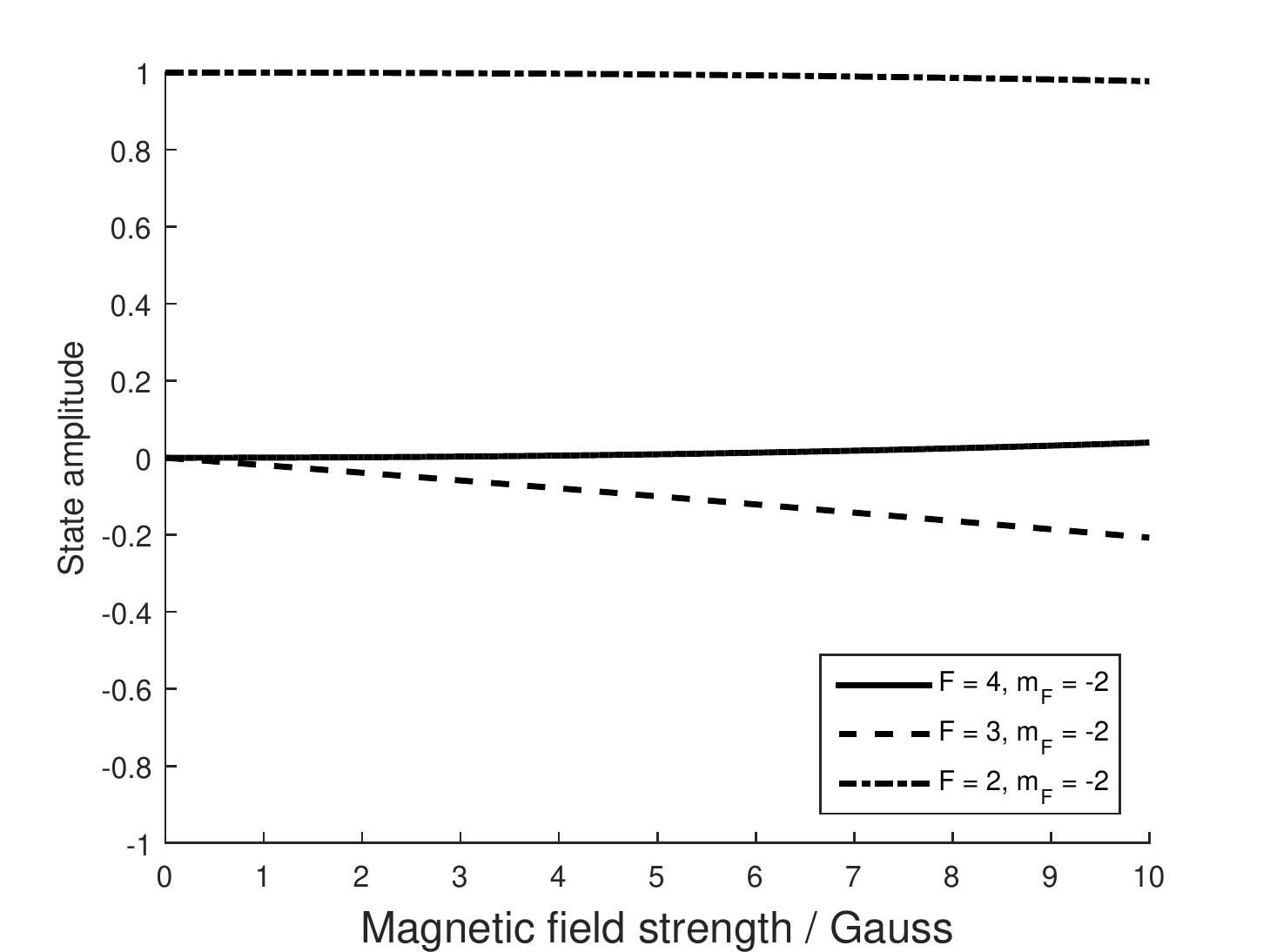}
    \caption{$\Tilde{F}=2, m_{\Tilde{F}}=-2$}
\end{subfigure}
\caption{Numerical estimations of the \ce{^{137}Ba^{+}} $\Tilde{F}=1$ and $\Tilde{F}=2$ energy eigenstates in the $5D_{5/2}$ level, expressed in the $\lvert F, m_F \rangle$ basis.}
\label{fig:SuppMat_EigenstatesEvol_F1F2}
\end{figure}

\begin{figure}[H]
\centering
\begin{subfigure}{0.3\linewidth}
    \includegraphics[width=\linewidth]{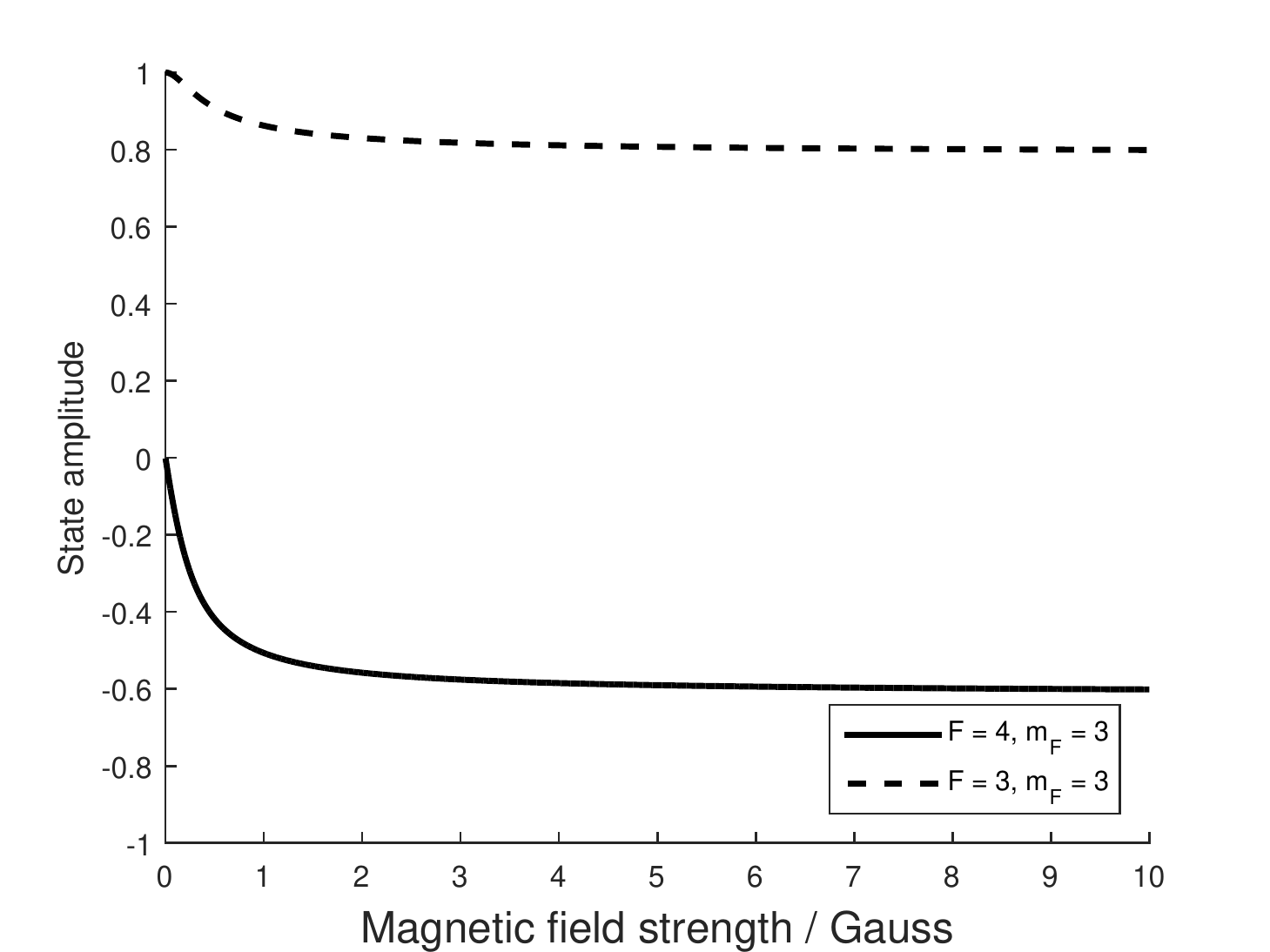}
    \caption{$\Tilde{F}=3, m_{\Tilde{F}}=3$}
\end{subfigure}
\begin{subfigure}{0.3\linewidth}
    \includegraphics[width=\linewidth]{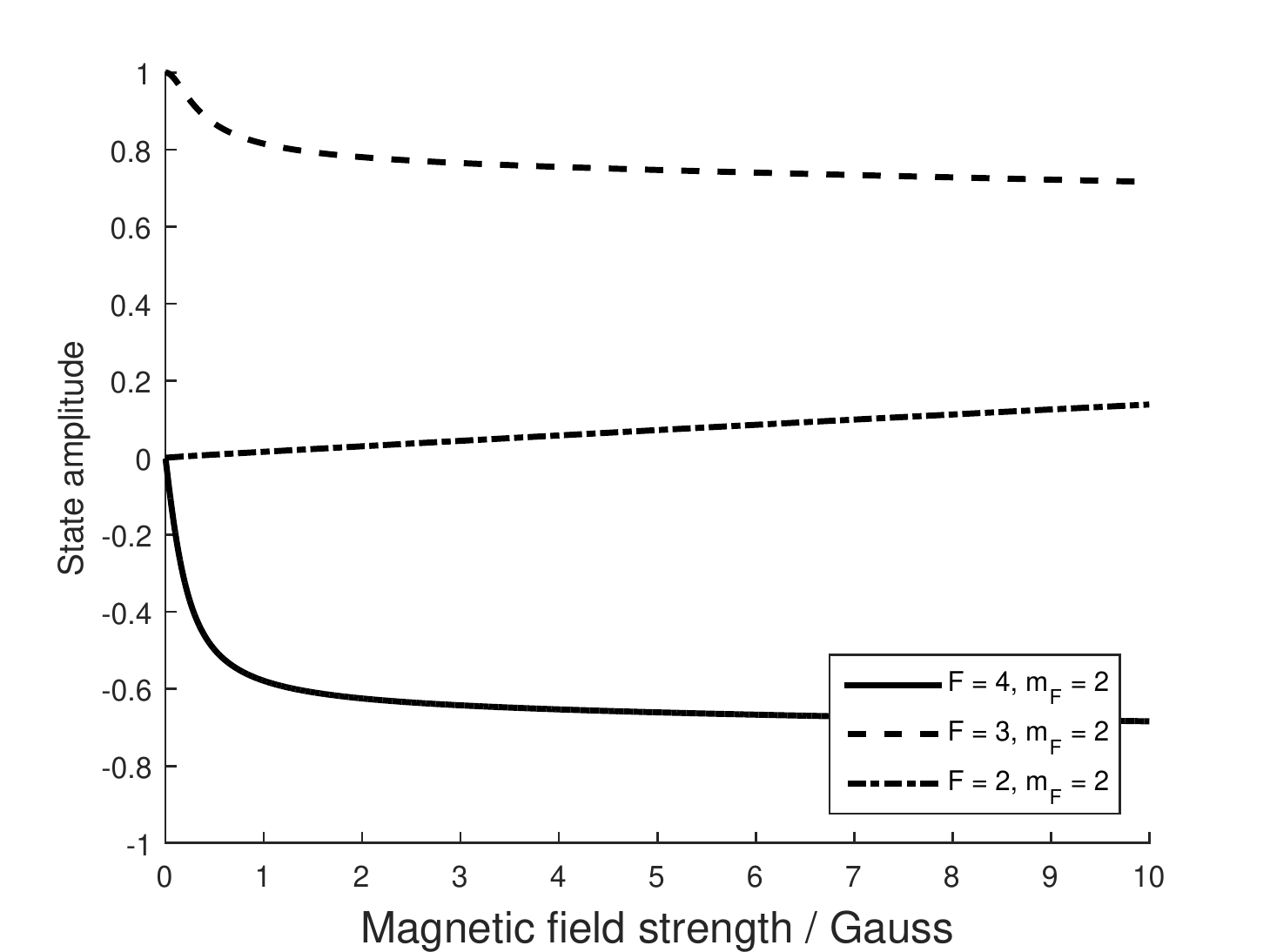}
    \caption{$\Tilde{F}=3, m_{\Tilde{F}}=2$}
\end{subfigure}
\begin{subfigure}{0.3\linewidth}
    \includegraphics[width=\linewidth]{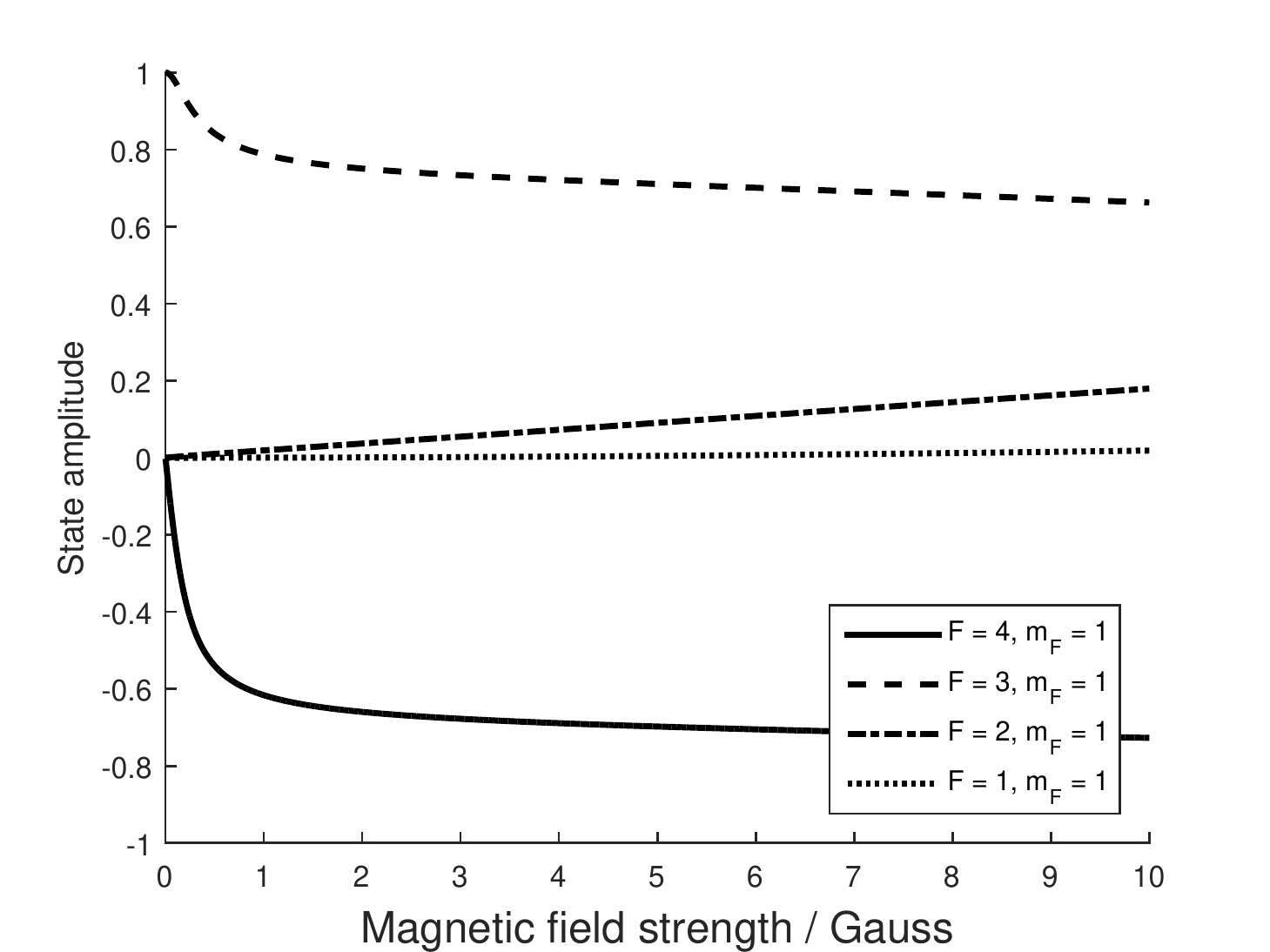}
    \caption{$\Tilde{F}=3, m_{\Tilde{F}}=1$}
\end{subfigure}
\begin{subfigure}{0.3\linewidth}
    \includegraphics[width=\linewidth]{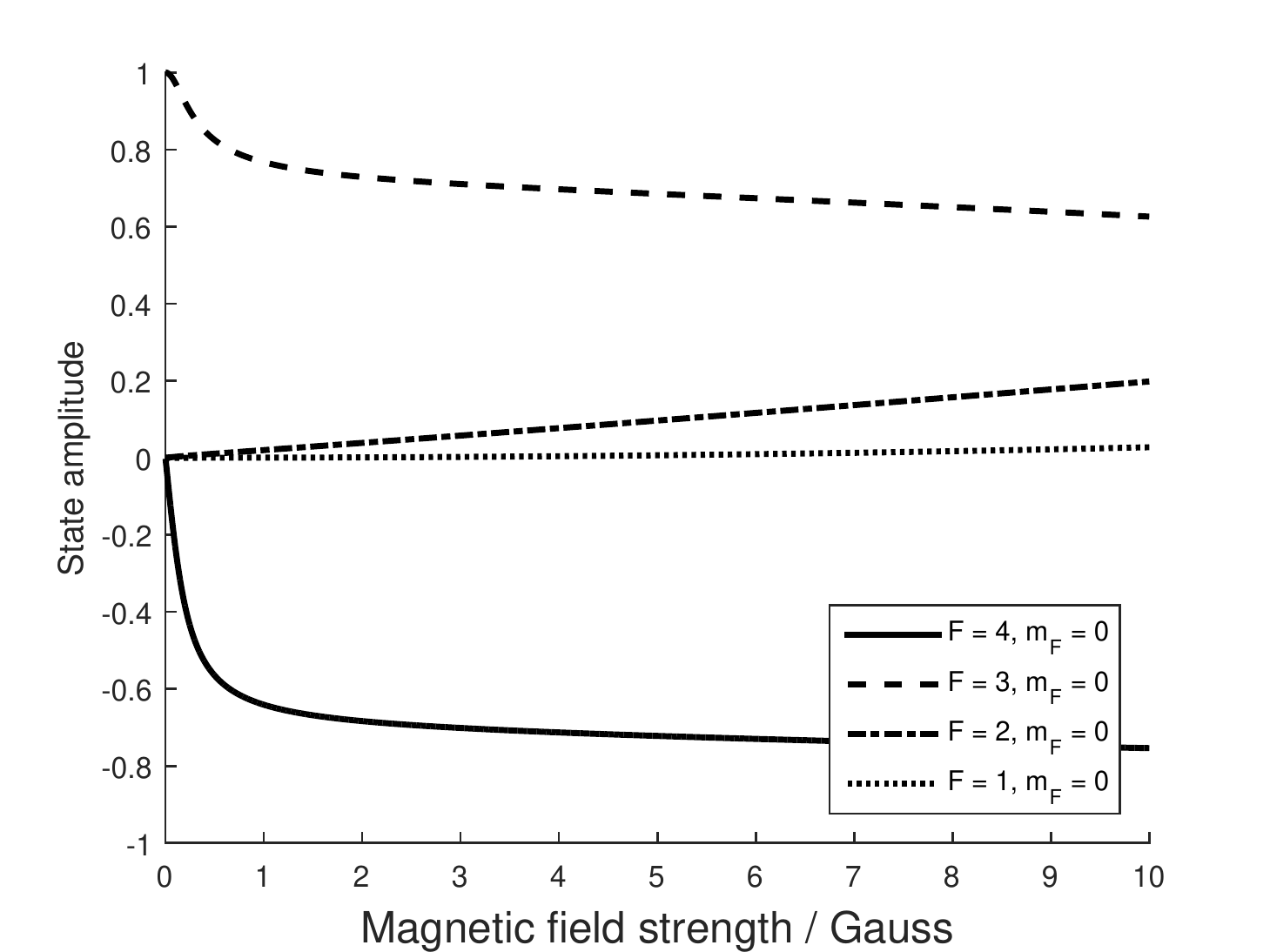}
    \caption{$\Tilde{F}=3, m_{\Tilde{F}}=0$}
\end{subfigure}
\begin{subfigure}{0.3\linewidth}
    \includegraphics[width=\linewidth]{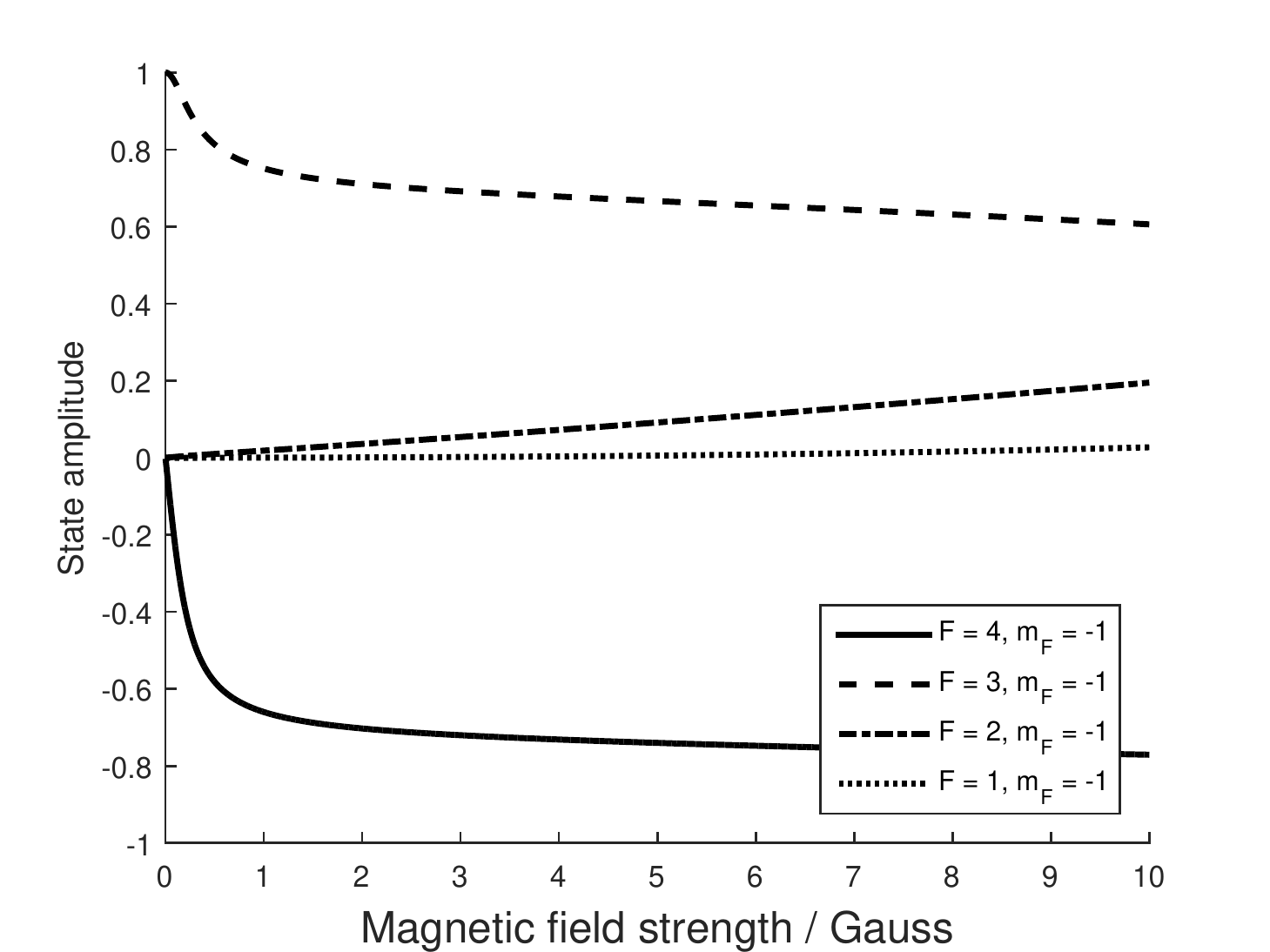}
    \caption{$\Tilde{F}=3, m_{\Tilde{F}}=-1$}
\end{subfigure}
\begin{subfigure}{0.3\linewidth}
    \includegraphics[width=\linewidth]{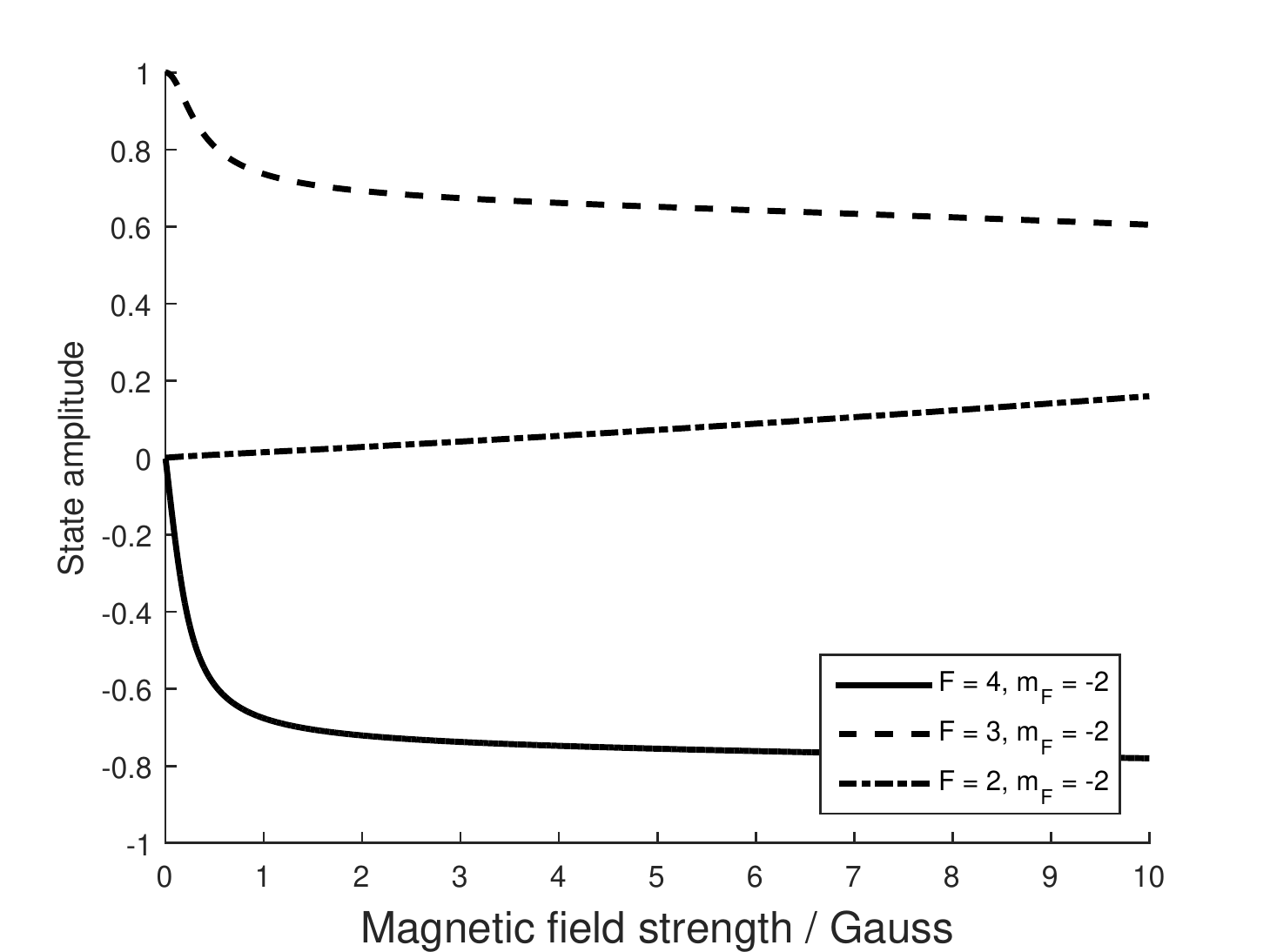}
    \caption{$\Tilde{F}=3, m_{\Tilde{F}}=-2$}
\end{subfigure}
\begin{subfigure}{0.3\linewidth}
    \includegraphics[width=\linewidth]{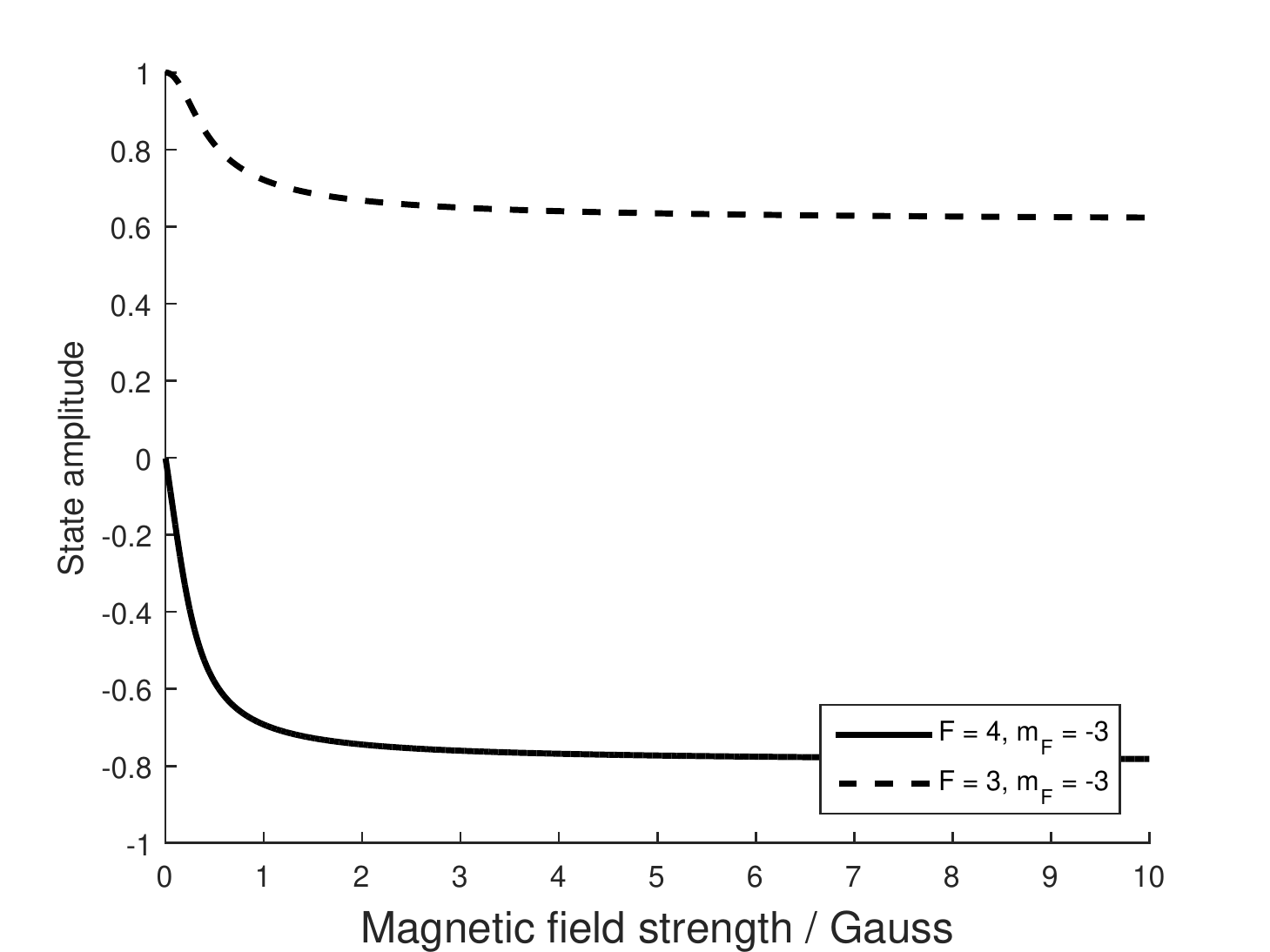}
    \caption{$\Tilde{F}=3, m_{\Tilde{F}}=-3$}
\end{subfigure}
\caption{Numerical estimations of the \ce{^{137}Ba^{+}} $\Tilde{F}=3$ energy eigenstates in the $5D_{5/2}$ level, expressed in the $\lvert F, m_F \rangle$ basis.}
\label{fig:SuppMat_EigenstatesEvol_F3}
\end{figure}

\begin{figure}[H]
\centering
\begin{subfigure}{0.3\linewidth}
    \includegraphics[width=\linewidth]{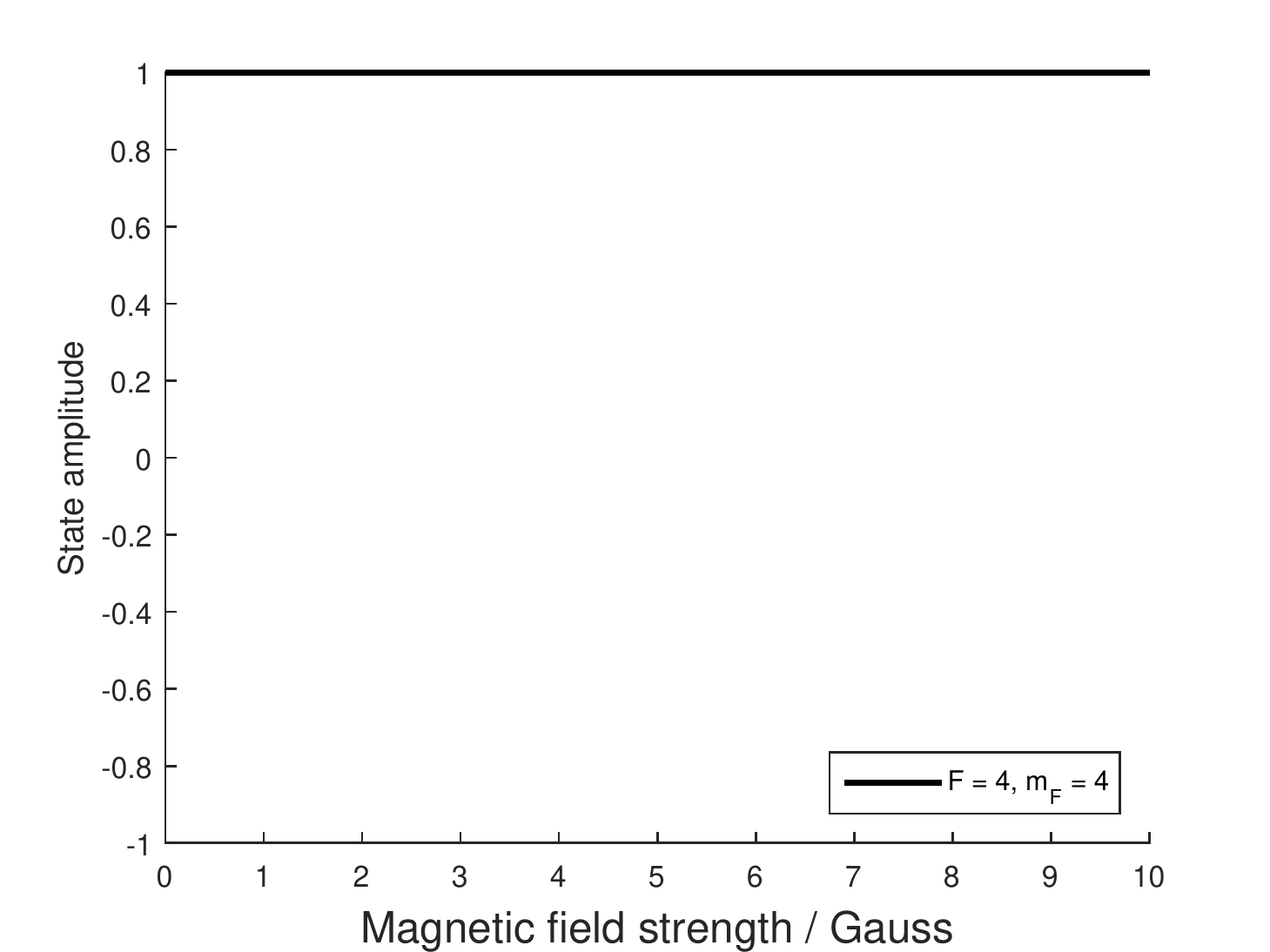}
    \caption{$\Tilde{F}=4, m_{\Tilde{F}}=4$}
\end{subfigure}
\begin{subfigure}{0.3\linewidth}
    \includegraphics[width=\linewidth]{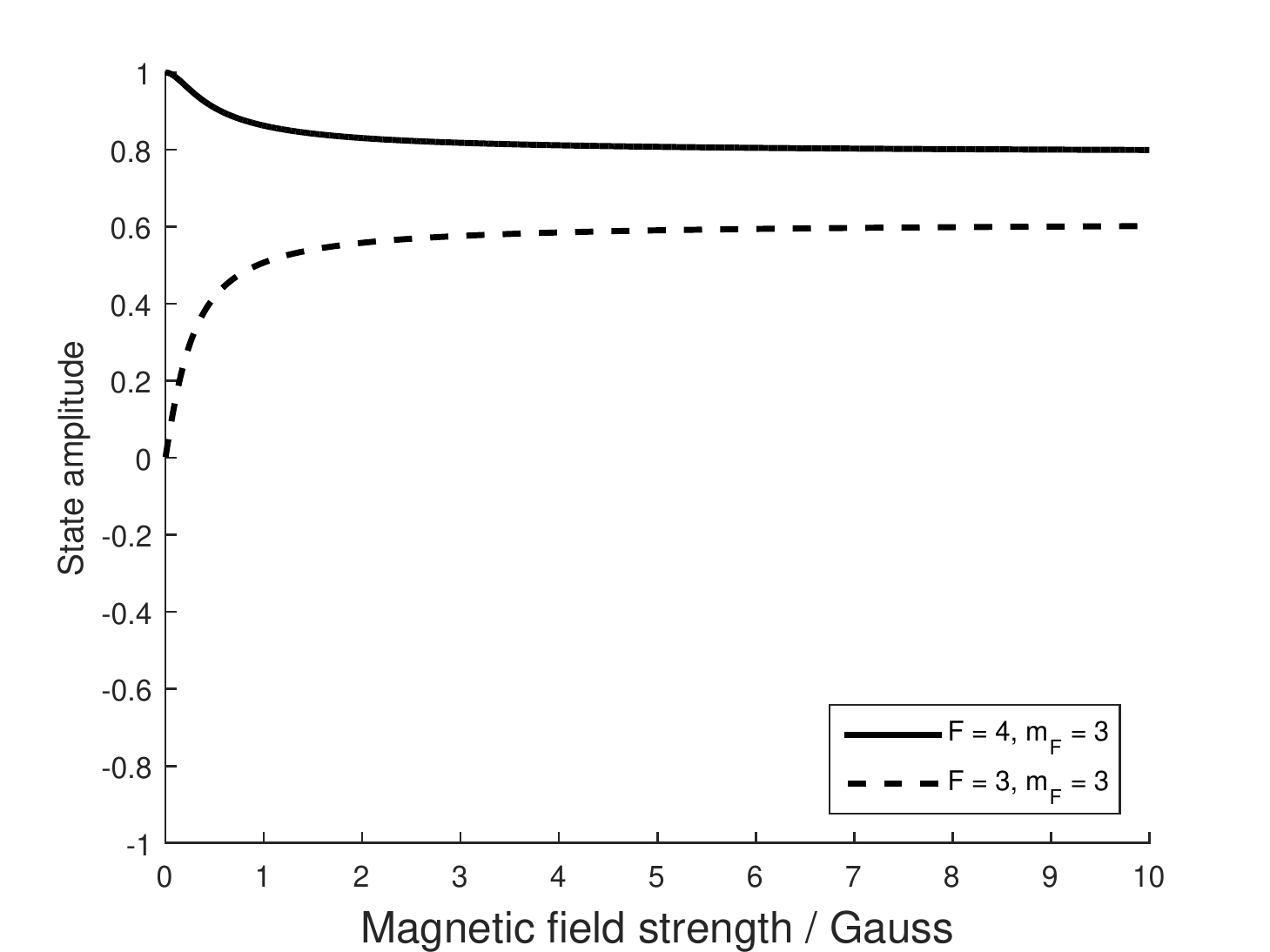}
    \caption{$\Tilde{F}=4, m_{\Tilde{F}}=3$}
\end{subfigure}
\begin{subfigure}{0.3\linewidth}
    \includegraphics[width=\linewidth]{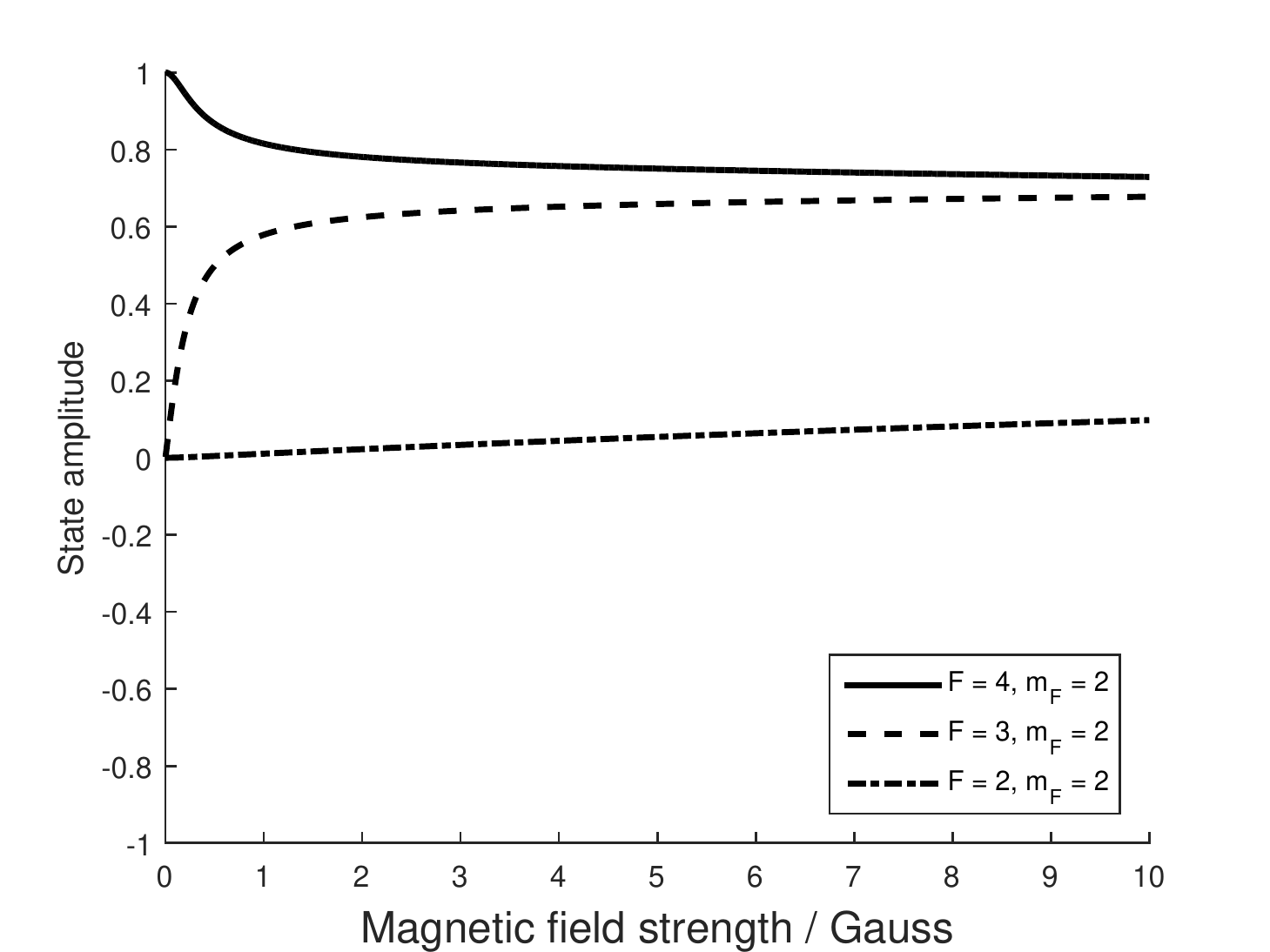}
    \caption{$\Tilde{F}=4, m_{\Tilde{F}}=2$}
\end{subfigure}
\begin{subfigure}{0.3\linewidth}
    \includegraphics[width=\linewidth]{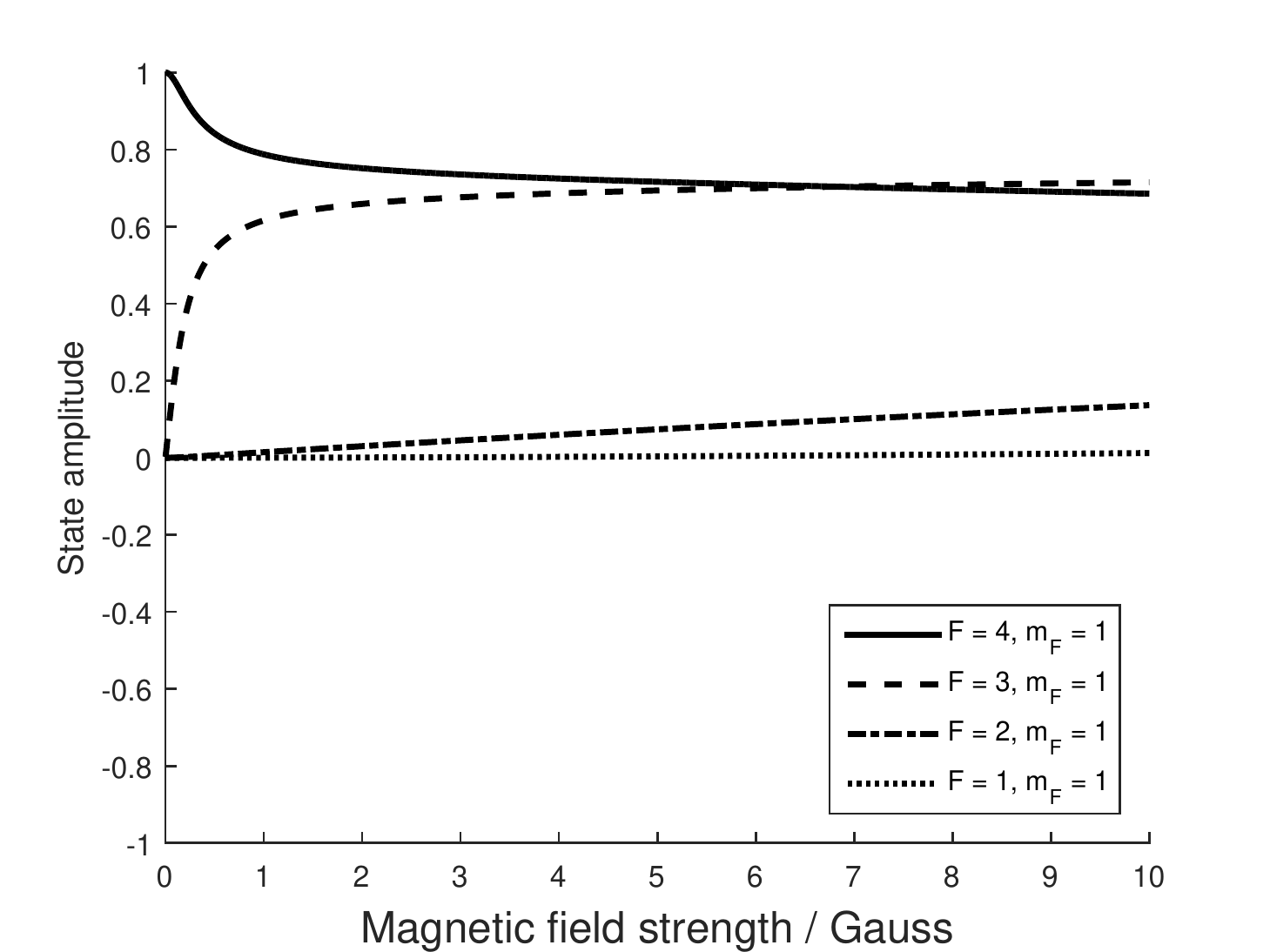}
    \caption{$\Tilde{F}=4, m_{\Tilde{F}}=1$}
\end{subfigure}
\begin{subfigure}{0.3\linewidth}
    \includegraphics[width=\linewidth]{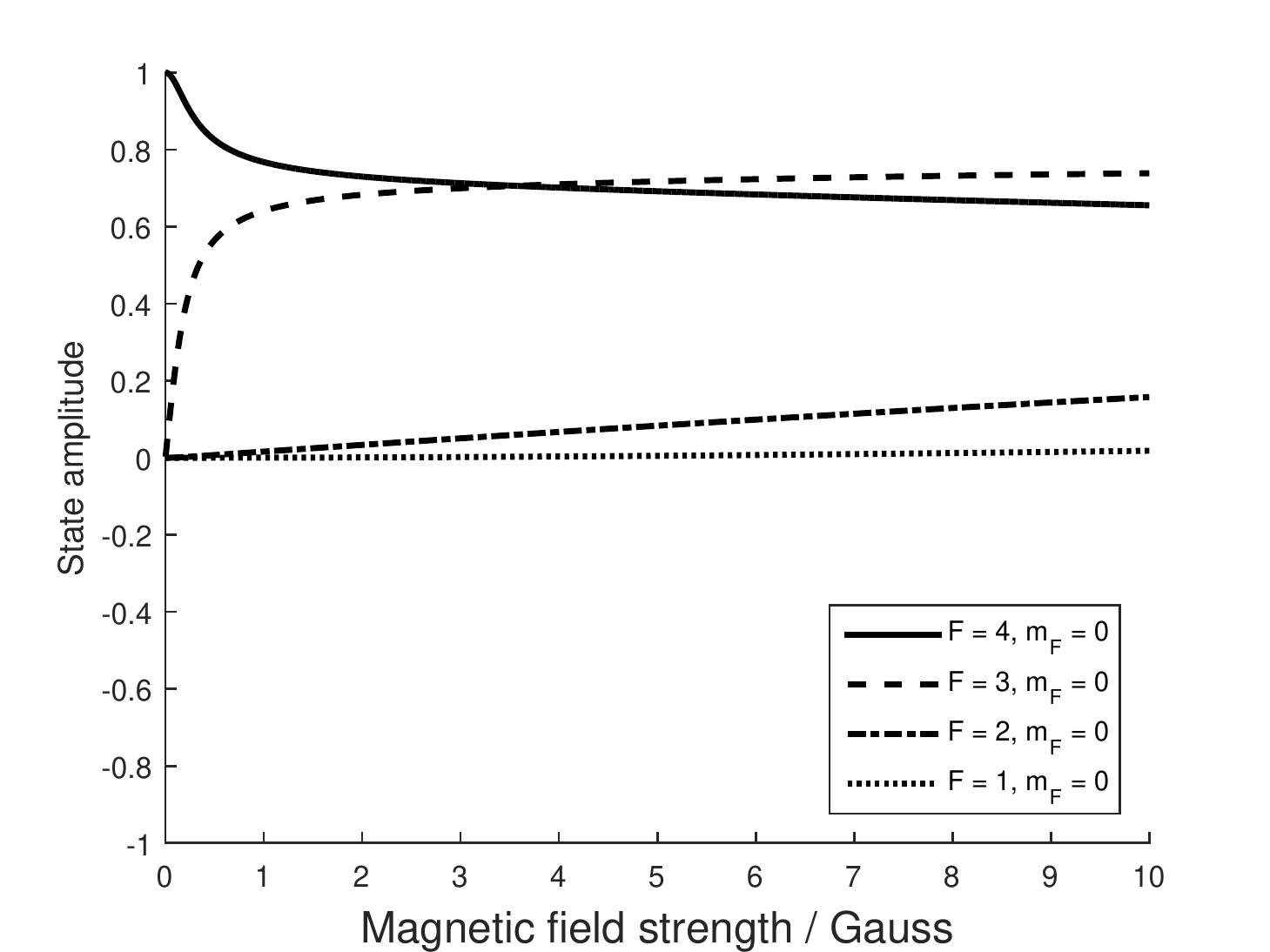}
    \caption{$\Tilde{F}=4, m_{\Tilde{F}}=0$}
\end{subfigure}
\begin{subfigure}{0.3\linewidth}
    \includegraphics[width=\linewidth]{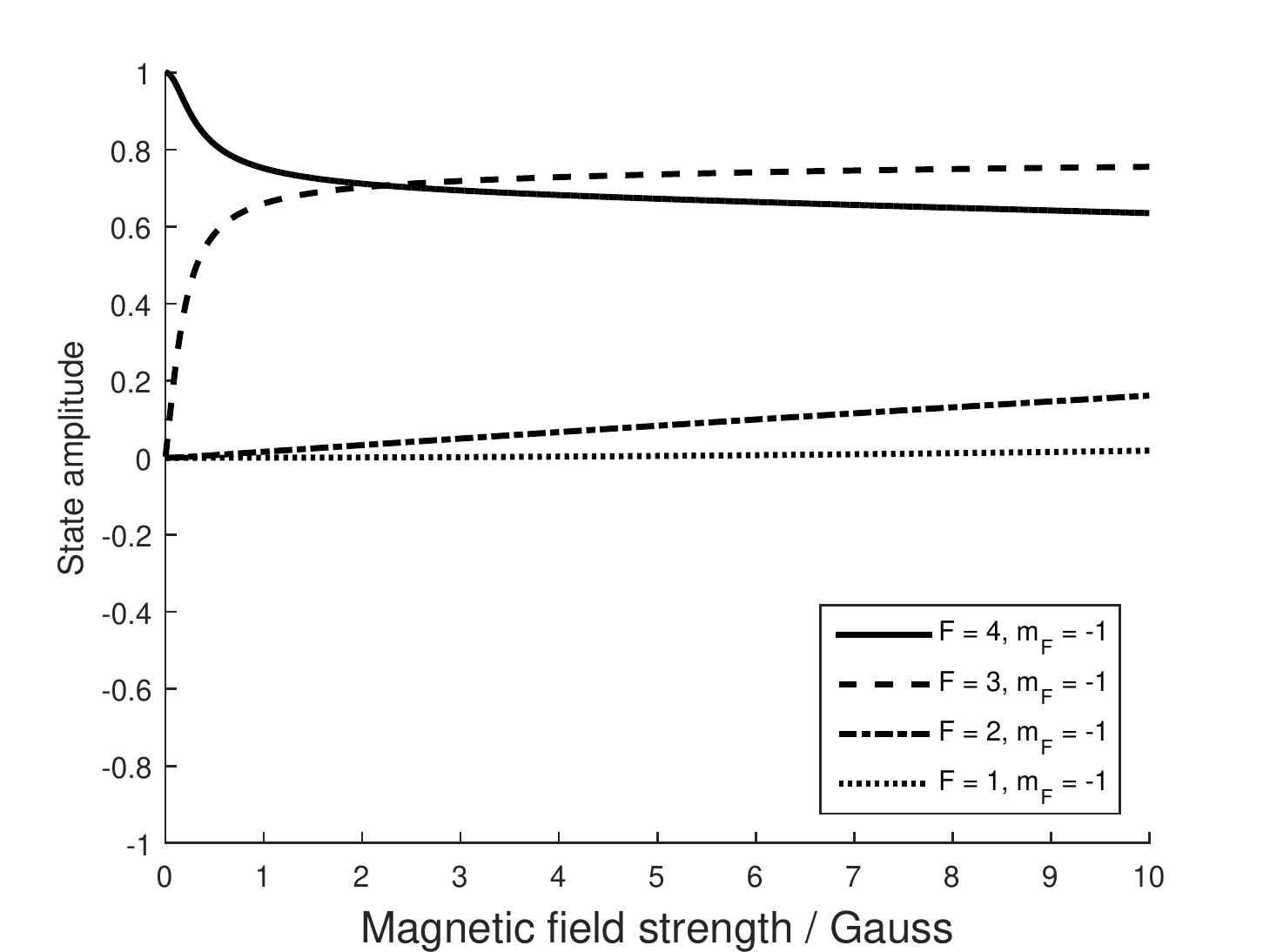}
    \caption{$\Tilde{F}=4, m_{\Tilde{F}}=-1$}
\end{subfigure}
\begin{subfigure}{0.3\linewidth}
    \includegraphics[width=\linewidth]{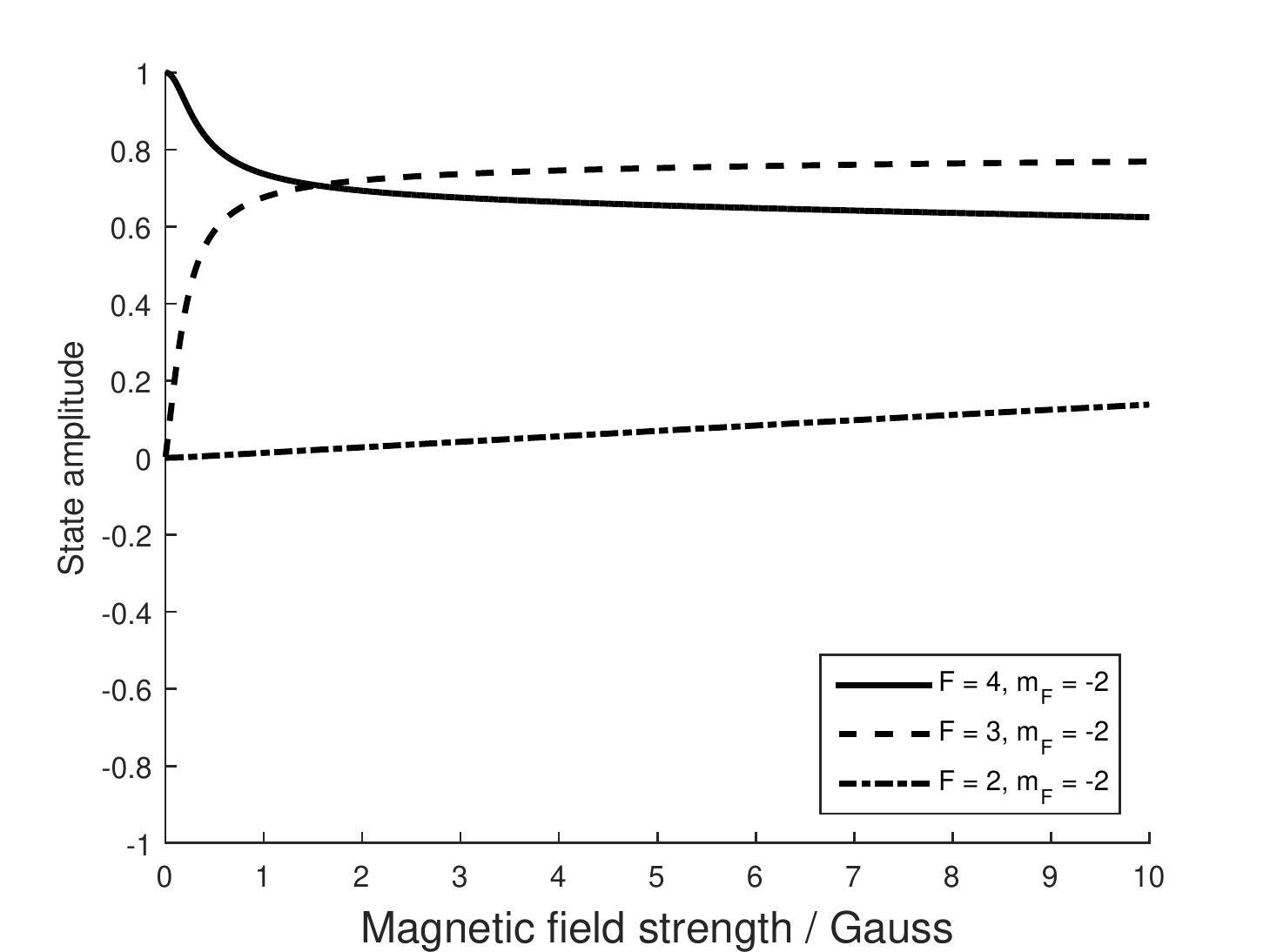}
    \caption{$\Tilde{F}=4, m_{\Tilde{F}}=-2$}
\end{subfigure}
\begin{subfigure}{0.3\linewidth}
    \includegraphics[width=\linewidth]{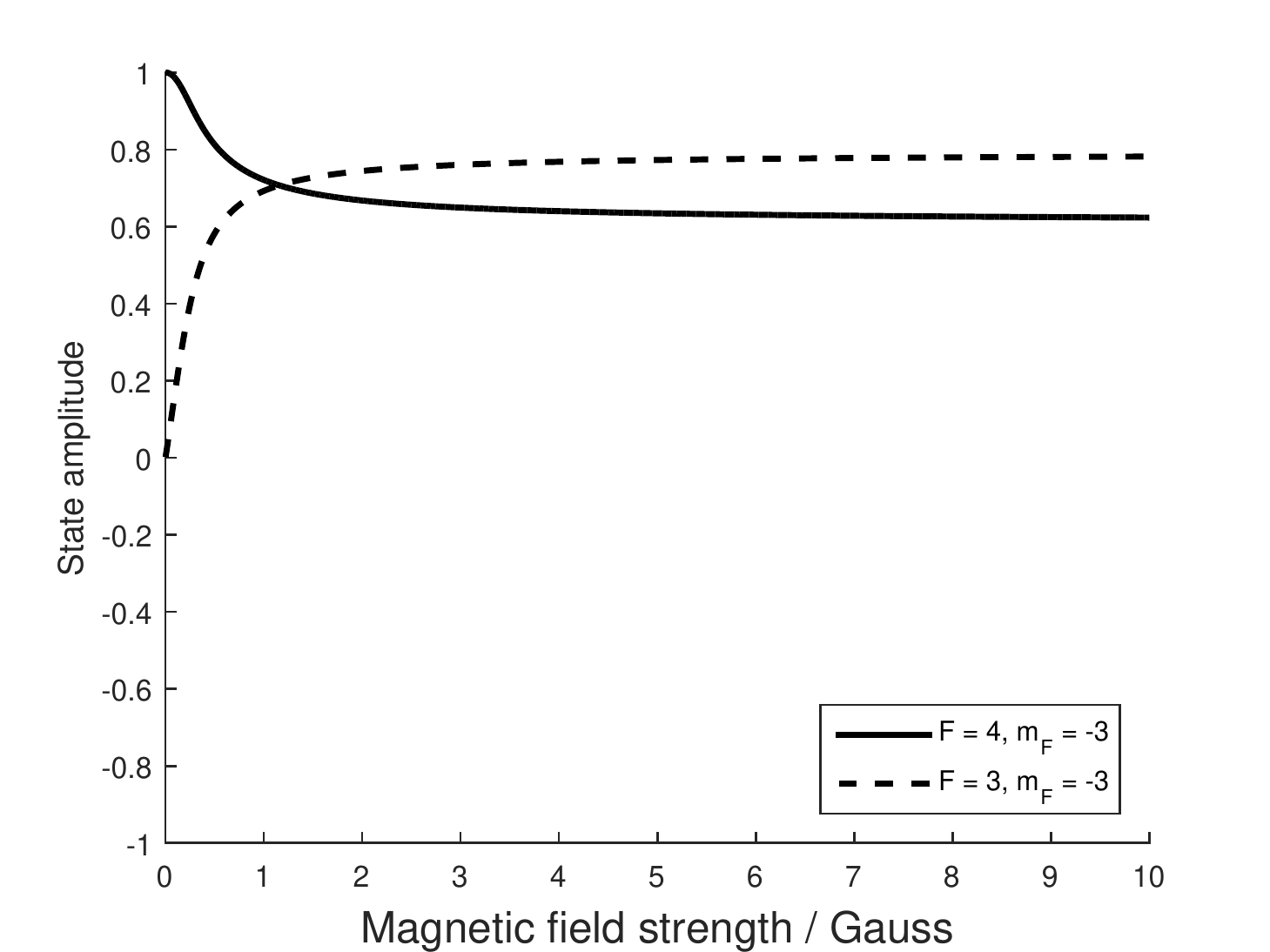}
    \caption{$\Tilde{F}=4, m_{\Tilde{F}}=-3$}
\end{subfigure}
\begin{subfigure}{0.3\linewidth}
    \includegraphics[width=\linewidth]{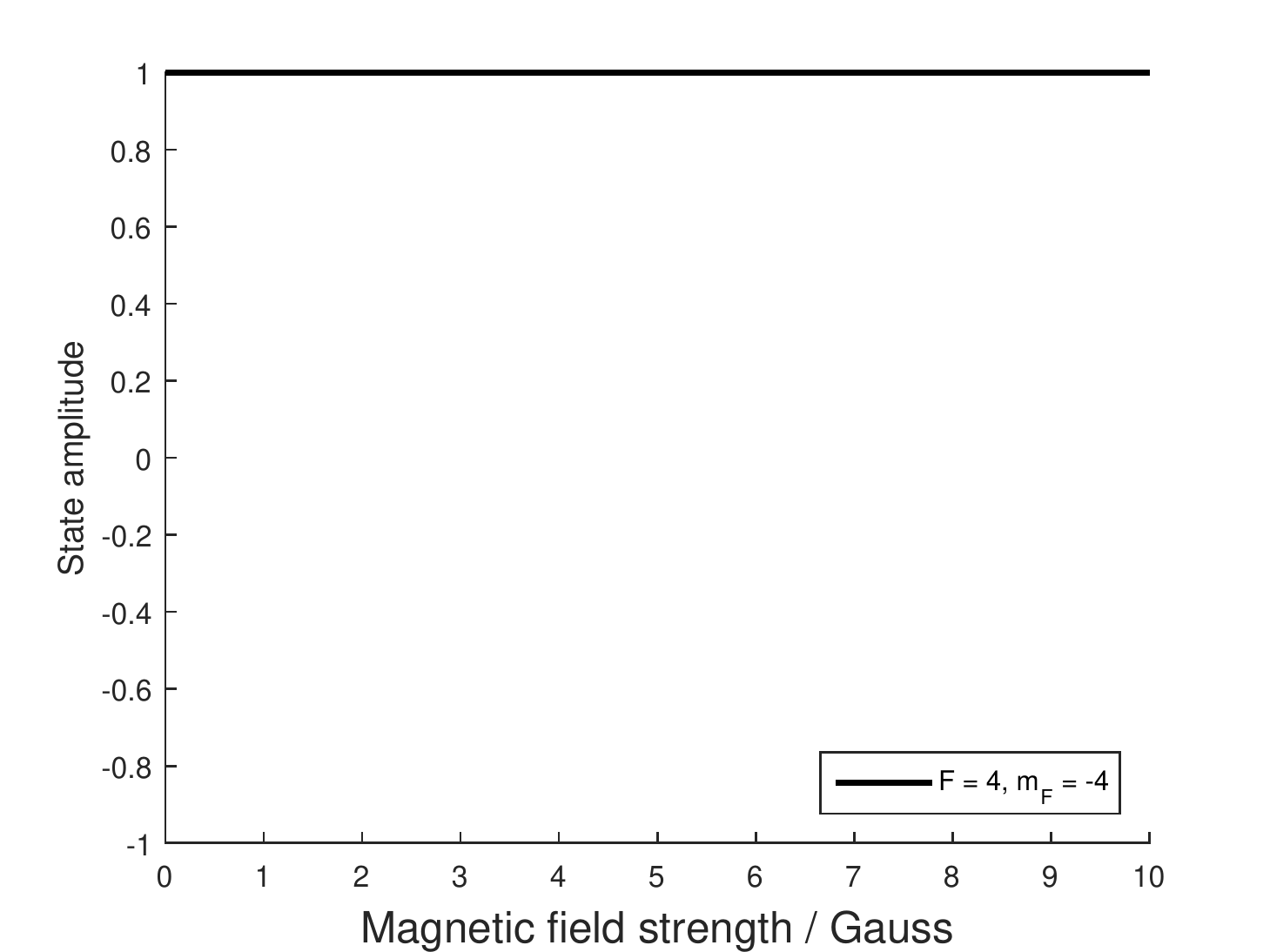}
    \caption{$\Tilde{F}=4, m_{\Tilde{F}}=-4$}
\end{subfigure}
\caption{Numerical estimations of the \ce{^{137}Ba^{+}} $\Tilde{F}=4$ energy eigenstates in the $5D_{5/2}$ level, expressed in the $\lvert F, m_F \rangle$ basis.}
\label{fig:SuppMat_EigenstatesEvol_F4}
\end{figure}

\begin{figure}[H]
\centering
\includegraphics[width=\linewidth]{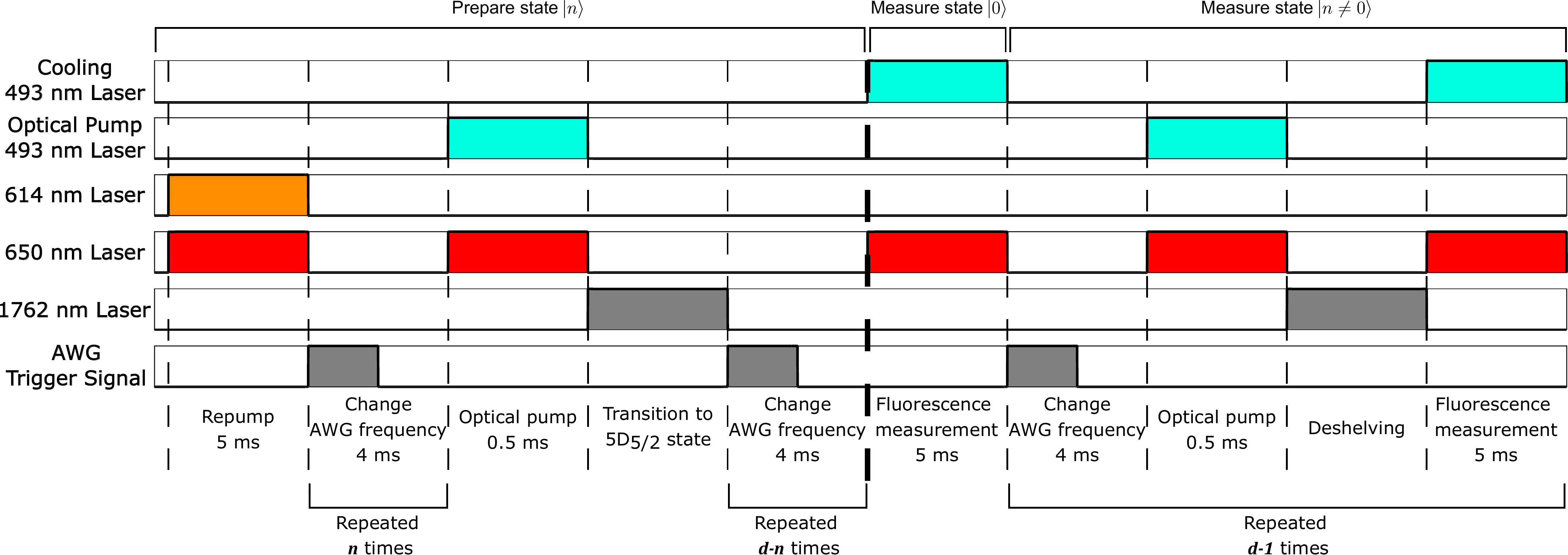}
\caption{Pulse sequence for one SPAM experiment for a prepared $\lvert n \rangle$ state of a $d$-dimensional qudit demonstrated in this work. 
The first five rows denote the on/off states of the laser beams as described in text. 
The last row denotes an electrical signal sent to the AWG to trigger the switch to the next waveform, effectively changing the frequency of the \qty{1762}{\nano\meter} laser. 
The time axis is not drawn to scale.
The pulse sequences to the left and right of the bold dashed line are the state preparation and state measurement processes respectively.
A list of $d$ individual frequencies corresponding to the transitions for each encoded state in a $d$-dimensional qudit is set in the AWG, and frequency switching is achieved by the AWG trigger signal.
For state $\lvert 0 \rangle$, the AWG frequency is set to a frequency that is off-resonant to any of the $6S_{1/2} \leftrightarrow 5D_{5/2}$ transitions.
The optical pumping step in the measurement process is technically unnecessary for the measurement procedure described in the main text.
It is present in this work for investigating another method of determining a measured state, which we have found to be less robust than what is described in the main text (see Supplementary Information).
}
\label{fig:PulseSequence}
\end{figure}

\begin{table}[H]
    \centering
    \begin{tabular}{|c|cccccccc|}
    \hline
        \multirow{3}{*}{$5D_{5/2}$ state} & \multicolumn{8}{c|}{$6S_{1/2}$ state} \\
        \cline{2-9}
         & \multicolumn{3}{c|}{$\Tilde{F}=1$} & \multicolumn{5}{c|}{$\Tilde{F}=2$} \\
         \cline{2-9}
         & \multicolumn{1}{c|}{$m_{\Tilde{F}}=-1$} & \multicolumn{1}{c|}{$m_{\Tilde{F}}=0$} & \multicolumn{1}{c|}{$m_{\Tilde{F}}=1$} & \multicolumn{1}{c|}{$m_{\Tilde{F}}=-2$} & \multicolumn{1}{c|}{$m_{\Tilde{F}}=-1$} & \multicolumn{1}{c|}{$m_{\Tilde{F}}=0$} & \multicolumn{1}{c|}{$m_{\Tilde{F}}=1$} & \multicolumn{1}{c|}{$m_{\Tilde{F}}=2$} \\
         \hline
         $\Tilde{F}=1, m_{\Tilde{F}}=1$ & \textbf{0.2139} & \textbf{0.1121} & \textbf{0.0569} & 0 & \textbf{0.0532} & \textbf{0.0462} & \textbf{0.0388} & 0.0226 \\
         $\Tilde{F}=1, m_{\Tilde{F}}=0$ & \textbf{0.1540} & \textbf{0.1535} & \textbf{0.0982} & \textbf{0.0689} & 0.0145 & 0.0149 & 0.0289 & \textbf{0.0386*} \\
         $\Tilde{F}=1, m_{\Tilde{F}}=-1$ & \textbf{0.1114} & \textbf{0.1390} & \textbf{0.1679} & \textbf{0.0537} & \textbf{0.0504} & \textbf{0.0327} & 0.0200 & 0 \\
         $\Tilde{F}=2, m_{\Tilde{F}}=2$ & 0 & \textbf{0.1948} & \textbf{0.0988} & 0 & 0 & \textbf{0.1010} & \textbf{0.0838} & \textbf{0.0581*} \\
         $\Tilde{F}=2, m_{\Tilde{F}}=1$ & \textbf{0.1049} & \textbf{0.1211} & \textbf{0.1437} & 0 & \textbf{0.1112} & 0.0080 & \textbf{0.0571} & \textbf{0.0724*} \\
         $\Tilde{F}=2, m_{\Tilde{F}}=0$ & \textbf{0.1287} & \textbf{0.0579} & \textbf{0.1487} & \textbf{0.0796} & \textbf{0.0683} & \textbf{0.0678} & 0.0083 & \textbf{0.0727*} \\
         $\Tilde{F}=2, m_{\Tilde{F}}=-1$ & \textbf{0.1562} & \textbf{0.0312} & \textbf{0.1540} & \textbf{0.1009} & 0.0028 & \textbf{0.0548} & \textbf{0.0782} & 0 \\
         $\Tilde{F}=2, m_{\Tilde{F}}=-2$ & \textbf{0.1382} & \textbf{0.1634} & 0 & \textbf{0.1046} & \textbf{0.0813} & \textbf{0.0551} & 0 & 0 \\
         $\Tilde{F}=3, m_{\Tilde{F}}=3$ & 0 & 0 & \textbf{0.1352} & 0 & 0 & 0 & \textbf{0.2309} & 0.0036 \\
         $\Tilde{F}=3, m_{\Tilde{F}}=2$ & 0 & \textbf{0.0493} & \textbf{0.1054} & 0 & 0 & \textbf{0.1813} & \textbf{0.1359} & \textbf{0.0312*} \\
         $\Tilde{F}=3, m_{\Tilde{F}}=1$ & 0.0127 & \textbf{0.0622} & \textbf{0.0965} & 0 & \textbf{0.1145} & \textbf{0.1777} & \textbf{0.0690} & \textbf{0.0371*} \\
         $\Tilde{F}=3, m_{\Tilde{F}}=0$ & 0.0232 & \textbf{0.0807} & \textbf{0.0709} & \textbf{0.0543} & \textbf{0.1584} & \textbf{0.1544} & 0.0043 & \textbf{0.0324*} \\
         $\Tilde{F}=3, m_{\Tilde{F}}=-1$ & \textbf{0.0426} & \textbf{0.0819} & \textbf{0.0516} & \textbf{0.1016} & \textbf{0.1952} & \textbf{0.0756} & \textbf{0.0326} & 0 \\
         $\Tilde{F}=3, m_{\Tilde{F}}=-2$ & \textbf{0.0630} & \textbf{0.0843} & 0 & \textbf{0.1749} & \textbf{0.1545} & 0.0052 & 0 & 0 \\
         $\Tilde{F}=3, m_{\Tilde{F}}=-3$ & \textbf{0.1058} & 0 & 0 & \textbf{0.2189} & \textbf{0.0561} & 0 & 0 & 0 \\
         $\Tilde{F}=4, m_{\Tilde{F}}=4$ & 0 & 0 & 0 & 0 & 0 & 0 & 0 & \textbf{0.2676*} \\
         $\Tilde{F}=4, m_{\Tilde{F}}=3$ & 0 & 0 & \textbf{0.1015} & 0 & 0 & 0 & \textbf{0.0636} & \textbf{0.2189*} \\
         $\Tilde{F}=4, m_{\Tilde{F}}=2$ & 0 & \textbf{0.0503} & \textbf{0.0952} & 0 & 0 & 0.0011 & \textbf{0.1285} & \textbf{0.1943*} \\
         $\Tilde{F}=4, m_{\Tilde{F}}=1$ & 0.0176 & \textbf{0.0694} & \textbf{0.0945} & 0 & 0.0200 & \textbf{0.0472} & \textbf{0.1805} & \textbf{0.1298*} \\
         $\Tilde{F}=4, m_{\Tilde{F}}=0$ & \textbf{0.0330} & \textbf{0.0933} & \textbf{0.0721} & 0.0169 & 0.0049 & \textbf{0.1161} & \textbf{0.1703} & \textbf{0.0806*} \\
         $\Tilde{F}=4, m_{\Tilde{F}}=-1$ & \textbf{0.0593} & \textbf{0.0949} & \textbf{0.0532} & 0.0241 & \textbf{0.0385} & \textbf{0.1617} & \textbf{0.1450} & 0 \\
         $\Tilde{F}=4, m_{\Tilde{F}}=-2$ & \textbf{0.0831} & \textbf{0.0963} & 0 & 0.0243 & \textbf{0.1070} & \textbf{0.1996} & 0 & 0 \\
         $\Tilde{F}=4, m_{\Tilde{F}}=-3$ & \textbf{0.1323} & 0 & 0 & 0.0037 & \textbf{0.2326} & 0 & 0 & 0 \\
         $\Tilde{F}=4, m_{\Tilde{F}}=-4$ & 0 & 0 & 0 & \textbf{0.2676} & 0 & 0 & 0 & 0 \\
         \hline
    \end{tabular}
    \caption{Relative transition strength factors $\langle \Tilde{F}_{S}, m_{\Tilde{F},S}; k = 2, q = m_{\Tilde{F},D} - m_{\Tilde{F},S} \lvert \Tilde{F}_{D}, m_{\Tilde{F},D} \rangle$ for $6S_{1/2} \leftrightarrow 5D_{5/2}$ transitions with $\phi = \qty{45}{\degree}$ and $\gamma = \qty{58}{\degree}$ at a magnetic field strength of $\qty{8.35}{G}$. 
    Values higher than $0.03$ are in bold texts.
    The value of 0.03 is determined by empirically observing that the transitions weaker than this have large errors due to decoherence from the SPAM experiments in this work, which is also a value that is approximately an order of magnitude lower than the strongest transitions.
    *Transitions used in this work.}
    \label{tab:TransitionStrengths}
\end{table}

\begin{table}[H]
    \centering
    \small
    \renewcommand{\arraystretch}{1.2}
    \begin{tabular}{|c|c|c|c|c|c|c|c|c|c|c|c|c|c|}
    \hline
        \multirow{2}{*}{\thead{Prepared \\ state}} & \multicolumn{13}{c|}{Measured state} \\
        \cline{2-14}
         & $\lvert 0 \rangle$ & $\lvert 1 \rangle$ & $\lvert 2 \rangle$ & $\lvert 3 \rangle$ & $\lvert 4 \rangle$ & $\lvert 5 \rangle$ & $\lvert 6 \rangle$ & $\lvert 7 \rangle$ & $\lvert 8 \rangle$ & $\lvert 9 \rangle$ & $\lvert 10 \rangle$ & $\lvert 11 \rangle$ & $\lvert 12 \rangle$ \\
         \hline
         $\lvert 0 \rangle$ & 0.999 & 0.001 & 0 & 0 & 0 & 0 & 0 & 0 & 0 & 0 & 0 & 0 & 0 \\
         \cline{1-1}
         $\lvert 1 \rangle$ & 0.061 & 0.939 & 0 & 0 & 0 & 0 & 0 & 0 & 0 & 0 & 0 & 0 & 0 \\
         \cline{1-1}
         $\lvert 2 \rangle$ & 0.038 & 0.001 & 0.958 & 0.001 & 0 & 0 & 0.002 & 0 & 0 & 0 & 0 & 0 & 0 \\
         \cline{1-1}
         $\lvert 3 \rangle$ & 0.029 & 0.001 & 0 & 0.970 & 0 & 0 & 0 & 0 & 0 & 0 & 0 & 0 & 0 \\
         \cline{1-1}
         $\lvert 4 \rangle$ & 0.045 & 0 & 0 & 0 & 0.953 & 0 & 0 & 0 & 0 & 0 & 0 & 0 & 0 \\
         \cline{1-1}
         $\lvert 5 \rangle$ & 0.087 & 0.001 & 0 & 0 & 0 & 0.911 & 0 & 0 & 0.001 & 0 & 0 & 0 & 0 \\
         \cline{1-1}
         $\lvert 6 \rangle$ & 0.318 & 0.001 & 0.003 & 0.003 & 0.003 & 0.001 & 0.662 & 0.003 & 0.001 & 0.001 & 0 & 0.004 & 0.001 \\
         \cline{1-1}
         $\lvert 7 \rangle$ & 0.061 & 0 & 0.002 & 0 & 0 & 0.001 & 0.004 & 0.932 & 0 & 0 & 0 & 0 & 0 \\
         \cline{1-1}
         $\lvert 8 \rangle$ & 0.076 & 0 & 0 & 0.001 & 0 & 0 & 0.002 & 0 & 0.920 & 0 & 0 & 0.001 & 0 \\
         \cline{1-1}
         $\lvert 9 \rangle$ & 0.145 & 0.002 & 0 & 0.003 & 0.001 & 0.005 & 0.003 & 0.002 & 0.001 & 0.829 & 0.007 & 0 & 0.001 \\
         \cline{1-1}
         $\lvert 10 \rangle$ & 0.029 & 0.003 & 0.002 & 0 & 0.001 & 0.003 & 0.003 & 0.002 & 0 & 0.002 & 0.952 & 0 & 0.002 \\
         \cline{1-1}
         $\lvert 11 \rangle$ & 0.039 & 0.002 & 0.003 & 0.001 & 0.001 & 0.001 & 0.001 & 0.004 & 0 & 0.004 & 0 & 0.942 & 0.001 \\
         \cline{1-1}
         $\lvert 12 \rangle$ & 0.044 & 0.001 & 0 & 0 & 0 & 0 & 0 & 0 & 0 & 0.001 & 0 & 0 & 0.954 \\
         \hline
    \end{tabular}
    \caption{Post-selected measurement probability of each prepared state from SPAM experiments used to plot Fig. \ref{fig:Fidel3Dbar}. See Extended Data Table \ref{tab:SPAMDataRaw} for the raw measurement probabilities.}
    \label{tab:SPAMDataNormalized}
\end{table}

\begin{table}[H]
    \centering
    \small
    \renewcommand{\arraystretch}{1.2}
    \begin{tabular}{|c|c|c|c|c|c|c|c|c|c|c|c|c|c|c|}
    \hline
        \multirow{2}{*}{\thead{Prepared \\ state}} & \multicolumn{14}{c|}{Measured state} \\
        \cline{2-15}
         & $\lvert 0 \rangle$ & $\lvert 1 \rangle$ & $\lvert 2 \rangle$ & $\lvert 3 \rangle$ & $\lvert 4 \rangle$ & $\lvert 5 \rangle$ & $\lvert 6 \rangle$ & $\lvert 7 \rangle$ & $\lvert 8 \rangle$ & $\lvert 9 \rangle$ & $\lvert 10 \rangle$ & $\lvert 11 \rangle$ & $\lvert 12 \rangle$ & Null \\
         \hline
         $\lvert 0 \rangle$ & 0.999 & 0.001 & 0 & 0 & 0 & 0 & 0 & 0 & 0 & 0 & 0 & 0 & 0 & 0 \\
         \cline{1-1}
         $\lvert 1 \rangle$ & 0.059 & 0.911 & 0 & 0 & 0 & 0 & 0 & 0 & 0 & 0 & 0 & 0 & 0 & 0.030 \\
         \cline{1-1}
         $\lvert 2 \rangle$ & 0.037 & 0.001 & 0.941 & 0.001 & 0 & 0 & 0.002 & 0 & 0 & 0 & 0 & 0 & 0 & 0.018 \\
         \cline{1-1}
         $\lvert 3 \rangle$ & 0.029 & 0.001 & 0 & 0.960 & 0 & 0 & 0 & 0 & 0 & 0 & 0 & 0 & 0 & 0.010 \\
         \cline{1-1}
         $\lvert 4 \rangle$ & 0.044 & 0 & 0 & 0 & 0.926 & 0 & 0 & 0 & 0.002 & 0 & 0 & 0 & 0 & 0.028 \\
         \cline{1-1}
         $\lvert 5 \rangle$ & 0.080 & 0.001 & 0 & 0 & 0 & 0.841 & 0 & 0 & 0 & 0.001 & 0 & 0 & 0 & 0.077 \\
         \cline{1-1}
         $\lvert 6 \rangle$ & 0.252 & 0.001 & 0.002 & 0.002 & 0.002 & 0.001 & 0.525 & 0.002 & 0.001 & 0.001 & 0 & 0.003 & 0.001 & 0.207\\
         \cline{1-1}
         $\lvert 7 \rangle$ & 0.058 & 0 & 0.002 & 0 & 0 & 0.001 & 0.004 & 0.887 & 0 & 0 & 0 & 0 & 0 & 0.048 \\
         \cline{1-1}
         $\lvert 8 \rangle$ & 0.071 & 0 & 0 & 0.001 & 0 & 0 & 0.002 & 0 & 0.861 & 0 & 0 & 0.001 & 0 & 0.064\\
         \cline{1-1}
         $\lvert 9 \rangle$ & 0.126 & 0.002 & 0 & 0.003 & 0.001 & 0.004 & 0.003 & 0.002 & 0.001 & 0.722 & 0.006 & 0 & 0.001 & 0.129\\
         \cline{1-1}
         $\lvert 10 \rangle$ & 0.027 & 0.003 & 0.002 & 0 & 0.001 & 0.003 & 0.003 & 0.002 & 0 & 0.002 & 0.897 & 0 & 0.002 & 0.058\\
         \cline{1-1}
         $\lvert 11 \rangle$ & 0.037 & 0.002 & 0.003 & 0.001 & 0.001 & 0.001 & 0.001 & 0.004 & 0 & 0.004 & 0 & 0.897 & 0.001 & 0.049\\
         \cline{1-1}
         $\lvert 12 \rangle$ & 0.043 & 0.001 & 0 & 0 & 0 & 0 & 0 & 0 & 0 & 0.001 & 0 & 0 & 0.929 & 0.026\\
         \hline
    \end{tabular}
    \caption{Raw measurement probability of each prepared state from SPAM experiments. The sample size is $1000$. 
    Null indicates that all fluorescence readouts during the measurement procedure are dark and no valid computational state output is obtained.}
    \label{tab:SPAMDataRaw}
\end{table}

\begin{table}[H]
    \centering
    \begin{tabular}{|c|c|c|c|c|c|}
    \hline
         \thead{Computational \\ state} & Atomic State & SPAM error & \thead{Relative magnetic field \\ sensitivity / \qty{}{\mega \hertz \  G^{-1}}} & $\pi$-pulse time, $\tau_{\pi}$ / \qty{}{\micro \second} & \thead{Estimated single \\ transition error} \\
         \hline
         $\lvert 0 \rangle$ & $\lvert 6S_{1/2}, \Tilde{F} = 2, m_{\Tilde{F}} = 2 \rangle$ & $0.001$ & NA & NA & NA \\
         $\lvert 1 \rangle$ & $\lvert 5D_{5/2}, \Tilde{F} = 4, m_{\Tilde{F}} = 4 \rangle$ & $0.061$ & $2.7992$ & $37.2$ & $0.075$ \\
         $\lvert 2 \rangle$ & $\lvert 5D_{5/2}, \Tilde{F} = 4, m_{\Tilde{F}} = 3 \rangle$ & $0.042$ & $1.1202$ & $29.7$ & $0.016$ \\
         $\lvert 3 \rangle$ & $\lvert 5D_{5/2}, \Tilde{F} = 4, m_{\Tilde{F}} = 2 \rangle$ & $0.030$ & $-0.3554$ & $37.5$ & $0.023$ \\
         $\lvert 4 \rangle$ & $\lvert 5D_{5/2}, \Tilde{F} = 4, m_{\Tilde{F}} = 1 \rangle$ & $0.047$ & $-1.7009$ & $54.9$ & $0.030$ \\
         $\lvert 5 \rangle$ & $\lvert 5D_{5/2}, \Tilde{F} = 4, m_{\Tilde{F}} = 0 \rangle$ & $0.09$ & $-2.9234$ & $63.0$ & $0.07$ \\
         $\lvert 6 \rangle$ & $\lvert 5D_{5/2}, \Tilde{F} = 3, m_{\Tilde{F}} = 2 \rangle$ & $0.34$ & $1.7050$ & $204.8$ & $0.23$ \\
         $\lvert 7 \rangle$ & $\lvert 5D_{5/2}, \Tilde{F} = 3, m_{\Tilde{F}} = 1 \rangle$ & $0.068$ & $0.5718$ & $188.3$ & $0.050$ \\
         $\lvert 8 \rangle$ & $\lvert 5D_{5/2}, \Tilde{F} = 3, m_{\Tilde{F}} = 0 \rangle$ & $0.080$ & $-0.5672$ & $151.2$ & $0.049$ \\
         $\lvert 9 \rangle$ & $\lvert 5D_{5/2}, \Tilde{F} = 2, m_{\Tilde{F}} = 2 \rangle$ & $0.17$ & $2.0094$ & $104.5$ & $0.11$ \\
         $\lvert 10 \rangle$ & $\lvert 5D_{5/2}, \Tilde{F} = 2, m_{\Tilde{F}} = 1 \rangle$ & $0.047$ & $0.3026$ & $87.3$ & $0.030$ \\
         $\lvert 11 \rangle$ & $\lvert 5D_{5/2}, \Tilde{F} = 2, m_{\Tilde{F}} = 0 \rangle$ & $0.058$ & $-1.3707$ & $64.1$ & $0.026$ \\
         $\lvert 12 \rangle$ & $\lvert 5D_{5/2}, \Tilde{F} = 1, m_{\Tilde{F}} = 0 \rangle$ & $0.046$ & $-0.7373$ & $105.4$ & $0.054$ \\
         NA & $\lvert 5D_{5/2}, \Tilde{F} = 1, m_{\Tilde{F}} = 1 \rangle$ & NA & $1.9462$ & $215$ & $0.31$ \\
         NA & $\lvert 5D_{5/2}, \Tilde{F} = 3, m_{\Tilde{F}} = 3 \rangle$ & NA & $2.7988$ & $\sim2000$* & NA \\
         \hline
    \end{tabular}
    \caption{Table summarizing the encoded states and their corresponding physical atomic states, parameters relevant to the SPAM experiments and post-selected SPAM errors. See Supplementary Information for how the single transition errors are estimated. *This value is estimated from relative transition strengths using Table \ref{tab:TransitionStrengths}, and not measured empirically.}
    \label{tab:ErrorBFieldSensitivityRabiFreq}
\end{table}

\bibliography{main}

\begin{thebibliography}{10}
\urlstyle{rm}
\expandafter\ifx\csname url\endcsname\relax
  \def\url#1{\texttt{#1}}\fi
\expandafter\ifx\csname urlprefix\endcsname\relax\def\urlprefix{URL }\fi
\expandafter\ifx\csname doiprefix\endcsname\relax\def\doiprefix{DOI: }\fi
\providecommand{\bibinfo}[2]{#2}
\providecommand{\eprint}[2][]{\url{#2}}

\bibitem{Ringbauer2022}
\bibinfo{author}{Ringbauer, M.} \emph{et~al.}
\newblock \bibinfo{journal}{\bibinfo{title}{A universal qudit quantum processor
  with trapped ions}}.
\newblock {\emph{\JournalTitle{Nature Physics}}} \textbf{\bibinfo{volume}{18}},
  \bibinfo{pages}{1053--1057}, \doiprefix\url{10.1038/s41567-022-01658-0}
  (\bibinfo{year}{2022}).

\bibitem{Campbell-Anwar-Browne-2012}
\bibinfo{author}{Campbell, E.~T.}, \bibinfo{author}{Anwar, H.} \&
  \bibinfo{author}{Browne, D.~E.}
\newblock
  \bibinfo{journal}{\bibinfo{title}{\href{https://doi.org/10.1103/PhysRevX.2.041021}{Magic-state
  distillation in all prime dimensions using quantum Reed-Muller codes}}}.
\newblock {\emph{\JournalTitle{Phys. Rev. X}}} \textbf{\bibinfo{volume}{2}},
  \bibinfo{pages}{041021}, \doiprefix\url{10.1103/PhysRevX.2.041021}
  (\bibinfo{year}{2012}).

\bibitem{Campbell-2014}
\bibinfo{author}{Campbell, E.~T.}
\newblock
  \bibinfo{journal}{\bibinfo{title}{\href{https://doi.org/10.1103/PhysRevLett.113.230501}{Enhanced
  fault-tolerant quantum computing in d-Level systems}}}.
\newblock {\emph{\JournalTitle{Phys. Rev. Lett.}}}
  \textbf{\bibinfo{volume}{113}}, \bibinfo{pages}{230501},
  \doiprefix\url{10.1103/PhysRevLett.113.230501} (\bibinfo{year}{2014}).

\bibitem{Andrist-Wootton-Katzgraber-2015}
\bibinfo{author}{Andrist, R.~S.}, \bibinfo{author}{Wootton, J.~R.} \&
  \bibinfo{author}{Katzgraber, H.~G.}
\newblock
  \bibinfo{journal}{\bibinfo{title}{\href{https://doi.org/10.1103/PhysRevA.91.042331}{Error
  thresholds for Abelian quantum double models: Increasing the bit-flip
  stability of topological quantum memory}}}.
\newblock {\emph{\JournalTitle{Phys. Rev. A}}} \textbf{\bibinfo{volume}{91}},
  \bibinfo{pages}{042331}, \doiprefix\url{10.1103/PhysRevA.91.042331}
  (\bibinfo{year}{2015}).

\bibitem{Hutter2015}
\bibinfo{author}{Hutter, A.}, \bibinfo{author}{Loss, D.} \&
  \bibinfo{author}{Wootton, J.~R.}
\newblock \bibinfo{journal}{\bibinfo{title}{Improved hdrg decoders for qudit
  and non-abelian quantum error correction}}.
\newblock {\emph{\JournalTitle{New Journal of Physics}}}
  \textbf{\bibinfo{volume}{17}}, \bibinfo{pages}{035017},
  \doiprefix\url{10.1088/1367-2630/17/3/035017} (\bibinfo{year}{2015}).

\bibitem{Watson2015}
\bibinfo{author}{Watson, F. H.~E.}, \bibinfo{author}{Anwar, H.} \&
  \bibinfo{author}{Browne, D.~E.}
\newblock \bibinfo{journal}{\bibinfo{title}{Fast fault-tolerant decoder for
  qubit and qudit surface codes}}.
\newblock {\emph{\JournalTitle{Phys. Rev. A}}} \textbf{\bibinfo{volume}{92}},
  \bibinfo{pages}{032309}, \doiprefix\url{10.1103/PhysRevA.92.032309}
  (\bibinfo{year}{2015}).

\bibitem{Senko2015}
\bibinfo{author}{Senko, C.} \emph{et~al.}
\newblock \bibinfo{journal}{\bibinfo{title}{Realization of a quantum
  integer-spin chain with controllable interactions}}.
\newblock {\emph{\JournalTitle{Phys. Rev. X}}} \textbf{\bibinfo{volume}{5}},
  \bibinfo{pages}{021026}, \doiprefix\url{10.1103/PhysRevX.5.021026}
  (\bibinfo{year}{2015}).

\bibitem{Barbara2022}
\bibinfo{author}{et~al., B.~A.}
\newblock \bibinfo{journal}{\bibinfo{title}{Engineering an effective three-spin
  hamiltonian in trapped-ion systems for applications in quantum simulation}}.
\newblock {\emph{\JournalTitle{Quantum Sci. Technol. 7 034001}}}
  \doiprefix\url{10.1088/2058-9565/ac5f5b} (\bibinfo{year}{2022}).

\bibitem{Lanyon-et-al-2008}
\bibinfo{author}{Lanyon, B.~P.} \emph{et~al.}
\newblock
  \bibinfo{journal}{\bibinfo{title}{\href{https://doi.org/10.1038/nphys1150}{Simplifying
  quantum logic using higher-dimensional Hilbert spaces}}}.
\newblock {\emph{\JournalTitle{Nature Phys.}}} \textbf{\bibinfo{volume}{5}},
  \bibinfo{pages}{134--140}, \doiprefix\url{10.1038/nphys1150}
  (\bibinfo{year}{2008}).

\bibitem{Ralph-Resch-Gilchrist-2007}
\bibinfo{author}{Ralph, T.~C.}, \bibinfo{author}{Resch, K.~J.} \&
  \bibinfo{author}{Gilchrist, A.}
\newblock
  \bibinfo{journal}{\bibinfo{title}{\href{https://link.aps.org/doi/10.1103/PhysRevA.75.022313}{Efficient
  Toffoli gates using qudits}}}.
\newblock {\emph{\JournalTitle{Phys. Rev. A}}} \textbf{\bibinfo{volume}{75}},
  \bibinfo{pages}{022313}, \doiprefix\url{10.1103/PhysRevA.75.022313}
  (\bibinfo{year}{2007}).

\bibitem{Ladd2010}
\bibinfo{author}{Ladd, T.~D.} \emph{et~al.}
\newblock \bibinfo{journal}{\bibinfo{title}{Quantum computers}}.
\newblock {\emph{\JournalTitle{Nature}}} \textbf{\bibinfo{volume}{464}},
  \bibinfo{pages}{45--53}, \doiprefix\url{10.1038/nature08812}
  (\bibinfo{year}{2010}).

\bibitem{Bruzewicz2019}
\bibinfo{author}{Bruzewicz, C.~D.}, \bibinfo{author}{Chiaverini, J.},
  \bibinfo{author}{McConnell, R.} \& \bibinfo{author}{Sage, J.~M.}
\newblock \bibinfo{journal}{\bibinfo{title}{Trapped-ion quantum computing:
  Progress and challenges}}.
\newblock {\emph{\JournalTitle{Applied Physics Reviews}}}
  \textbf{\bibinfo{volume}{6}}, \bibinfo{pages}{021314},
  \doiprefix\url{10.1063/1.5088164} (\bibinfo{year}{2019}).
\newblock \eprint{https://doi.org/10.1063/1.5088164}.

\bibitem{Gaebler2016}
\bibinfo{author}{Gaebler, J.~P.} \emph{et~al.}
\newblock \bibinfo{journal}{\bibinfo{title}{High-fidelity universal gate set
  for ${^{9}\mathrm{Be}}^{+}$ ion qubits}}.
\newblock {\emph{\JournalTitle{Phys. Rev. Lett.}}}
  \textbf{\bibinfo{volume}{117}}, \bibinfo{pages}{060505},
  \doiprefix\url{10.1103/PhysRevLett.117.060505} (\bibinfo{year}{2016}).

\bibitem{Philips2022}
\bibinfo{author}{Philips, S. G.~J.} \emph{et~al.}
\newblock \bibinfo{journal}{\bibinfo{title}{Universal control of a six-qubit
  quantum processor in silicon}}.
\newblock {\emph{\JournalTitle{Nature}}} \textbf{\bibinfo{volume}{609}},
  \bibinfo{pages}{919--924}, \doiprefix\url{10.1038/s41586-022-05117-x}
  (\bibinfo{year}{2022}).

\bibitem{Kielpinski2002}
\bibinfo{author}{Kielpinski, D.}, \bibinfo{author}{Monroe, C.} \&
  \bibinfo{author}{Wineland, D.~J.}
\newblock \bibinfo{journal}{\bibinfo{title}{Architecture for a large-scale
  ion-trap quantum computer}}.
\newblock {\emph{\JournalTitle{Nature}}} \textbf{\bibinfo{volume}{417}},
  \bibinfo{pages}{709--711}, \doiprefix\url{10.1038/nature00784}
  (\bibinfo{year}{2002}).

\bibitem{Monroe2014}
\bibinfo{author}{Monroe, C.} \emph{et~al.}
\newblock \bibinfo{journal}{\bibinfo{title}{Large-scale modular
  quantum-computer architecture with atomic memory and photonic
  interconnects}}.
\newblock {\emph{\JournalTitle{Phys. Rev. A}}} \textbf{\bibinfo{volume}{89}},
  \bibinfo{pages}{022317}, \doiprefix\url{10.1103/PhysRevA.89.022317}
  (\bibinfo{year}{2014}).

\bibitem{Smith2013}
\bibinfo{author}{Smith, A.} \emph{et~al.}
\newblock \bibinfo{journal}{\bibinfo{title}{Quantum control in the cs
  $6{S}_{1/2}$ ground manifold using radio-frequency and microwave magnetic
  fields}}.
\newblock {\emph{\JournalTitle{Phys. Rev. Lett.}}}
  \textbf{\bibinfo{volume}{111}}, \bibinfo{pages}{170502},
  \doiprefix\url{10.1103/PhysRevLett.111.170502} (\bibinfo{year}{2013}).

\bibitem{Leupold2018}
\bibinfo{author}{Leupold, F.~M.} \emph{et~al.}
\newblock \bibinfo{journal}{\bibinfo{title}{Sustained state-independent quantum
  contextual correlations from a single ion}}.
\newblock {\emph{\JournalTitle{Phys. Rev. Lett.}}}
  \textbf{\bibinfo{volume}{120}}, \bibinfo{pages}{180401},
  \doiprefix\url{10.1103/PhysRevLett.120.180401} (\bibinfo{year}{2018}).

\bibitem{Malinowski2018}
\bibinfo{author}{Malinowski, M.} \emph{et~al.}
\newblock \bibinfo{journal}{\bibinfo{title}{Probing the limits of correlations
  in an indivisible quantum system}}.
\newblock {\emph{\JournalTitle{Phys. Rev. A}}} \textbf{\bibinfo{volume}{98}},
  \bibinfo{pages}{050102}, \doiprefix\url{10.1103/PhysRevA.98.050102}
  (\bibinfo{year}{2018}).

\bibitem{Hrmo2023}
\bibinfo{author}{Hrmo, P.} \emph{et~al.}
\newblock \bibinfo{journal}{\bibinfo{title}{Native qudit entanglement in a
  trapped ion quantum processor}}.
\newblock {\emph{\JournalTitle{Nature Communications}}}
  \textbf{\bibinfo{volume}{14}}, \bibinfo{pages}{2242},
  \doiprefix\url{10.1038/s41467-023-37375-2} (\bibinfo{year}{2023}).

\bibitem{Madej1990}
\bibinfo{author}{Madej, A.~A.} \& \bibinfo{author}{Sankey, J.~D.}
\newblock \bibinfo{journal}{\bibinfo{title}{Quantum jumps and the single
  trapped barium ion: Determination of collisional quenching rates for the
  5${d}^{2}$${D}_{5/2}$ level}}.
\newblock {\emph{\JournalTitle{Phys. Rev. A}}} \textbf{\bibinfo{volume}{41}},
  \bibinfo{pages}{2621--2630}, \doiprefix\url{10.1103/PhysRevA.41.2621}
  (\bibinfo{year}{1990}).

\bibitem{Silverans1986}
\bibinfo{author}{Silverans, R.~E.}, \bibinfo{author}{Borghs, G.},
  \bibinfo{author}{De~Bisschop, P.} \& \bibinfo{author}{Van~Hove, M.}
\newblock \bibinfo{journal}{\bibinfo{title}{Hyperfine structure of the 5d
  $^{2}\mathrm{D}_{\mathrm{j}}$ states in the alkaline-earth ba ion by
  fast-ion-beam laser-rf spectroscopy}}.
\newblock {\emph{\JournalTitle{Phys. Rev. A}}} \textbf{\bibinfo{volume}{33}},
  \bibinfo{pages}{2117--2120}, \doiprefix\url{10.1103/PhysRevA.33.2117}
  (\bibinfo{year}{1986}).

\bibitem{Berkeland2002}
\bibinfo{author}{Berkeland, D.~J.} \& \bibinfo{author}{Boshier, M.~G.}
\newblock \bibinfo{journal}{\bibinfo{title}{Destabilization of dark states and
  optical spectroscopy in zeeman-degenerate atomic systems}}.
\newblock {\emph{\JournalTitle{Phys. Rev. A}}} \textbf{\bibinfo{volume}{65}},
  \bibinfo{pages}{033413}, \doiprefix\url{10.1103/PhysRevA.65.033413}
  (\bibinfo{year}{2002}).

\bibitem{GIT}
\bibinfo{title}{Qudit 13level spam scripts and data}.
\newblock \bibinfo{howpublished}{\url{https://doi.org/10.5281/zenodo.8000676}}
  (\bibinfo{year}{2023}).

\bibitem{Ball2015}
\bibinfo{author}{Ball, H.} \& \bibinfo{author}{Biercuk, M.~J.}
\newblock \bibinfo{journal}{\bibinfo{title}{Walsh-synthesized noise filters for
  quantum logic}}.
\newblock {\emph{\JournalTitle{EPJ Quantum Technology}}}
  \textbf{\bibinfo{volume}{2}}, \bibinfo{pages}{11},
  \doiprefix\url{10.1140/epjqt/s40507-015-0022-4} (\bibinfo{year}{2015}).

\bibitem{Day2022}
\bibinfo{author}{Day, M.~L.}, \bibinfo{author}{Low, P.~J.},
  \bibinfo{author}{White, B.}, \bibinfo{author}{Islam, R.} \&
  \bibinfo{author}{Senko, C.}
\newblock \bibinfo{journal}{\bibinfo{title}{Limits on atomic qubit control from
  laser noise}}.
\newblock {\emph{\JournalTitle{npj Quantum Information}}}
  \textbf{\bibinfo{volume}{8}}, \bibinfo{pages}{72},
  \doiprefix\url{10.1038/s41534-022-00586-4} (\bibinfo{year}{2022}).

\bibitem{An2022}
\bibinfo{author}{An, F.~A.} \emph{et~al.}
\newblock \bibinfo{journal}{\bibinfo{title}{High fidelity state preparation and
  measurement of ion hyperfine qubits with ${I} > \frac{1}{2}$}}.
\newblock {\emph{\JournalTitle{Phys. Rev. Lett.}}}
  \textbf{\bibinfo{volume}{129}}, \bibinfo{pages}{130501},
  \doiprefix\url{10.1103/PhysRevLett.129.130501} (\bibinfo{year}{2022}).

\bibitem{Ruster2016}
\bibinfo{author}{Ruster, T.} \emph{et~al.}
\newblock \bibinfo{journal}{\bibinfo{title}{A long-lived zeeman trapped-ion
  qubit}}.
\newblock {\emph{\JournalTitle{Applied Physics B}}}
  \textbf{\bibinfo{volume}{122}}, \bibinfo{pages}{254},
  \doiprefix\url{10.1007/s00340-016-6527-4} (\bibinfo{year}{2016}).

\bibitem{Hu2022}
\bibinfo{author}{Hu, H.}, \bibinfo{author}{Xie, Y.} \& \bibinfo{author}{chao
  Zhang~et al., M.}
\newblock \bibinfo{journal}{\bibinfo{title}{Compensation of low-frequency noise
  induced by power line in trapped-ion system}}.
\newblock {\emph{\JournalTitle{PREPRINT (Version 1) available at Research
  Square}}} \doiprefix\url{10.21203/rs.3.rs-1608974/v1} (\bibinfo{year}{2022}).

\bibitem{Low2020}
\bibinfo{author}{Low, P.~J.}, \bibinfo{author}{White, B.~M.},
  \bibinfo{author}{Cox, A.~A.}, \bibinfo{author}{Day, M.~L.} \&
  \bibinfo{author}{Senko, C.}
\newblock \bibinfo{journal}{\bibinfo{title}{Practical trapped-ion protocols for
  universal qudit-based quantum computing}}.
\newblock {\emph{\JournalTitle{Phys. Rev. Res.}}} \textbf{\bibinfo{volume}{2}},
  \bibinfo{pages}{033128}, \doiprefix\url{10.1103/PhysRevResearch.2.033128}
  (\bibinfo{year}{2020}).

\bibitem{Blatt1982}
\bibinfo{author}{Blatt, R.} \& \bibinfo{author}{Werth, G.}
\newblock \bibinfo{journal}{\bibinfo{title}{Precision determination of the
  ground-state hyperfine splitting in $^{137}\mathrm{Ba}^{+}$ using the
  ion-storage technique}}.
\newblock {\emph{\JournalTitle{Phys. Rev. A}}} \textbf{\bibinfo{volume}{25}},
  \bibinfo{pages}{1476--1482}, \doiprefix\url{10.1103/PhysRevA.25.1476}
  (\bibinfo{year}{1982}).

\bibitem{Roos2000}
\bibinfo{author}{Roos, C.}
\newblock \emph{\bibinfo{title}{Controlling the quantum state of trapped
  ions}}.
\newblock \bibinfo{type}{{PhD} dissertation}, \bibinfo{school}{University of
  Innsbruck} (\bibinfo{year}{2000}).
\newblock \bibinfo{note}{Available at
  \url{https://www.quantumoptics.at/images/publications/dissertation/roos-diss.pdf}}.

\bibitem{Bramman2019}
\bibinfo{author}{Bramman, B.}
\newblock \emph{\bibinfo{title}{Measuring Trapped Ion Qudits}}.
\newblock \bibinfo{type}{{MSc} thesis}, \bibinfo{school}{University of
  Waterloo} (\bibinfo{year}{2019}).
\newblock \bibinfo{note}{Available at
  \url{https://uwspace.uwaterloo.ca/handle/10012/15165}}.

\end{thebibliography}


\begin{thebibliography}{1}

\bibitem{Day2022}
M.~L. Day, P.~J. Low, B.~White, R.~Islam, and C.~Senko, ``Limits on atomic
  qubit control from laser noise,'' {\em npj Quantum Information}, vol.~8,
  p.~72, Jun 2022.

\bibitem{Hu2022}
H.~Hu, Y.~Xie, and M.~chao Zhang~et al., ``Compensation of low-frequency noise
  induced by power line in trapped-ion system,'' {\em PREPRINT (Version 1)
  available at Research Square}, May 2022.

\bibitem{Gaebler2016}
J.~P. Gaebler, T.~R. Tan, Y.~Lin, Y.~Wan, R.~Bowler, A.~C. Keith, S.~Glancy,
  K.~Coakley, E.~Knill, D.~Leibfried, and D.~J. Wineland, ``High-fidelity
  universal gate set for ${^{9}\mathrm{Be}}^{+}$ ion qubits,'' {\em Phys. Rev.
  Lett.}, vol.~117, p.~060505, Aug 2016.

\end{thebibliography}

\section*{Acknowledgements}

We thank Yvette de Sereville for assisting with preliminary calibrations, and we thank Nicholas Zutt and Noah Greenberg for reviewing the manuscript.
This research was supported, in part, by the Natural Sciences and Engineering Research Council of Canada
(NSERC), RGPIN-2018-05253 and the Canada First Research Excellence Fund (CFREF) (Transformative Quantum
Technologies), CFREF-2015-00011. C.S. is also supported by a Canada Research Chair.

\section*{Author contributions statement}

P.J.L., B.W. and C.S. conceived the experiment(s), P.J.L. and B.W. conducted the experiment(s), P.J.L., B.W. and C.S. analyzed the results. All authors reviewed the manuscript.

\section*{Additional information}

The authors report no competing interests in the submission of this article. 

\end{document}

\preprint{APS/123-QED}

\title{Supplementary information for: Control and Readout of a 13-level Trapped Ion Qudit}

\author{Pei Jiang Low}
\author{Brendan White}
\author{Crystal Senko}
\affiliation{%
 Institute for Quantum Computing and Department of Physics and Astronomy, University of Waterloo, Waterloo, N2L 3R1, Canada\\
}%

\date{\today}% It is always \today, today,

\maketitle

\section{$6S_{1/2} \leftrightarrow 5D_{5/2}$ Relative Transition Strengths Estimation} \label{sec:SuppMat_RelativeTransitionStrength}

In the intermediate magnetic field regime, the energy eigenstates of the $5D_{5/2}$ level are not the $\lvert F, m_{F} \rangle$ states.
Thus, the relative transition strengths of the $\lvert 6S_{1/2}, \Tilde{F}=2, m_{\Tilde{F}}=2 \rangle$ state to the $\lvert 6S_{1/2}, \Tilde{F}, m_{\Tilde{F}} \rangle$ states cannot be evaluated using Wigner-Eckart theorem assuming $\lvert F, m_{F} \rangle$ states.
We work in the $\lvert m_I, m_J \rangle$ basis and express the energy eigenstates as
\begin{equation}
    \lvert \Tilde{F}, m_{\Tilde{F}} \rangle = \sum_{m_I,m_J} c_{m_I, m_J} \lvert m_I, m_J \rangle.
\end{equation}
The transition matrix element in a linearly polarized light is
\begin{equation}
    \begin{aligned}
    & g^{(q)}\left( \gamma, \phi \right) \langle \Tilde{F}_{D}, m_{\Tilde{F},D} \rvert \hat{Q}_{q=m_{\Tilde{F},D}-m_{\Tilde{F},S}} \lvert \Tilde{F}_{S}, m_{\Tilde{F},S} \rangle \\
    &= g^{(q)}\left( \gamma, \phi \right) \sum_{m_{I,D}, m_{J,D}} \sum_{m_{I,S}, m_{J,S}} c^*_{m_{I,D}, m_{J,D}} c_{m_{I,S}, m_{J,S}} \langle m_{I,D}, m_{J,D} \rvert \hat{Q}_{q=m_{\Tilde{F},D}-m_{\Tilde{F},S}} \lvert m_{I,S}, m_{J,S} \rangle.
    \end{aligned}
    \label{eq:SuppMat_SDTransMatrixElem_mImJ}
\end{equation}
Since the electric quadrupole operator $\hat{Q}_q$ only acts on the electronic component of the state, Equation \ref{eq:SuppMat_SDTransMatrixElem_mImJ} can be rewritten as
\begin{equation}
    \begin{aligned}
    &g^{(q)}\left( \gamma, \phi \right) \langle \Tilde{F}_{D}, m_{\Tilde{F},D} \rvert \hat{Q}_{q=m_{\Tilde{F},D}-m_{\Tilde{F},S}} \lvert \Tilde{F}_{S}, m_{\Tilde{F},S} \rangle \\
    &=  g^{(q)}\left( \gamma, \phi \right) \sum_{m_{I,D}, m_{J,D}} \sum_{m_{I,S}, m_{J,S}} c^*_{m_{I,D}, m_{J,D}} c_{m_{I,S}, m_{J,S}} \langle J_D = 5/2, m_{J,D} \rvert \hat{Q}_{q=m_{J,D}-m_{J,S}} \lvert J_S = 1/2, m_{J,S} \rangle \delta_{m_{I,S} m_{I,D}}
    \end{aligned}
\end{equation}
where $\delta_{m_{I,S} m_{I,D}}$ is the Kronecker delta function.
The factors $\langle J_D = 5/2, m_{J,D} \rvert \hat{Q}_{q=m_{J,D}-m_{J,S}} \lvert J_S = 1/2, m_{J,S} \rangle$ can be reduced using Wigner-Eckart theorem to
\begin{equation}
    \begin{aligned}
    &g^{(q)}\left( \gamma, \phi \right) \langle J_D = 5/2, m_{J,D} \rvert \hat{Q}_{q=m_{J,D}-m_{J,S}} \lvert J_S = 1/2, m_{J,S} \rangle \\
    &=g^{(q)}\left( \gamma, \phi \right) \langle J_S = 1/2, m_{J,S} = 1/2; k = 2, q=m_{J,D}-m_{J,S} \lvert J_D = 5/2, m_{J,D} \rangle \langle J_{D} \lvert \rvert \hat{Q} \rvert \lvert J_{S} \rangle
    \end{aligned}
    \label{eq:SuppMat_TransMatrixReduction}
\end{equation}
where $\langle J_S = 1/2, m_{J,S} = 1/2; k = 2, q=m_{J,D}-m_{J,S} \lvert J_D = 5/2, m_{J,D} \rangle$ is the Clebsch-Gordan coefficient for coupling $J_S = 1/2$ and $k$ to get $J_D = 5/2$ and $\langle J_{D} \lvert \rvert \hat{Q} \rvert \lvert J_{S} \rangle$ is the reduced transition matrix element.
The relative transition strength (pre-factor of $\langle J_{D} \lvert \rvert \hat{Q} \rvert \lvert J_{S} \rangle$) can then be calculated using
\begin{equation}
    \begin{aligned}
    &g^{(q)}\left( \gamma, \phi \right) \langle \Tilde{F}_{S}, m_{\Tilde{F},S}; k = 2, q = m_{\Tilde{F},D} - m_{\Tilde{F},S} \lvert \Tilde{F}_{D}, m_{\Tilde{F},D} \rangle \\
    &= g^{(q)}\left( \gamma, \phi \right) \times \\
    &\sum_{m_{I,D},m_{J,D}} \sum_{m_{I,S},m_{J,S}} c^*_{m_{I,D}, m_{J,D}} c_{m_{I,S}, m_{J,S}} \langle J_S = 1/2, m_{J,S}; k = 2, q=m_{J,D}-m_{J,S} \lvert J_D = 5/2, m_{J,D} \rangle \delta_{m_{I,S} m_{I,D}}
    \end{aligned}
    \label{eq:SupPMat_MatrixReductionFactorBreakdown}
\end{equation}
which is what is used to produce Table E1 in the main text.

\section{Estimation of $6S_{1/2} \leftrightarrow 5D_{5/2}$ Single $\pi$-pulse Transition Errors} \label{sec:SuppMat_PiPulseError}

To estimate the single $\pi$-pulse error of  $6S_{1/2} \leftrightarrow 5D_{5/2}$ transitions free from calibration error, Rabi flopping experiments are performed for each transition to the encoded states.
The Rabi flopping experimental procedure is as follows:
\begin{enumerate}
    \item The waveform frequency from the AWG for the $\qty{1762}{\nano\meter}$ laser EOM sideband is set to the resonant frequency of a chosen 2-state transition.
    \item The pulse sequence to prepare $^{137}Ba^+$ in the $\lvert 6S_{1/2}, \Tilde{F}=2, m_{\Tilde{F}}=2 \rangle$ state is sent to the ion (see main text).
    \item Once the $\lvert 6S_{1/2}, \Tilde{F}=2, m_{\Tilde{F}}=2 \rangle$ state is prepared, the \qty{1762}{\nano \meter} laser is turned on for some time $t$.
    \item The fluorescence lasers are turned on and the ion fluorescence is collected in the PMT for \qty{5}{\milli \second}.
    \item Steps 2-4 are repeated 100 to obtain a large enough sample size to determine the probability of the ion being dark.
    \item Step 5 is repeated for $t = \qty{0}{\micro \second}$ to $t = \qty{500}{\micro \second}$ in steps of \qty{1}{\micro \second}.
\end{enumerate}
The data points around the first transition probability peak of the Rabi flopping data are used to fit to the function as shown in Equation \ref{eq:SuppMat_RabiFlopEq},
\begin{equation}
    p \left( t \right) = A \cos^2{\left( \frac{\pi \left( t - t_{peak} \right)}{2t_{scale}} \right)} + C
    \label{eq:SuppMat_RabiFlopEq}
\end{equation}
where $A$, $C$, $t_{peak}$ and $t_{scale}$ are scalar parameters that are allowed to vary to fit the function to the data points.
Specifically, data points from $t = t_{peak,r}/2$ to $t = 3t_{peak,r}/2$, where $t_{peak,r}$ is the time where the data point has the highest measured transition probability in the first Rabi flopping peak, are selected for functional fitting.
From the functional fit, the single $\pi$-pulse transition error is estimated to be $\epsilon_{\pi} = 1 - A - C$.

For small $\epsilon_{\pi}$, $\epsilon_{SPAM} \approx \epsilon_{\pi}$ from the main text.
By comparing $\epsilon_{SPAM}$ with the empirically measured $\epsilon_{\pi}$, for $\epsilon_{\pi}<0.1$, $\epsilon_{SPAM}$ is larger than $\epsilon_{\pi}$ by $1.5 \pm 2 \%$ on average, which we attribute to be errors due to parameter drifts.

\begin{comment}
\begin{figure}
    \centering
    \includegraphics{Figures/SuppMat_RabiFloppingFitF4m1_20221205.pdf}
    \caption{A representative plot of a Rabi flopping experiment for estimating the singple $\pi$-pulse transition error. The data shown is the transition to the $\lvert 5D_{5/2}, \Tilde{F}=4, m_{\Tilde{F}} = 1 \rangle$ state. Circles denote the experimental data points, and the red (color online) solid line is the functional fit using Equation \ref{eq:SuppMat_RabiFlopEq}.}
    \label{fig:my_label}
\end{figure}
\end{comment}

\section{Derivation of the Theoretical Form of Magnetic Field Noise Error} \label{sec:SuppMat_BFieldErrorDerivation}

The error for a single $\pi$-pulse transition can be expressed as
\begin{equation}
    \epsilon_{\pi} = \frac{1}{2} \left( 1 - e^{-\chi} \right),
    \label{eq:SuppMat_Error_General}
\end{equation}
and $\chi$ is
\begin{equation}
    \chi = \frac{1}{\pi} \int_0^{\infty} \frac{1}{\omega^2} S\left( \omega \right) F \left( \omega \right) d\omega,
    \label{eq:SuppMat_Chi_General}
\end{equation}
where $S\left( \omega \right)$ is the power spectral density (PSD) of the transition-frequency noise of the 2-state transition, $F \left( \omega \right)$ is the filter function of the applied operation and $\omega$ is the angular frequency in the Fourier space.

We model the PSD of the magnetic field noise, which we equate to the transition-frequency noise, to be a $1/f$-noise with a prominent peak at the mains electricity alternating current (AC) frequency and a baseline white noise,
\begin{equation}
    S \left( \omega \right) = \begin{cases} h_a/\omega_0, & \omega<\omega_0 \\ h_{peak}, & \omega_{AC} - \Delta \omega_{AC}/2 < \omega < \omega_{AC} + \Delta \omega_{AC}/2 \\ h_a/\omega + h_b, & \mathrm{Otherwise} \end{cases},
    \label{eq:SuppMat_FreqNoisePSDFunctionalForm}
\end{equation}
where $h_a$ is the scaling coefficient of the $1/f$-noise component, $h_b$ is the baseline white noise, $h_{peak}$ is the noise PSD peak at the mains electricity frequency, $\omega_0$ is some threshold Fourier frequency such that noise PSD does not go to infinity at low Fourier frequencies to keep the model physical, $\omega_{AC}$ is the mains electricity frequency and $\Delta \omega_{AC}$ is the width of the mains electricity noise peak.
For a $\pi$-pulse transition, the filter function has the form \cite{Day2022}
\begin{equation}
    F \left( \omega \right) = \begin{cases} 4\frac{\omega^2}{\Omega^2}, & \omega < \Omega \\ 4, & \omega \geq \Omega \end{cases}
    \label{eq:SuppMat_FilterFunctionForm}
\end{equation}
Assuming that $\omega_{AC} < \Omega$, evaluating Equation \ref{eq:SuppMat_Chi_General} with Equations \ref{eq:SuppMat_FreqNoisePSDFunctionalForm} and \ref{eq:SuppMat_FilterFunctionForm} gives
\begin{equation}
    \begin{aligned}
    \chi &= \frac{4}{\pi} \left( \frac{3h_a}{2\Omega^2} + \frac{h_a}{\Omega^2} \ln{ \left( \frac{\Omega}{\omega_0} \right)} + \frac{h_b}{\Omega^2} \left( \Omega - \omega_0 \right) - \frac{h_a}{\Omega^2} \ln{ \left( \frac{\omega_{AC} + \Delta \omega_{AC}/2}{\omega_{AC} - \Delta \omega_{AC}/2} \right)} - \frac{h_b}{\Omega^2} \Delta \omega_{AC} + \frac{h_b}{\Omega} + \frac{h_{peak}}{\Omega^2} \Delta \omega_{AC} \right) \\
    &\approx \frac{4}{\pi\Omega^2} \left( \frac{3h_a}{2} + h_a \ln{ \left( \frac{\Omega}{\omega_0} \right)} + h_{peak} \Delta \omega_{AC} \right) + \frac{8h_b}{\pi\Omega},
    \end{aligned}
    \label{eq:SuppMat_ChiDerivedForm}
\end{equation}
where we have made the approximations $\omega_0 \ll \Omega$, $\Delta \omega_{AC} \ll \omega_{AC}$ and $h_b \ll \Omega^2 \Delta \omega_{AC}$.
From Equation \ref{eq:SuppMat_ChiDerivedForm}, there are 2 asymptotic behavior.
In the regime where $\Omega \ll \frac{K}{2h_b}$, where $K = \frac{3h_a}{2} + h_a \ln{ \left( \frac{\Omega}{\omega_0} \right)} + h_{peak} \Delta \omega_{AC}$, $\chi \propto \frac{1}{\Omega^2}$.
In the regime where $\Omega \gg \frac{K}{2h_b}$, $\chi \propto \frac{1}{\Omega}$.
Since it is commonly reported that the mains electricity AC noise is a dominant error source \cite{Hu2022}, it is reasonable to assume that we are in the $\Omega \ll \frac{K}{2h_b}$ regime in this work as well.

In the time domain, assuming that the fluctuation in the magnetic field strength is sufficiently small, the transition frequency shift of 2 energy states can be estimated to be linear, i.e.
\begin{equation}
    \Delta \nu \left( t \right) \propto \kappa B_e \left( t \right),
    \label{eq:SuppMat_FreqNoiseTimeDomain}
\end{equation}
where $\Delta \nu \left( t \right)$ denotes the transition frequency shift, $\kappa$ is the linear magnetic field sensitivity and $B_e$ is the magnetic field strength.
This implies that the power spectral density of the transition frequency shift, which is the transition-frequency noise to be proportional to $\kappa^2$, i.e.
\begin{equation}
    S\left( \omega \right) = \left| \Delta \hat{\nu} \left( \omega \right) \right|^2 \propto \kappa^2.
    \label{eq:SuppMat_FreqNoisePowerSpectralDensity}
\end{equation}

With $\tau_{\pi} \propto 1/\Omega$, Equations \ref{eq:SuppMat_ChiDerivedForm} and \ref{eq:SuppMat_FreqNoisePowerSpectralDensity}, in the $\Omega \ll \frac{K}{2h_b}$ regime, we have
\begin{equation}
    \chi \propto \kappa^2\tau_{\pi}^2,
    \label{eq:SuppMat_Chi_Propto}
\end{equation}
which is what we have in the main text.

\section{Error Estimations of Other Known Error Sources}

The longest time a state stays shelved in $5D_{5/2}$ level during the SPAM process in this work is around $\qty{120}{\milli \second}$.
This leads to an estimated upper bound error of around $e^{0.12/35} \approx 0.3\%$ due to spontaneous emission from the $5D_{5/2}$ level.
The transitions with transition frequencies closest to each other are the transitions to the $\lvert 5D_{5/2}, \Tilde{F}=4, m_{\Tilde{F}}=2 \rangle$ and $\lvert 5D_{5/2}, \Tilde{F}=3, m_{\Tilde{F}}=0 \rangle$ states, which are separated by around $\Delta = \qty{475}{\kilo \hertz}$.
With $6S_{1/2} \leftrightarrow 5D_{5/2}$ transition Rabi frequencies of around $\Omega = \qty{10}{\kilo \hertz}$, the estimated error due to off-resonant transition is around $\frac{\Omega^2}{\Omega^2 + \Delta^2} = 0.04\%$.
The photon count threshold to determine bright or dark ion states is set to 11 counts for the SPAM data set in this work.
The average photon count of the data points below this threshold, which we estimate to be the dark state average photon count, is $0.651 \pm 0.002$.
The average photon count of the data points above this threshold, which we estimate to be the bright state average photon count, is $27.87 \pm 0.05$.
Assuming Poisson distributions for dark and bright state photon counts, an upper bound of $0.025 \%$ for bright/dark state discrimination error.

\section{Avenues for Reduction of SPAM Time}

The measurement time of around \qty{100}{\milli \second} is relatively long for typical trapped ion quantum operations \cite{Gaebler2016}.
The main contributors to the measurement time are the AWG trigger and fluorescence steps.
The AWG trigger time is an artificially limitation from the method we employed in this work for switching the \qty{1762}{\nano \meter} laser EOM frequency from the AWG (see Methods in the main text).
To speed up this process, in principle, it is possible to program and generate a single arbitrary waveform which changes in frequency according to the timings of the deshelving steps. 
This bypasses the \qty{4}{\milli \second} AWG external signal trigger step in this work. 
The fluorescence time in this work is set to be comparable to the AWG signal trigger times to maximize collected photon counts of bright ions without significant compromise to the overall measurement time.
If the \qty{1762}{\nano \meter} laser frequency switching time is reduced, the fluorescence time can be reduced as well, as long as it does not significantly impact the error from distinguishing bright and dark states.

With the measurement time reduced, the main contributor to the SPAM experimental time could be the state preparation procedure, which is dominated by the repump time.
The repump time is long due to the dipole transition from the $\lvert 5D_{5/2}, \Tilde{F} = 4, m_{\Tilde{F}} = 4 \rangle$ state to the $\lvert 6P_{3/2}, F = 2\rangle$ state being forbidden.
Thus, to access the $6P_{3/2}$ level from the $\lvert 5D_{5/2}, \Tilde{F} = 4, m_{\Tilde{F}} = 4 \rangle$ state, it has to go to the $\lvert 6P_{3/2}, F = 3\rangle$ state, which is roughly \qty{400}{\mega \hertz} detuned, and therefore requiring a long transition time.
The transition frequencies of the rest of the encoded states to the $\lvert 6P_{3/2}, F = 2\rangle$ state are only detuned in the order of \qty{10}{\mega \hertz}, and a repump time in the order of \qty{100}{\micro \second} is sufficient.
To reduce the repump time to the order of \qty{100}{\micro \second}, an EOM can be set up to generate an additional frequency that is resonant to the $\lvert 5D_{5/2}, \Tilde{F} = 4, m_{\Tilde{F}} = 4 \rangle \leftrightarrow \lvert 6P_{3/2}, F = 3\rangle$ transition, and further reduction in repump time is possible with more effort towards increasing \qty{614}{\nano \meter} laser intensity at the ion.
It is also interesting to note that although $\lvert F=4 \rangle \leftrightarrow \lvert F=2 \rangle$ is dipole forbidden, $\lvert \Tilde{F}=4 \rangle \leftrightarrow \lvert F=2 \rangle$ is not in general, as $\lvert \Tilde{F}=4 \rangle$ states are superposition states of $F$ hyperfine states.

\section{Transition Frequency Calibration Parameters}
\label{sec:SuppMat_TransFreqCalParameters}

From the main text, transition frequencies are calibrated using the Equation
\begin{equation}
    f_n = a_{n,1} \Delta f + f_{offset} + a_{n,2}
    \label{eq:SuppMat_TransitionFreqCal}
\end{equation}
To determine the parameters $a_{n,1}$ and $a_{n,2}$, the transition frequency of each transition to the encoded $5D_{5/2}$ states are determined empirically.
The experimental procedure to determine a transition frequency for the $\qty{1762}{\nano\meter}$ laser is as follows:
\begin{enumerate}
    \item The waveform amplitude from the AWG for the $\qty{1762}{\nano\meter}$ laser EOM sideband is lowered to the point where the magnitude of power broadening is in the order of \qty{1}{\kilo \hertz}, so that we can resolve the resonant frequencies to a resolution of \qty{1}{\kilo \hertz}.
    \item The pulse sequence to prepare $^{137}Ba^+$ in the $\lvert 6S_{1/2}, \Tilde{F}=2, m_{\Tilde{F}}=2 \rangle$ state is sent to the ion (see main text).
    \item Once the $\lvert 6S_{1/2}, \Tilde{F}=2, m_{\Tilde{F}}=2 \rangle$ state is prepared, the \qty{1762}{\nano \meter} laser is turned on for a time that is longer than the coherence time in our system. We set it to \qty{3}{\milli \second} in this work.
    \item The fluorescence lasers are turned on and the ion fluorescence is collected in the PMT for \qty{5}{\milli \second}. If the \qty{1762}{\nano \meter} laser frequency is resonant, there is around $0.5$ probability of the ion not fluorescing.
    \item Steps 2-4 are repeated 400 to obtain a large enough sample size to determine the probability of the ion being dark.
    \item Steps 5 is repeated for the EOM frequencies from \qty{-50}{\kilo \hertz} to \qty{+50}{\kilo \hertz} from the previously determined resonant frequency in steps of \qty{10}{\kilo \hertz}, assuming that the frequencies have not drifted more than $\pm \qty{50}{\kilo \hertz}$.
    \item The AWG frequency where the average PMT count is the lowest from Step 6 is determined. Fine frequency scan in steps of \qty{1}{\kilo \hertz} is scanned from \qty{-10}{\kilo \hertz} to \qty{+10}{\kilo \hertz} of this frequency.
    \item From the fine frequency scan, the transition probability against EOM frequency data is fit with a Lorentzian function, and the centre of the Lorentzian function is set as the resonant frequency.
\end{enumerate}

We rely on natural drifts of our setup and collect sets of empirically determined $6S_{1/2} \leftrightarrow 5D_{5/2}$ transitions on different days.
With a large enough data sets, linear regression is performed for each transition frequency, $f_n$, to obtain the parameters $a_{n,1}$ and $a_{n,2}$.

\begin{comment}
\begin{figure}
    \centering
    \begin{subfigure}{0.45\linewidth}
        \includegraphics[width=\linewidth]{Figures/SuppMat_DetermineResonantFreqF4m1.pdf}
        \caption{}
        \label{fig:SuppMat_DetermineResonantFreq}
    \end{subfigure}
    \begin{subfigure}{0.45\linewidth}
        \includegraphics[width=\linewidth]{Figures/SuppMat_FreqCalF4m1.pdf}
        \caption{}
        \label{fig:SuppMat_FreqCal}
    \end{subfigure}
    \caption{(a) A representative plot of transition probability against EOM frequency for the transition to $\lvert 5D_{5/2}, \Tilde{F}=4, m_{\Tilde{F}}=1 \rangle$. The data points are fit to a Lorentzian function to determine the resonant frequency of a $6S_{1/2} \leftrightarrow 5D_{5/2}$ transition. (b) A representative plot of shifted transition frequencies against $\Delta f$ for the transition to $\lvert 5D_{5/2}, \Tilde{F}=4, m_{\Tilde{F}}=1 \rangle$. Each data point is an empirically determined transition frequency on different days or different times. Linear regression is performed to determine the parameters $a_{n,1}$ and $a_{n,2}$.}
    \label{fig:SuppMat_FreqCalPlots}
\end{figure}
\end{comment}

\section{Efficient $\pi$-pulse Time Calibration}
The resonant Rabi frequency of a quadrupole transition between 2 states, $\lvert i \rangle$ in the $5D_{5/2}$ level and $\lvert j \rangle$ in the $6S_{1/2}$ level, can be written as
\begin{equation}
    \begin{aligned}
        \Omega_{ij} &= g^{(q)}\left( \gamma, \phi \right) \frac{\langle i \rvert \hat{Q}_q \lvert j \rangle}{\hbar} \\
        &= g^{(q)}\left( \gamma, \phi \right) K_{ij,q} \frac{\langle J_D \rvert \rvert \hat{Q} \lvert \lvert J_S \rangle}{\hbar}
    \end{aligned}
\end{equation}
where $K_{ij,q}$ is a proportionality constant obtained from reducing the transition matrix element as shown in Equations \ref{eq:SuppMat_TransMatrixReduction} and \ref{eq:SupPMat_MatrixReductionFactorBreakdown}.
The ratio of the Rabi frequencies between two transitions with the same $\Delta m = q$ is then a constant value of
\begin{equation}
    R_{i'j',ij} = \frac{\Omega_{i'j'}}{\Omega_{ij}} = \frac{K_{i'j',q}}{K_{ij,q}}
\end{equation}
which is independent of the laser parameters.
The resonant Rabi frequencies and thus the $R_{i'j',ij}$ ratios can be empirically measured in the laboratory.
During the calibration process to determine $\pi$-pulse times, for each initial state $\lvert j \rangle$ and each $\Delta m = q$, only Rabi frequency measurement of one transition, let it be $\Omega_{ij}$, needs to be performed.
Then, the Rabi frequencies of the rest of the transitions with the same $\Delta m = q$ can be computed using
\begin{equation}
    \Omega_{i'j'} = R_{i'j',ij} \Omega_{ij}
\end{equation}
For a quadrupole transition, $q \in \{-2,-1,0,1,2\}$.
Thus, at most Rabi frequencies of five transitions, one for each $q$ value, need to be measured empirically during the calibration process, regardless of the qudit dimension $d$.

The scheme discussed here assumes that the actual Rabi frequencies are sufficiently close to the resonant Rabi frequencies, which does not necessarily hold true in an experimental setup.
With some laser frequency detuning $\delta$ from resonance, the actual Rabi frequency is
\begin{equation}
    \Omega' = \sqrt{\Omega^2 + \delta^2}
\end{equation}
In this work, the resolution of the laser frequency is restricted to \qty{1}{\kilo \hertz} due to the limitation of the AWG memory size, and the weakest transition used in this work has a Rabi frequency of \qty{2.5}{\kilo \hertz}.
Thus, the assumption $\Omega' \approx \Omega$ does not hold true consistently, and this $\pi$-pulse calibration scheme is not used in this work, as mentioned in the main text.

\section{Alternative Interpretation of SPAM Error} \label{sec:SuppMat_AltSPAMError}

In the SPAM experiment performed in this work, during the measurement process, after checking for fluorescence of the $\lvert 0 \rangle$ state, each encoded state in the $5D_{5/2}$ level, $\lvert n \rangle$, is de-shelved by sending a $\pi$-pulse corresponding to the $\lvert 6S_{1/2}, \Tilde{F} = 2, m_{\Tilde{F}} = 2 \rangle \leftrightarrow \lvert n \rangle$ transition.
During the de-shelving $\pi$-pulse transition, what is effectively performed on the ion is a re-shelving process that is happening simultaneously with the de-shelving process.
Thus, there is an alternative method to interpret the measurement fidelity that we initially employed, that turns out to be less robust to ultimately what is used in the main text.
In the main text, the measured state is set to be the first instance the ion is bright during the measurement sequence, and disregarding whether the ion is bright or dark for the rest of the sequence in the measurement process.
The alternative interpretation treats the cases where the ion is still detected to be bright after the first to be failures of the measurement process, as the ion is supposed to be re-shelved after the first bright event and no other bright events are supposed to be detected.
This places an unnecessarily stricter condition for a successful measurement process, and is not discussed in the main text.
Table \ref{tab:SuppMat_SPAMDataRawAlt} summarizes the SPAM fidelity for each prepared state with the same data set presented in the main text, but with the alternative interpretation of a successful measurement event as presented in this section.
The overall fidelity is lower than the interpretation used in the main text, as expected.
Table \ref{tab:SuppMat_SPAMDataNormedAlt} is the post-selected fidelity of Table \ref{tab:SuppMat_SPAMDataRawAlt}.
Qualitatively, the post-selected fidelities do not differ from the interpretation employed in the main text.

\begin{table}[H]
    \centering
    \begin{tabular}{|c|c|c|c|c|c|c|c|c|c|c|c|c|c|c|}
    \hline
        \multirow{2}{*}{Prepared state} & \multicolumn{14}{c|}{Measured state} \\
        \cline{2-15}
         & $\lvert 0 \rangle$ & $\lvert 1 \rangle$ & $\lvert 2 \rangle$ & $\lvert 3 \rangle$ & $\lvert 4 \rangle$ & $\lvert 5 \rangle$ & $\lvert 6 \rangle$ & $\lvert 7 \rangle$ & $\lvert 8 \rangle$ & $\lvert 9 \rangle$ & $\lvert 10 \rangle$ & $\lvert 11 \rangle$ & $\lvert 12 \rangle$ & Null \\
         \hline
         $\lvert 0 \rangle$ & 0.951 & 0 & 0 & 0 & 0 & 0 & 0 & 0 & 0 & 0 & 0 & 0 & 0 & 0.049 \\
         \cline{1-1}
         $\lvert 1 \rangle$ & 0.059 & 0.787 & 0 & 0 & 0 & 0 & 0 & 0 & 0 & 0 & 0 & 0 & 0 & 0.154 \\
         \cline{1-1}
         $\lvert 2 \rangle$ & 0.042 & 0.002 & 0.922 & 0 & 0 & 0 & 0.001 & 0 & 0 & 0 & 0.001 & 0 & 0 & 0.032 \\
         \cline{1-1}
         $\lvert 3 \rangle$ & 0.025 & 0.001 & 0 & 0.916 & 0.001 & 0 & 0 & 0.001 & 0 & 0.002 & 0 & 0 & 0 & 0.054 \\
         \cline{1-1}
         $\lvert 4 \rangle$ & 0.031 & 0 & 0 & 0 & 0.852 & 0 & 0 & 0.001 & 0 & 0 & 0 & 0.001 & 0 & 0.115 \\
         \cline{1-1}
         $\lvert 5 \rangle$ & 0.103 & 0 & 0 & 0 & 0 & 0.585 & 0 & 0 & 0 & 0 & 0 & 0 & 0 & 0.312 \\
         \cline{1-1}
         $\lvert 6 \rangle$ & 0.220 & 0 & 0.001 & 0.002 & 0.002 & 0.002 & 0.537 & 0.001 & 0.002 & 0.001 & 0 & 0.001 & 0 & 0.231 \\
         \cline{1-1}
         $\lvert 7 \rangle$ & 0.057 & 0.001 & 0 & 0.001 & 0.001 & 0.001 & 0 & 0.861 & 0 & 0 & 0.001 & 0 & 0 & 0.077 \\
         \cline{1-1}
         $\lvert 8 \rangle$ & 0.040 & 0 & 0.002 & 0.002 & 0.001 & 0 & 0.003 & 0.002 & 0.803 & 0 & 0 & 0 & 0.001 & 0.146 \\
         \cline{1-1}
         $\lvert 9 \rangle$ & 0.147 & 0.004 & 0.008 & 0.003 & 0 & 0.003 & 0.003 & 0.004 & 0.005 & 0.645 & 0.005 & 0.004 & 0.001 & 0.168 \\
         \cline{1-1}
         $\lvert 10 \rangle$ & 0.027 & 0.001 & 0 & 0.002 & 0 & 0.002 & 0.002 & 0.003 & 0.001 & 0.002 & 0.870 & 0 & 0 & 0.090 \\
         \cline{1-1}
         $\lvert 11 \rangle$ & 0.029 & 0.003 & 0.003 & 0.001 & 0.002 & 0.001 & 0.001 & 0.001 & 0 & 0.003 & 0.004 & 0.867 & 0.001 & 0.084 \\
         \cline{1-1}
         $\lvert 12 \rangle$ & 0.030 & 0.001 & 0 & 0 & 0.001 & 0 & 0.001 & 0 & 0 & 0.001 & 0 & 0 & 0.938 & 0.028\\
         \hline
    \end{tabular}
    \caption{Raw measurement probability of each prepared state from the SPAM data set in the main text, but with the alternative interpretation of a successful measurement. The sample size is $1000$. Null indicates not only one bright fluorescence readout during the measurement procedure is obtained.}
    \label{tab:SuppMat_SPAMDataRawAlt}
\end{table}

\begin{table}[H]
    \centering
    \begin{tabular}{|c|c|c|c|c|c|c|c|c|c|c|c|c|c|}
    \hline
        \multirow{2}{*}{Prepared state} & \multicolumn{13}{c|}{Measured state} \\
        \cline{2-14}
         & $\lvert 0 \rangle$ & $\lvert 1 \rangle$ & $\lvert 2 \rangle$ & $\lvert 3 \rangle$ & $\lvert 4 \rangle$ & $\lvert 5 \rangle$ & $\lvert 6 \rangle$ & $\lvert 7 \rangle$ & $\lvert 8 \rangle$ & $\lvert 9 \rangle$ & $\lvert 10 \rangle$ & $\lvert 11 \rangle$ & $\lvert 12 \rangle$ \\
         \hline
         $\lvert 0 \rangle$ & 1 & 0 & 0 & 0 & 0 & 0 & 0 & 0 & 0 & 0 & 0 & 0 & 0 \\
         \cline{1-1}
         $\lvert 1 \rangle$ & 0.070 & 0.930 & 0 & 0 & 0 & 0 & 0 & 0 & 0 & 0 & 0 & 0 & 0 \\
         \cline{1-1}
         $\lvert 2 \rangle$ & 0.043 & 0.002 & 0.952 & 0 & 0 & 0 & 0.001 & 0 & 0 & 0 & 0.001 & 0 & 0 \\
         \cline{1-1}
         $\lvert 3 \rangle$ & 0.026 & 0.001 & 0 & 0.968 & 0.001 & 0 & 0 & 0.001 & 0 & 0.002 & 0 & 0 & 0 \\
         \cline{1-1}
         $\lvert 4 \rangle$ & 0.035 & 0 & 0 & 0 & 0.963 & 0 & 0 & 0.001 & 0 & 0 & 0 & 0.001 & 0 \\
         \cline{1-1}
         $\lvert 5 \rangle$ & 0.150 & 0 & 0 & 0 & 0 & 0.850 & 0 & 0 & 0 & 0 & 0 & 0 & 0 \\
         \cline{1-1}
         $\lvert 6 \rangle$ & 0.286 & 0 & 0.001 & 0.003 & 0.003 & 0.003 & 0.698 & 0.001 & 0.003 & 0.001 & 0 & 0.001 & 0 \\
         \cline{1-1}
         $\lvert 7 \rangle$ & 0.062 & 0.001 & 0 & 0.001 & 0.001 & 0.001 & 0 & 0.933 & 0 & 0 & 0.001 & 0 & 0 \\
         \cline{1-1}
         $\lvert 8 \rangle$ & 0.047 & 0 & 0.002 & 0.002 & 0.001 & 0 & 0.004 & 0.002 & 0.940 & 0 & 0 & 0 & 0.001 \\
         \cline{1-1}
         $\lvert 9 \rangle$ & 0.177 & 0.005 & 0.010 & 0.004 & 0 & 0.004 & 0.004 & 0.005 & 0.006 & 0.775 & 0.006 & 0.005 & 0.001 \\
         \cline{1-1}
         $\lvert 10 \rangle$ & 0.030 & 0.001 & 0 & 0.002 & 0 & 0.002 & 0.002 & 0.003 & 0.001 & 0.002 & 0.956 & 0 & 0 \\
         \cline{1-1}
         $\lvert 11 \rangle$ & 0.032 & 0.003 & 0.003 & 0.001 & 0.002 & 0.001 & 0.001 & 0.001 & 0 & 0.003 & 0.004 & 0.947 & 0.001 \\
         \cline{1-1}
         $\lvert 12 \rangle$ & 0.031 & 0.001 & 0 & 0 & 0.001 & 0 & 0.001 & 0 & 0 & 0.001 & 0 & 0 & 0.965 \\
         \hline
    \end{tabular}
    \caption{Post-selected measurement probability of each prepared state from the SPAM data set in the main text, but with the alternative interpretation of a successful measurement.}
    \label{tab:SuppMat_SPAMDataNormedAlt}
\end{table}

\bibliographystyle{ieeetr}
\bibliography{supplementary}% Produces the bibliography via BibTeX.